\documentclass[%
 reprint,
 amsmath,amssymb,
 aps,
prb,
]{revtex4-1}

\usepackage{graphicx}% Include figure files
\usepackage{dcolumn}% Align table columns on decimal point
\usepackage{bm}% bold math
\usepackage{hyperref}% add hypertext capabilities
\usepackage[caption=false]{subfig} % enable subfiure
\usepackage{placeins}
\DeclareMathOperator{\Tr}{Tr}

\begin{document}

% \preprint{APS/123-QED}

\title{Tensor network renormalization:  application to dynamic correlation  functions and non-hermitian systems}

\author{Ying-Jie Wei}
\affiliation{
Department of Physics, The Chinese University of Hong Kong, Shatin, New Territories, Hong Kong, China
} 
\author{Zheng-Cheng Gu}
\email{zcgu@phy.cuhk.edu.hk}
\affiliation{
 Department of Physics, The Chinese University of Hong Kong, Shatin, New Territories, Hong Kong, China
}

\date{\today}% It is always \today, today,

%%%%%%%%%%%%%%%% Abstract
\begin{abstract}
In recent years, tensor network renormalization (TNR) has emerged as an efficient and accurate method for studying  (1+1)D quantum systems or 2D classical systems using real-space renormalization group (RG) techniques. One notable application of TNR is its ability to extract central charge and conformal scaling dimensions for critical systems. In this paper, we present the implementation of the Loop-TNR algorithm, which allows for the computation of dynamical correlation functions. Our algorithm goes beyond traditional approaches by not only calculating correlations in the spatial direction, where the separation is an integer, but also in the temporal direction, where the time difference can contain decimal values. Our algorithm is designed to handle both imaginary-time and real-time correlations, utilizing a tensor network representation constructed from a path-integral formalism. Additionally, we highlight that the Loop-TNR algorithm can also be applied to investigate critical properties of non-Hermitian systems, an area that was previously inaccessible using density matrix renormalization group(DMRG) and matrix product state(MPS) based algorithms.
\end{abstract}

\maketitle

%\tableofcontents
%%%%%%%%%%%%%%%%%%%%%%%% introduction % the version polished by chat-gpt 3.5
\section{Introduction}
In the past two decades, tensor network methods have become a powerful tool for investigating strongly correlated many-body systems.
It is well known that the matrix product state (MPS)\cite{Fannes92,Klumper91,Klumper93,Ostlund95,Rommer97,White92_1,White92_2,Orus14} has been widely used as an efficient and faithful representation for ground states of (1+1)D gapped quantum systems with local Hamiltonians.
MPS methods also allow for straightforward computation of physical measurements, including (equal-time) two-point correlations.
However, MPS with finite bond dimensions cannot accurately represent gapless states \cite{Schuch08,Eisert10,Schollwock05,Orus14}.

% On the other hand, real-time correlations can also be computed, by sandwiching operators with the evolved state.
On the other hand, real-time correlations can also be computed, where the real-time evolution of MPS is involved.
For example, one scheme applies the time evolution operator $e^{-iHt}$ to an underlying MPS, with $e^{-iHt}$ decomposed using Suzuki-Trotter expansion\cite{Suzuki76_1,Suzuki76_2}.
Specifically, time-evolving block decimation (TEBD)\cite{Vidal04,Vidal07,Daley04}, which truncates the evolved state based on singular value decomposition (SVD), and the time-dependent density matrix renormalization group (tDMRG)\cite{white04,Karrasch12,Bruognolo15}, which updates the evolved state using the variational principle, fall under this scheme. Another class of methods for time evolution is the tangent-space methods\cite{Laurens19,vanhecke21}, which project the evolved state onto the manifold of MPS. The time-dependent variational principle (TDVP)\cite{Haegeman11,Haegeman16} and the variational uniform matrix product state (VUMPS)\cite{Stauber18} follow these ideas. There are also other algorithms for time evolution, such as Krylov-based methods\cite{garcia06,wall15}, which approximate the state using linear combinations of Krylov vectors and methods based on Chebyshev expansions\cite{Weisse06,Xie18}.

Nevertheless, there is an intrinsic difficulty in real-time evolution: 
% even for gapped quantum systems (in thermodynamic limit) whose ground state can be efficiently described by an MPS, the entanglement entropy of the evolved state through real-time evolution still grows linearly with the evolution time, i.e., $ S(t) \propto t $. 
the entanglement entropy of a (1+1)D quantum state grows linearly with the time evolved.
Moreover, linear growth of the entanglement entropy requires an exponential increase of bond dimension of MPS to describe the evolved state\cite{Schuch08,Schuch08_2,Ba23}, making MPS calculations inefficient. This is the notorious entanglement barrier problem.
On the other hand, for a finite-size system, the entanglement entropy grows to a constant value over a sufficiently long time\cite{Pasquale05}. This enables MPS-based methods to simulate real-time evolution in limited cases, such as finite system size and high-temperature cases\cite{Daley04,Bruognolo15,Huang11,Karrasch12}. % needs modification
In recent years, attempts have been made to circumvent or overcome the problem of entanglement barriers. These include evolving operators in the Heisenberg picture\cite{Hartmann09}, the folding method\cite{Banuls09,Hermes12}, and biorthonormal transfer matrix DMRG (BTMRG)\cite{Huang11}.

In contrast to the MPS-based method for describing quantum states, the tensor network renormalization (TNR) based algrithms\cite{Levin07,Gu08,Zcgu09,Xie12,Evenbly15,Ys17,Evenbly17} offer a direct approach to deal with partition functions in $1+1$D. This means that physical measurements, such as correlation functions at equal times, can be derived straightforwardly using the ensemble method. The simplest TNR algorithm, specifically the tensor renormalization group (TRG)\cite{Levin07}, has already been presented for spatial separations with $r = 2^N$\cite{Gu08,Nakamoto16,Kadoh19,Judah14}, where $N$ is a nonnegative integer. However, it is important to note that TRG also faces challenges in accurately calculating correlation functions for critical systems. This is primarily due to significant truncation errors caused by its local truncation scheme and its inability to filter out short-range entanglement, as pointed out in Ref.\cite{Zcgu09}. 
%As a result, the accuracy of correlation function calculations for critical systems is compromised. 
%To address these limitations, further advancements and refinements in the TNR-based method are necessary. By developing improved techniques to mitigate truncation errors and effectively account for short-range entanglement, we can enhance the accuracy of correlation function calculations, particularly for critical systems.
%It should be noted that the tensor network representation constructed from the path-integral formalism is suitable not only for TRG-based methods like Loop-TNR, TNR\cite{Evenbly15,Evenbly17}, and Higher Order TRG (HOTRG)\cite{Xie12} but also for other methods, including VUMPS\cite{Fishman18,Li20}, infinite-TEBD (iTEBD)\cite{Orus08}, and Corner Transfer Matrix (CTM) methods\cite{Orus09}.

Recently, a groundbreaking advancement in the TNR algorithm called Loop-TNR has emerged\cite{Ys17}. This algorithm focuses on minimizing truncation errors within a small patch and effectively eliminates short-range entanglement, making it particularly suitable for investigating critical models. As a result, the application of loop-TNR to compute correlation functions for critical systems shows great potential for accurately extracting critical exponents. This paper begins with an overview of the algorithm for computing spatial correlation functions with fractions $r = M \times 2^N$, where $M$ is an odd and positive integer. Both the TRG algorithm and the loop-TNR algorithm are introduced to illustrate the computational process.
Furthermore, we extend our efforts to compute imaginary-time correlation functions. The time difference, denoted as $\tau$, can generally be expressed as $\tau = r + k \epsilon$, where $r$ represents the integer part of the separation and $0 \le k \epsilon < 1$ represents the decimal part. Finally, to overcome the problem of the entanglement barrier, we also employ the path-integral formalism for real-time correlation. 
%This approach proves to be advantageous for the application of Loop-TNR. 
Our method enables the computation of real-time correlation functions in the thermodynamic limit, even for critical systems.
Furthermore, we stress that Loop-TNR exhibits its superiority in studying non-hermitian systems by also allowing for the computation of conformal data. This is an aspect that cannot be accessed through the MPS/DMRG-based method, further highlighting the advantages of Loop-TNR in the field of non-Hermitian critical system analysis.
%This paper provides a detailed introduction to the method of extracting the central charge, scaling dimension, and conformal spin of a critical system using Loop-TNR. Specifically, such method is applied in the study of non-Hermitian models, which corresponds to the non-unitary conformal field theory (CFT).

The remainder of this paper is organized as follows. 
In Section~\ref{sec:TN_rep}, we introduce the method to construct the tensor network representation for physical measurements, which is the starting point of our TNR calculations. 
In Section~\ref{sec:iniT}, we present the way to initialize the tensors a priori to the application of Loop-TNR, where the impurity tenor is introduced. 
The computing of a single-body operator is also introduced. 
In Section~\ref{sec:two_body}, we illustrate the algorithms for two-body correlations where the separation is along the spatial direction. 
In Section~\ref{sec:corre_imag}, the algorithm for imaginary time correlation is introduced. 
In Section~\ref{sec:corre_real}, the algorithm is generalized to real-time correlation, which is based on path-integral formalism. 
In Section~\ref{sec:conformal}, we present the method to extract conformal data from the fixed point tensor obtained by Loop-TNR in detail.
%In Section~\ref{sec:results}, the results of correlations obtained by the algorithms illustrated above are displayed. Additionally, 
Finally, we applied Loop-TNR in the study of a non-Hermitian system and computed the scaling dimensions and topological spins for Yang-Lee edge singularity. 
In Appendix~\ref{sec:comp}, we provide a detailed explanation of compressing two layers of tensors into one layer.  
In Appendix~\ref{sec:App_VUMPS}, we provide a brief introduction to the VUMPS algorithm and demonstrate its application in computing correlation functions.
%%%%%%%%%%%%%%%%%%%%%%%%%%%%%%%%%%%%%%%%%%%%%

%%%%%%%%%%%%%%%%%%%%%%%%%%%% algorithm
\section{Tensor-network representation for physical observables}
\label{sec:TN_rep}
We start with the computation of a general physical observable $O$:
\begin{equation}
\langle O \rangle = \frac{1}{ \mathcal{Z} } \Tr \left( O e^{ -\beta H } \right),
\label{equ:obser}
\end{equation}
where $H$ is the Hamiltonian of the system and $\mathcal{Z}$ is the partition function, defined as:
\begin{equation} 
\mathcal{Z} = \Tr \left( e^{ -\beta H } \right)
\label{equ:partition}
\end{equation}
For simplicity, we assume that $H$ contains nearest neighbour interactions only, $H= \sum_{i} h_{i,i+1} $. As a result, the terms in Hamiltonian can be regrouped as follows:
\begin{equation}
\begin{aligned}
H = H_{ \mathrm{even} } + H_{\mathrm{odd} } = \sum_{i \in \mathrm{even} \, \mathbb{Z}} h_{i,i+1} + \sum_{i \in \mathrm{odd} \, \mathbb{Z}} h_{i,i+1}
\end{aligned}
\label{equ:split_gates}
\end{equation}
Note that terms within even or odd part commute. We can then %(apply Suzuki-Trotter expansion to  )
decompose the evolution operator as:
\begin{equation}
\begin{aligned}
\mathcal{Z} &= \Tr ( e^{-\delta \tau (H_{\mathrm{even} } + H_{\mathrm{odd} } )} )^{M} \\
            &\approx \Tr ( e^{ -\delta \tau H_{\mathrm{even} } } e^{ -\delta \tau H_{\mathrm{odd} } } )^{M} \\
            &=\Tr (\prod_{i\in \mathrm{even} \, \mathbb{Z}} e^{ -\delta \tau h_{i,i+1} } \prod_{i\in \mathrm{odd} \, \mathbb{Z}} e^{ -\delta \tau h_{i,i+1} } )^{M}
\end{aligned}
\label{equ:Trotter_expan}
\end{equation}
where $\delta \tau = \frac{\beta}{M} $. 
For the second line in Eq.~(\ref{equ:Trotter_expan}), we apply Suzuki-Trotter expansion \cite{Suzuki76_1,Suzuki76_2},  and an error of order $\mathcal{O}( \delta \tau^{2} )$ is introduced. We usually choose a small $\delta \tau$ or large $M$ (for a fixed $\beta$) to reduce the error.

\begin{figure}[hbt]
  \centering
  \includegraphics[width=\linewidth]{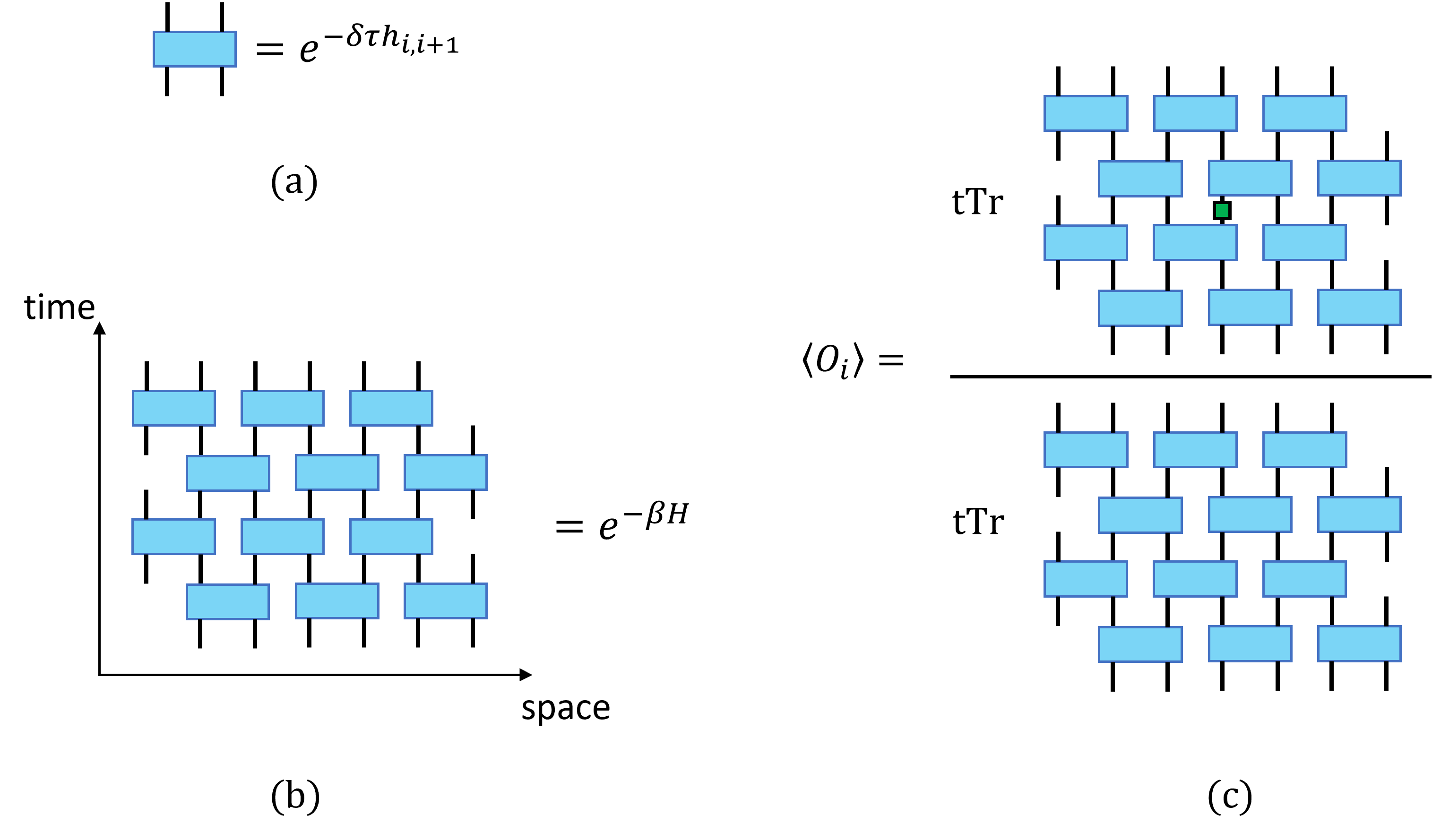} % check
    \caption{ (a) Rank-4 tensor defined in Eq.~(\ref{equ:tensor_gate}) as a two-body gate. (b) Tensor network representation for the partition function in Eq.~(\ref{equ:Trotter_expan}), which consists of only two-body gates. (c) The tensor network representation of Eq.~(\ref{equ:obser}). For the numerator, the single-body operator is represented as a green square. }
  \label{fig:iniTensor}
\end{figure}

The two-body gate $e^{ -\delta \tau h_{i,i+1} } $ in Eq.~(\ref{equ:Trotter_expan}):
\begin{equation}
T_{ \sigma_{i},\sigma_{i+1},\sigma'_{i},\sigma'_{i+1} } = \langle \sigma'_{i}, \sigma'_{i+1} | e^{ -\delta \tau h_{i,i+1} } | \sigma_{i}, \sigma_{i+1} \rangle
\label{equ:tensor_gate}
\end{equation}
can be expressed as a rank-4 tensor under a complete set of local basis $\{ \sigma_{i} \} $. See Fig.~\ref{fig:iniTensor}(a) for a graphical representation.
%% the exponential of e^A
%Note that the matrix exponential of the form $ e^{ A } $ in Eq.~(\ref{equ:tensor_gate}) can be calculated by performing an eigen-decomposition of the matrix $A$, $A= X \lambda X^{-1} $, where $X$ is the eigenvector and $\lambda$ is the eigenvalue (which is a diagonal matrix) of $A$. Therefore the matrix exponential can be readily written as $e^{A} = X e^{\lambda} X^{-1} $. 
% The equation can be verified by performing Taylor expansion on both sides and compare each term.
After we define the two-body gate as a tensor, the partition function in Eq.~(\ref{equ:Trotter_expan}) can be represented as the trace of a tensor network, see Fig.~\ref{fig:iniTensor}(b).

For the numerator of Eq.~(\ref{equ:obser}), after Suzuki-Trotter expansion, we have,
\begin{equation}
\begin{aligned}
\Tr \left( O e^{ -\beta H } \right)=& \Tr \left[ O_{i} \left( \prod_{j} e^{ -\epsilon h_{j,j+1} } \right)^{M} \right] \\
=& \Tr \left[ \left( \prod_{j<i-1} e^{ -\epsilon h_{j,j+1} } O_{i} \prod_{j \ge i-1} e^{ -\epsilon h_{j,j+1} } \right) \times \right. \\
& \left. \left( \prod_{j} e^{ -\epsilon h_{j,j+1} } \right)^{M-1} \right]
\end{aligned}
\label{equ:numerator}
\end{equation}
For simplicity, we assume $O$ is a local operator $O_{i} $ at site $i$. 
Accordingly, a tensor network representation for the numerator in Eq.~(\ref{equ:obser}) is also obtained, as shown in Fig.~\ref{fig:iniTensor}(c), where the single-body operator is represented by a green square.

So far, we have obtained the tensor network representations for both numerator and denominator in Eq.~(\ref{equ:obser}). As a result, the physical measurement $\langle O\rangle$ is expressed as the ratio of two tensor network traces, as shown in Fig.~\ref{fig:iniTensor}(c). 
 % should be modified
%\FloatBarrier

However, a straightforward contraction of the entire tensor network in Fig.~\ref{fig:iniTensor}(c) is exponentially hard. In this paper, we will perform a real-space RG scheme to approximate the original tensor network contraction and reduce the computational cost to polynomial time.

\begin{figure}[htb]
  \centering
  \includegraphics[width=\linewidth]{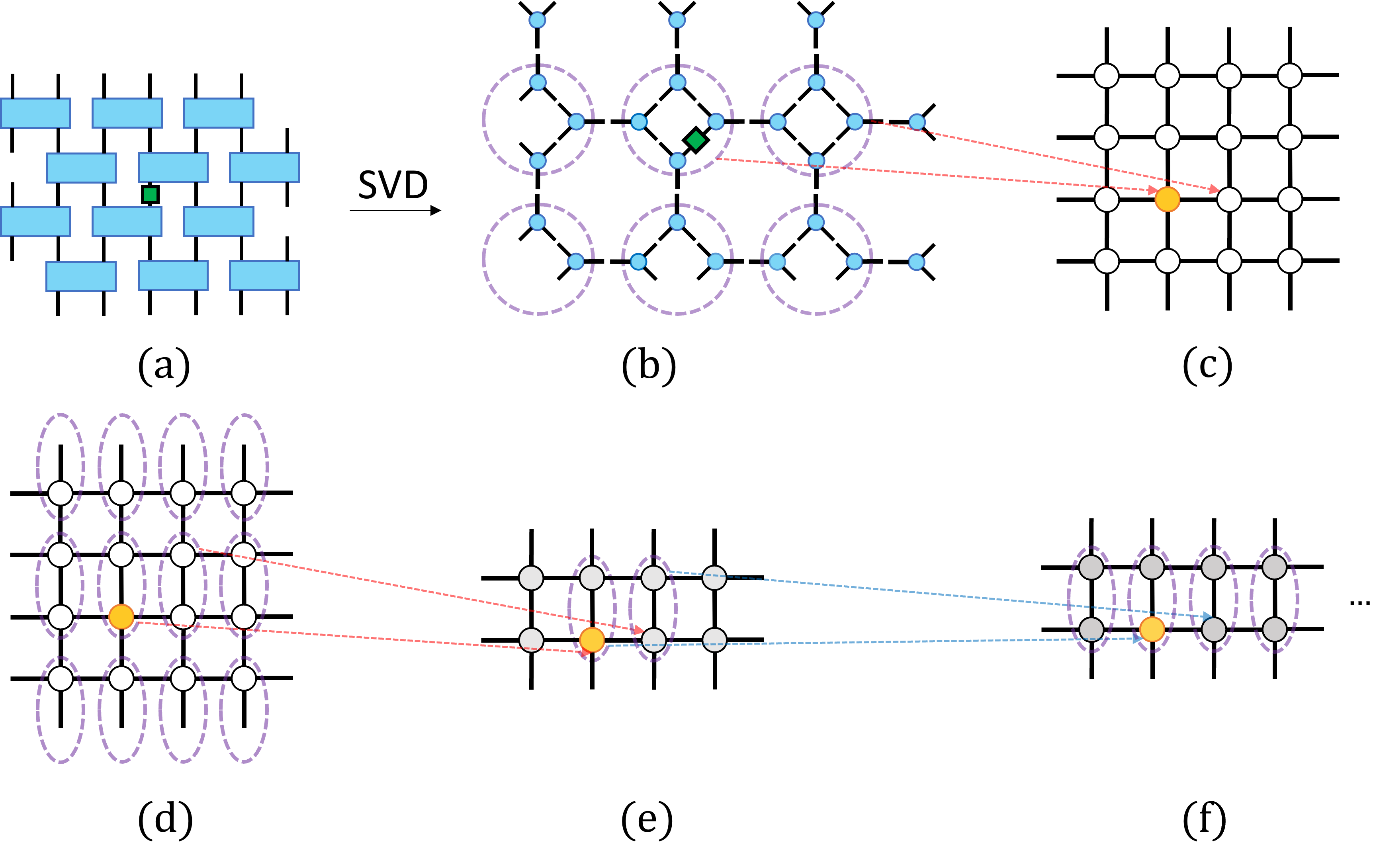}
    \caption{Tensor initialization to make tensors isotropic. (a) to (c) A TRG step is performed on the original tensor network in the numerator of Fig.~\ref{fig:iniTensor}(c). For the denominator, we just remove the green square and apply TRG. After the TRG step, the impurity tensor is produced and the tensor network is rotated by 45 degrees. (d) to (f) Compression step is performed iteratively until the tensor is isotropic.  }
  \label{fig:TRG_comp}
\end{figure}

\begin{figure}[htb]
  \centering
  \includegraphics[width=\linewidth]{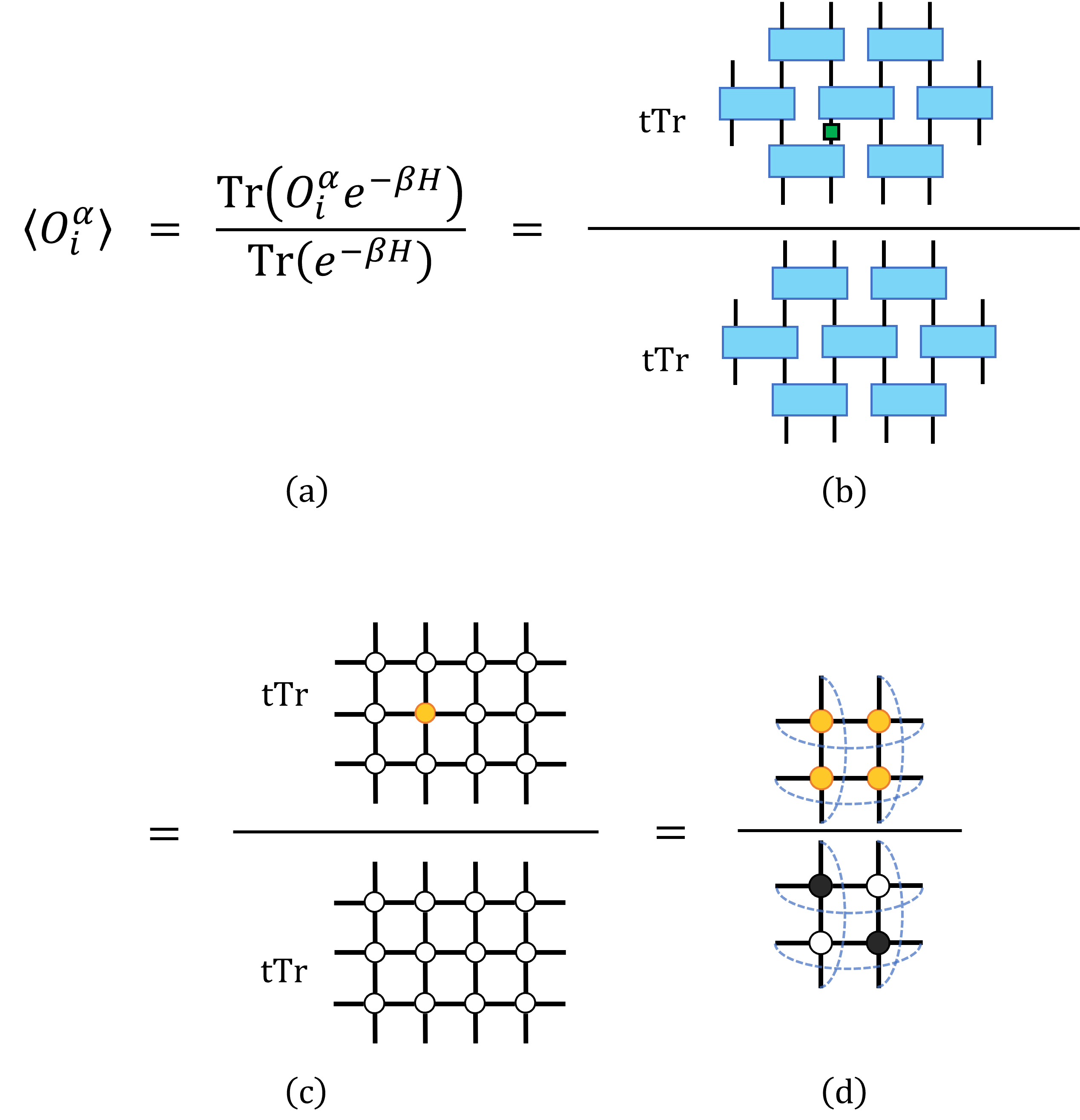}
    \caption{ (a) Mathematical expression of a single-body operator at site $i$. (b) The tensor network representation of the single-body operator. The green square represents the operator $O_{i}^{\alpha}. $ (c) The tensor network that consists of isotropic tensors after the steps introduced in Section~\ref{sec:iniT}. (d) After performing enough TRG or Loop-TNR steps, the original tensor network is reduced to a 2-by-2 one, the trace of which is direct to compute. For the denominator, if we perform TRG, the black and white circles both represent the same kind of uniform tensor. For Loop-TNR, they represent different uniform tensors. }
  \label{fig:sin_op}
\end{figure}

\section{Initialization of tensors with a single-body operator}
\label{sec:iniT}
% anisotropy of the tensor
It turns out that the tensor network representations constructed in the last section are not appropriate to be applied with real-space based RG algorithms directly. This is because we have to choose a small $\delta \tau$ in the Suzuki-Trotter expansion to reduce the error. However, the small $\delta \tau$ makes the tensor anisotropic, and it is not suitable for real-space RG scheme which is isotropic. 
% in the sense that, the effective lengths each tensor represent differ a lot in space and (imaginary) time directions, respectively (the anisotropy can also be seen by the spectrum of singular values in horizontal and vertical directions of the two-body gate by SVD). %% more check is needed for this part TRG or Loop-TNR is a fully isotropic RG scheme. 
The anisotropy would make the tensor have inconsistent truncation errors in different directions.

Let us first introduce the simplest case, which is the computation of a single-body operator. 
%We first construct a tensor network representation of it (see Fig.~\ref{fig:sin_op}(b)). 
%Then we initialize the tensors so that the tensors are isotropic and ready for the TRG or Loop-TNR steps.
To make the tensors isotropic, we first need to perform one TRG step (without truncation of the bond dimensions) to rotate the tensor network by 45 degrees. After that, the 'impurity' tensor comes out, which is represented by an orange circle, as shown in Fig.~\ref{fig:TRG_comp}(c). For the tensors that are represented by white circles, they are referred as 'uniform tensors'.

Next, we perform compressions of the tensors along the time (vertical) direction, which is shown in Fig.~\ref{fig:TRG_comp}(d)-(f). The method is the same as iMPS-based method introduced in Ref.\cite{wang11}(see Appendix.~\ref{sec:comp} for full details). After we compress two layers of tensors into one, the effective length that a tensor represents in the time direction doubles. Such operations are performed until the effective length in time direction becomes of order one for each tensor.
%(the isotropy of a tensor is determined by the modular parameter $\tau$, see Section~\ref{sec:conformal} for more details).
% (see Section~\ref{sec:conformal} for a detailed illustration about the choice of initial $\beta$ in Eq.~(\ref{equ:partition})). 
%As a result, the tensors are isotropic and ready to be applied with TRG or Loop-TNR.

%\FloatBarrier

% (\textbf{Yingjie, you need to explain more for the compress step, e.g., using the standard $iMPS$ algorithm, and add citation accordingly.})

\begin{figure}[tb]
  \centering
  \includegraphics[width=\linewidth]{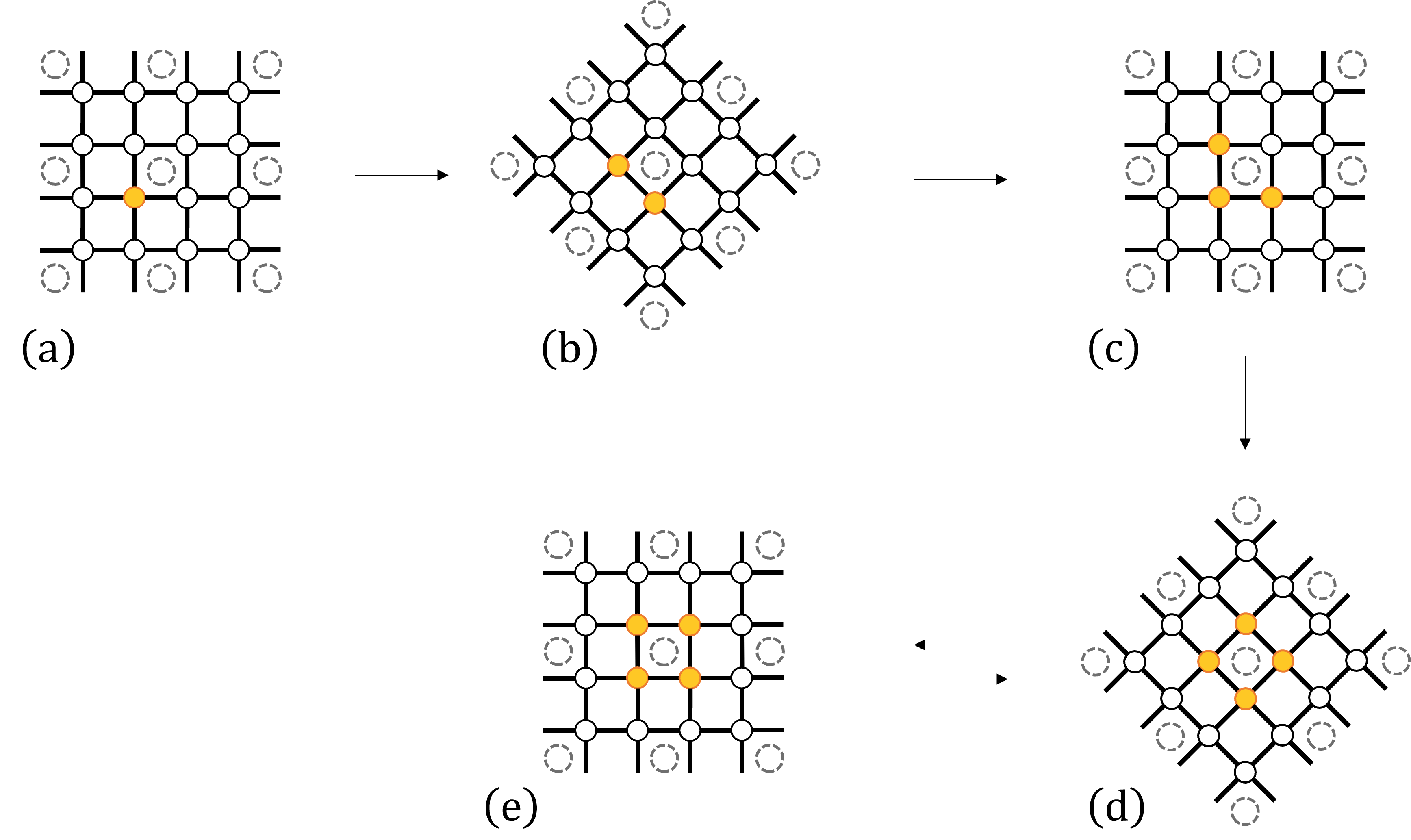}
    \caption{ The configurations of impurity tensors under successive TRG steps. Uniform tensors are represented by white circles. While for impurity tensors that can be different from each other, they are all represented by orange circles. (a) to (d) Impurity tensors increase when TRG scheme is applied. (d) and (e) The number of impurity tensors becomes four and is saturated, thus reaching stable configurations under successive TRG. }
  \label{fig:sin_RG}
\end{figure}

The change in configuration for the impurity tensor network under successive TRG operations is shown in Fig.~\ref{fig:sin_RG}. Note that we use orange circles to represent impurity tensors. 
%that can be mutually different. 
If we perform TRG within the patch of the tensor network marked by the gray circle, there will be at most four impurity tensors and they are confined within a 2-by-2 tensor network, making the computation of a single-body operator very simple (Fig.~\ref{fig:sin_op}(d)). 

For Loop-TNR, the number of impurity tensors becomes four after the first RG step. However, the size of the impurity tensor network will remain to be 2-by-2 if we perform Loop-TNR around the same square tensor network, similarly as Fig.~\ref{fig:sin_RG}(d) and (e).
%\FloatBarrier

% should ref looptnr part, choices of normalization
Note that we usually perform normalization on the newly obtained tensors at the end of each TRG or Loop-TNR step to prevent the elements of the tensors from overflow or underflow. 
As a result, the trace of the tensor network at the next iteration is no longer the same as the original one. Therefore, we need to consider the normalization effect\cite{Zcgu09}.
However, by noting that the expression of a single-body operator is in the form of a ratio, such an effect can be avoided by normalizing all the tensors with the trace of the 2-by-2 tensor network which consists of uniform tensors only. % maybe modified for loop-TNR
% Another reason we choose such way to normalize all the tensors is, the magnitudes of the elements in purity and impurity tenors are comparable, while the trace of the 2-by-2 impurity tensor network can be very small (which is about $\mathcal{O}(10^{-7})$). The elements in the impurity tensors would be very large and the loop-optimization part would be unstable, if we choose to normalize impurity tensors by their own trace. So that the way to normalize impurity tensors by trace of the purity ones is chosen.

\begin{figure}[htb]
  \centering
  \includegraphics[width=0.8\linewidth]{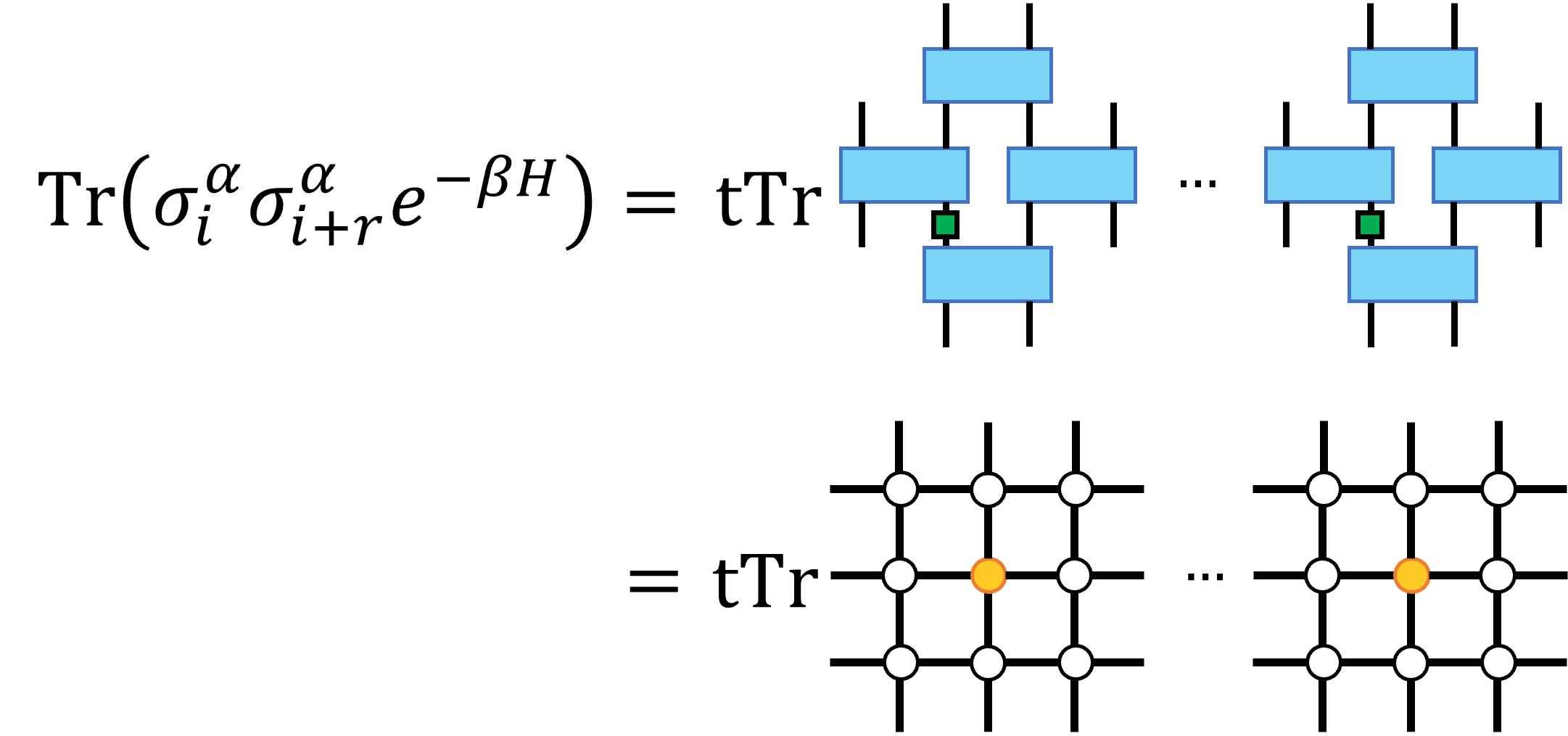}
  \caption{ The numerator of two-point correlation functions in horizontal (spatial) direction with separation $r$. Right-hand-side of the first equality: the tensor network representations obtained directly from Eq.~(\ref{equ:two_point_general}) with $\tau=0$. RHS of the second equality: the tensor network after tensor initialization as introduced in Section~\ref{sec:iniT}. }
  \label{fig:spatial_tnp}
\end{figure}
%\FloatBarrier

\begin{figure}[htb]
  \centering
  \includegraphics[width=\linewidth]{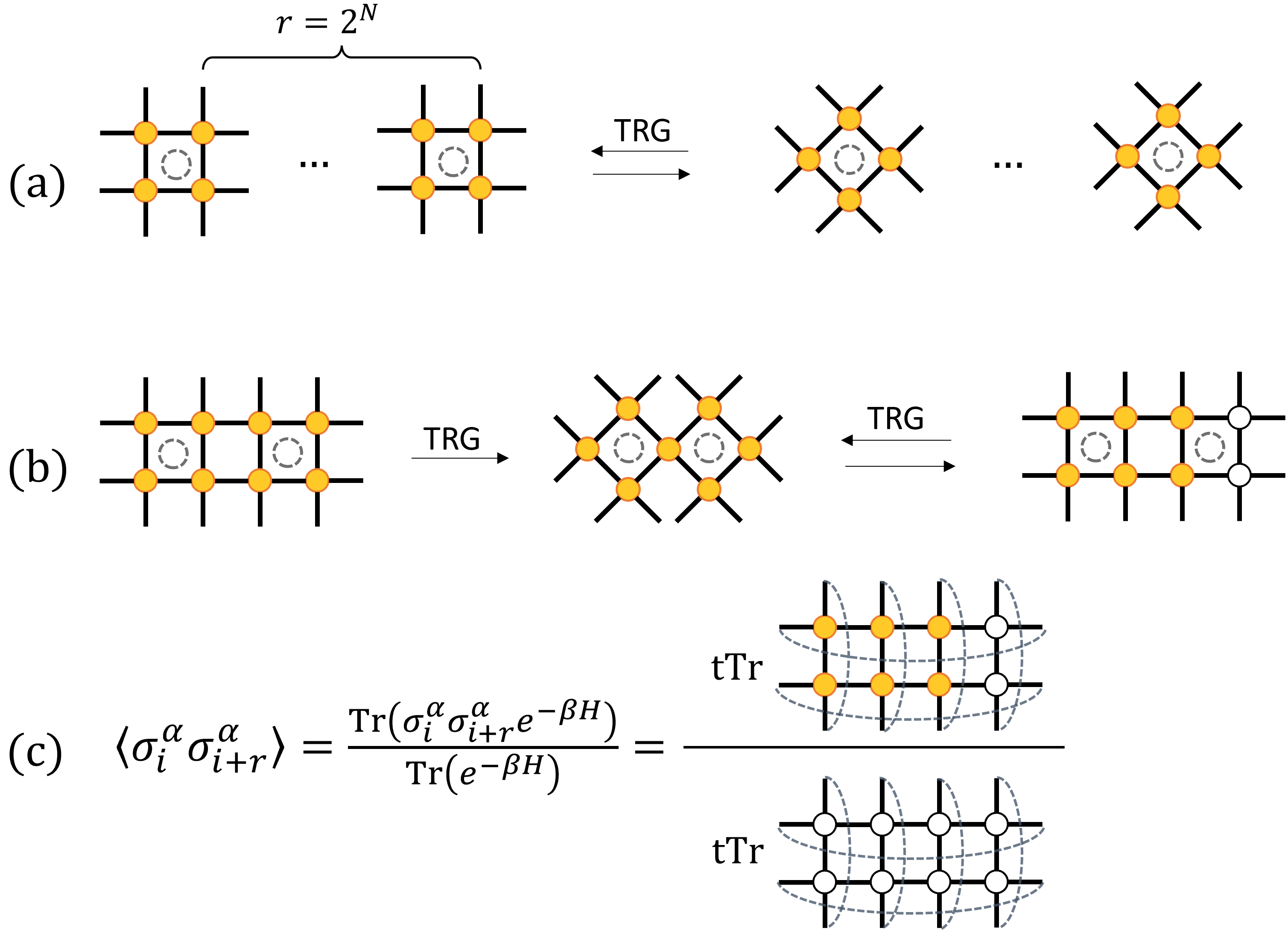}
    \caption{ The algorithm to compute two-point correlation by TRG with separation $r=2^{N} $. (a) At first, the two impurity tensor networks are separated. The change of configurations in the two impurity tensor networks is independent of each other. The distance between the impurity tensor networks will be reduced by half after two TRG steps. (b) The change in configuration when the two impurity tensor networks are combined through TRG. The stable configuration of impurity tensor network will be constrained to a 2-by-3 rectangular. (c) The computation of two-point correlation Fig.~\ref{fig:spatial_tnp} is reduced to tracing out two 2-by-4 tensor networks after successive applications of TRG.  }
  \label{fig:two_TRGr2N}
\end{figure}

\section{ Correlations in spatial direction }
\label{sec:two_body}
%                                 needs modification
Following the procedure of computing single-body operators, we proceed to introduce the algorithms for two-point correlations using TRG and Loop-TNR. We will see that the configurations of impurity tensor network become more complicated than the single-body operator case.
First, we construct the tensor network representations from the definition of two-point correlations, which generally can be expressed as: 
\begin{equation}
\langle \sigma^{\alpha}_{i}(\tau) \sigma^{\alpha}_{i+r} \rangle = \frac{ \Tr ( \sigma^{\alpha}_{i}(\tau) \sigma^{\alpha}_{i+r} e^{-\beta H} ) }{ \Tr (e^{-\beta H} ) }
\label{equ:two_point_general}
\end{equation}
In Fig.~\ref{fig:spatial_tnp} we show tensor network representations for the numerator of the correlation functions in the horizontal (spatial) direction. For the denominator, its tensor network representation has already been obtained previously.

Note that if the temporal difference $\tau$ is an integer such as $r$, the algorithm to compute the correlations in the two directions is almost the same, except for a rotation of the tensor network by 90 degrees. Therefore, in this section, we only introduce the method to compute correlations in the horizontal direction and the correlation in time direction will be discussed in the next section with generic time difference $\tau = r + k \delta \tau $ where $r$ is the integer part and $0 \leq k\delta \tau <1 $ is the decimal part of the time difference.

With the tensor network representations, we can perform the initializations of the tensor as introduced in Section~\ref{sec:iniT}. The resultant tensors are suitable for the TRG or Loop-TNR algorithm.
For different separations $r$, the algorithms will be different. In the following, we will start with the simplest case, $r=2^{N}$ where $N$ is a nonnegative integer, and then generalize it to $r=M \times 2^{N} $ where $M$ is a positive odd integer, for both TRG and Loop-TNR.

\subsection{ $r=2^{N} $ case }
\label{sec:r2N}

\subsubsection{ TRG }
For an impurity tensor network in which the separation of the corresponding impurity tensors is $r=2^{N} $, the distance between the impurity tensors will be reduced by half after two successive TRG steps, as marked by the gray dashed circles in Fig.~\ref{fig:two_TRGr2N}(a). We can call this stage a 'separated stage' since the two impurity tensor networks change their configurations independently. Note that we use orange circles to represent the impurity tensors. 

After $2 \left( N-1 \right) $ TRG steps, the distance is reduced to two and the two impurity tensor networks start to combine. We call this stage the 'combined stage'. Fig.~\ref{fig:two_TRGr2N}(b) shows the way that the combined impurity tensor network changes, and the stable configuration contains six impurity tensors. After enough steps of the TRG $N_{rg} $, the initial tensor network is reduced to a 2-by-4 one, whose trace is easy to compute (of course, we can also trace a 2-by-3 tensor network, which suggests that the initial size of the tensor network is $2 \times 2^{N_{rg}/2} $ by $3 \times 2^{N_{rg}/2} $).

Note that in Fig.~\ref{fig:two_TRGr2N} and the figures below, we only show impurity tensors. 
%In fact, the impurity tensors are surrounded by the uniform ones for the numerator. 
For the denominator, it consists of uniform tensors only and we just perform the usual TRG or Loop-TNR to reduce the size of the tensor network and compute the trace. In addition, we represent all the impurity tensors with orange circles for simplicity, actually all of them can be different.
%\FloatBarrier

\subsubsection{ Loop-TNR }
\begin{figure}[tb]
  \centering
  \includegraphics[width=\linewidth]{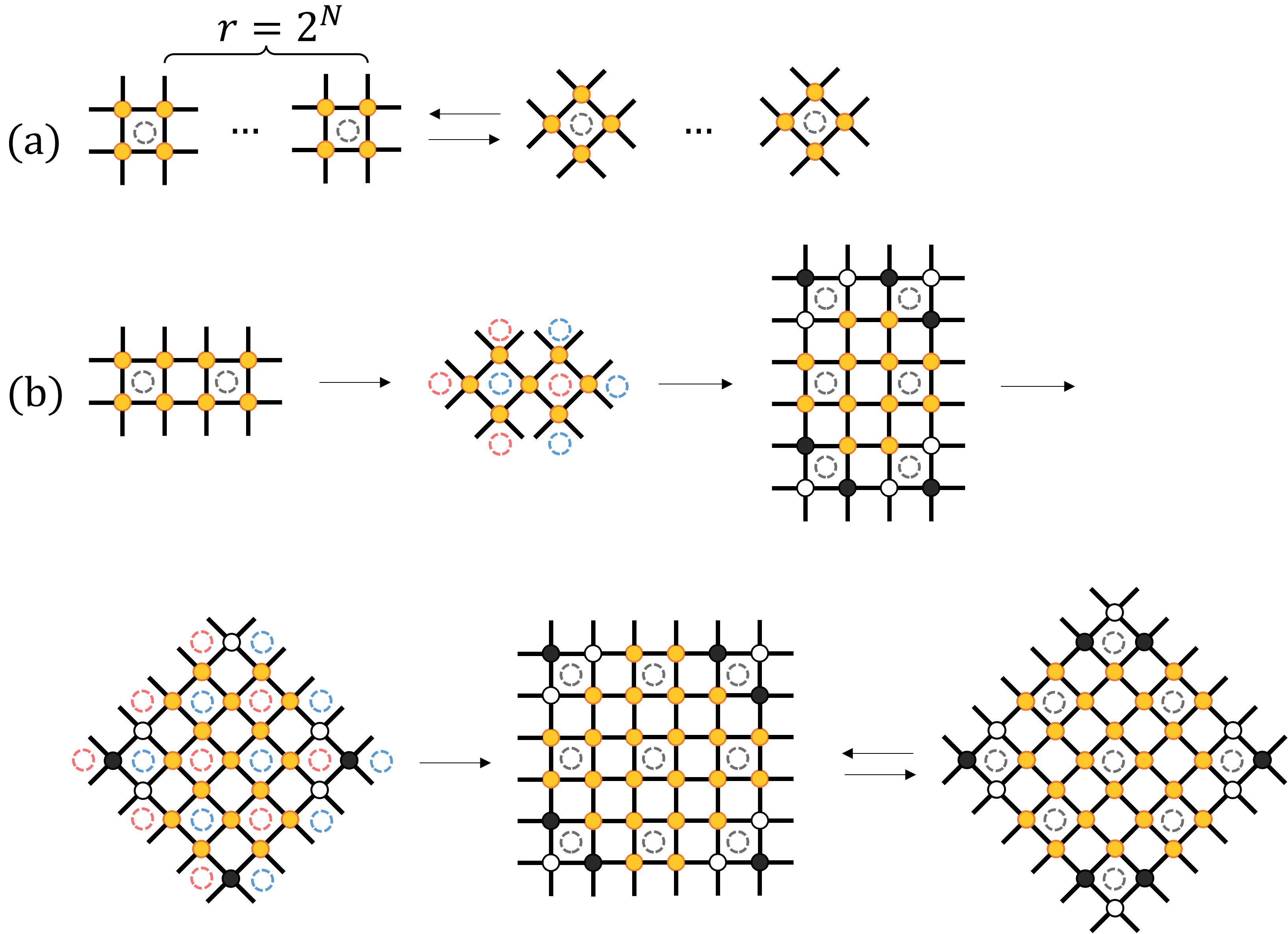}
    \caption{ The algorithm to compute two-point correlation by Loop-TNR with separation $r=2^{N} $. (a) The two impurity tensor networks are separated. At this separated stage, the change of configurations in the two impurity tensor networks is independent on each other. The distance between the impurity tensor networks will be reduced by half after two iterations. (b) At the combined stage, the number of impurity tensors increases step by step. Finally, a 6-by-6 network is reached, where 24 impurity tensors are included. Note that for some steps where the tensor network is rotated by 45 degrees, there are two equivalent ways to decompose the network such that the configurations for the next step are the same, as marked by blue and red circles, respectively.
    }
  \label{fig:two_loopr2N}
\end{figure}

For TRG, the algorithm to compute the two-point correlation is not much more complicated than the algorithm for the single-body operator. However, when we try to apply Loop-TNR for two-point correlations, it is rather different. As seen in Fig.~\ref{fig:two_loopr2N} for the change in the configurations of the impurity tensor network under successive Loop-TNR steps. In the combined stage, the number of impurity tensors is increasing step by step until it reaches a stable configuration with 24 impurity tensors in a 6-by-6 tensor network. 
For such configuration, we should perform Loop-TNR around ten 2-by-2 tensor networks (one for uniform tensor network and nine for impurity network) to finish one iteration, as marked by dashed gray circles in Fig.~\ref{fig:two_loopr2N}(b). 

The large size of the impurity tensor network also gives rise to the difficulty in computing its trace. The direct computation of a trace for a 6-by-6 tensor network is, of course, with high computational cost. A practical solution is to put the 6-by-6 tensor network into an 8-by-8 one, so that after four iterations the size of the original tensor network is reduced to 2-by-2, as can be seen in Fig.~\ref{fig:trace8by8}. % maybe the statement for high computational cost should be postpone later

%\FloatBarrier

\begin{figure}[htb]
  \centering
  \includegraphics[width=\linewidth]{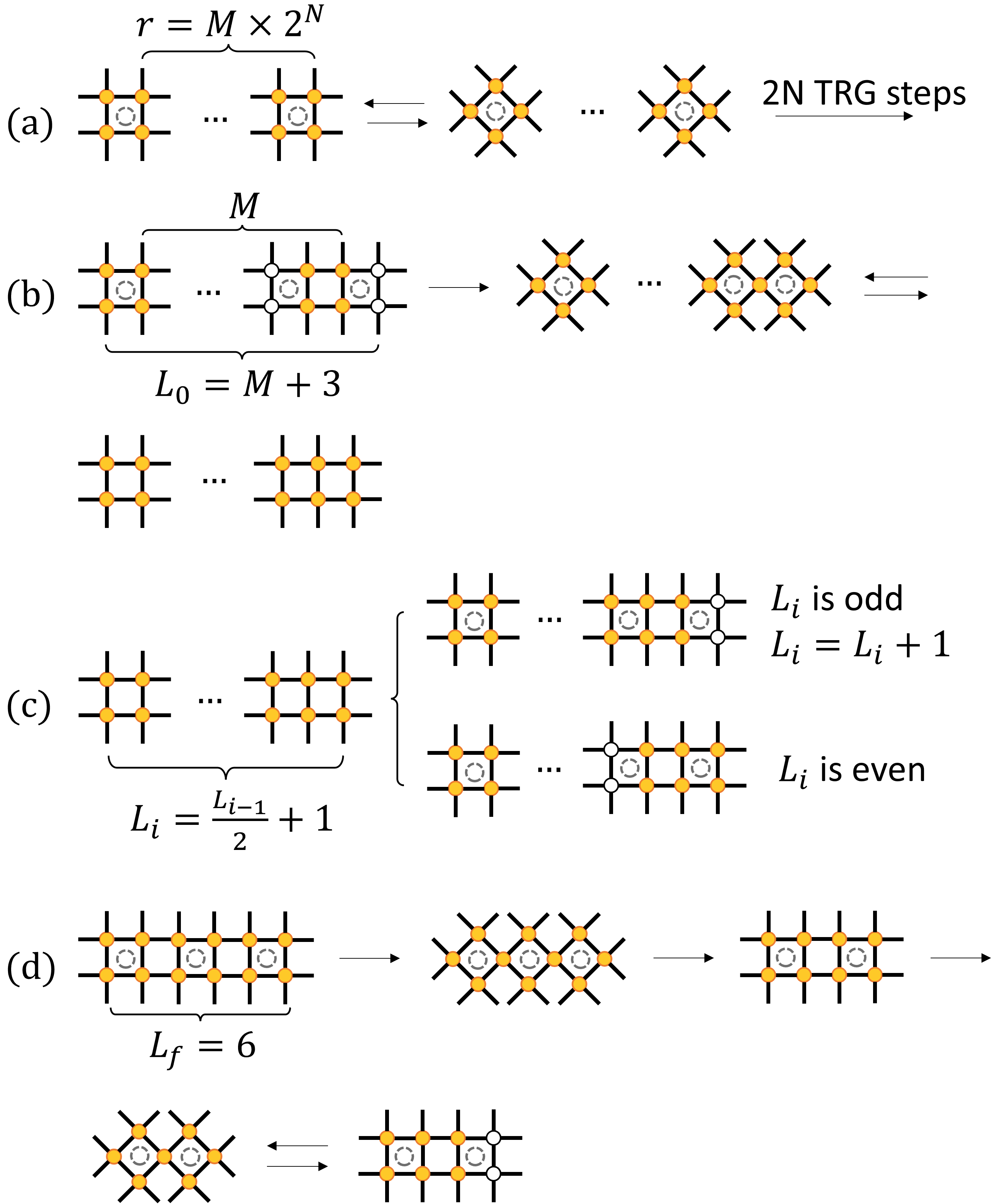}
    \caption{ Change of configurations of impurity tensors under successive TRG steps for $r=M \times 2^{N}$. (a) For the first $2N$ steps, the two impurity tensor networks change their configurations independently. (b) After $2N$ steps, the distance between the impurity tensors is $M$ and the corresponding columns $L_{0} $ become $M+3$. The stable configuration in the left part of the impurity tensor network is the same as the previous step. However, for the right part, the number of impurity tensors increases and the configurations change like the combined stage for $r=2^{N}$ case. (c) The change of $L_{k}$ after two TRG steps. (d) $L_{f}=6 $ indicates the end of the intermediate stage, where the impurity tensor networks begin to combine. After two more TRG steps, the change in the configurations of the tensor network is the same as the combined stage for $r=2^{N} $ case. }  % maybe the caption should be shortened, some adding to main text
  \label{fig:TRG_M2N}
\end{figure}
%\FloatBarrier

\begin{figure}[htb]
  \centering
  \includegraphics[width=\linewidth]{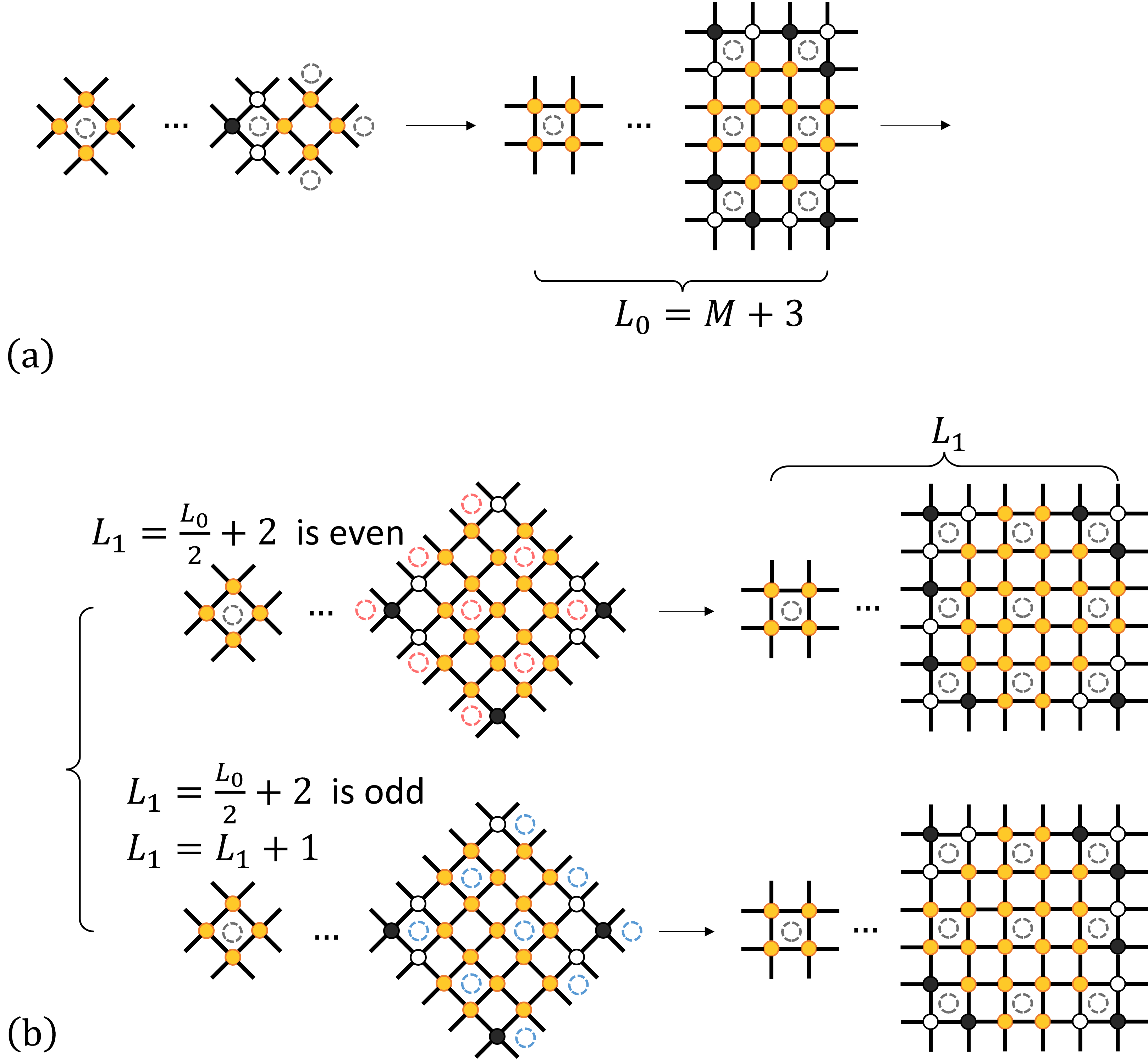}
    \caption{ The change in configurations of impurity tensor network after $2N-1$ RG steps, for separation $r=M \times 2^{N}$ applied with Loop-TNR. (a) The number of impurity tensors on the right part of the tensor network increases, which behaves differently from the TRG case. (b) The size of the impurity tensor network on the right becomes 6-by-6 after two more Loop-TNR iterations. Note that for different parity of $L_{1} $, the decompositions on the right part are different, as marked by red and blue circles respectively. }
  \label{fig:loop_rM2N_1}
\end{figure}

\subsection{ $r= M \times 2^{N} $ case }
\label{sec:rM2N}

Applying the methods in the previous subsection only gives rise to a limited set of data. Here, we try to generalize the algorithm to enable the computation of correlation function with separation $ r= M \times 2^{N} $ for odd $M > 1$ and nonnegative integer $N$, that is, any integer separation except $ r = 2^{N} $ case.

% subsection ? 
\subsubsection{ TRG }
For simplicity, we first introduce the algorithm for TRG, as shown in Fig.~\ref{fig:TRG_M2N}. The first stage is the same as in the $r=2^{N} $ case, where the two impurity tensor networks change their configurations independently under TRG operations. 
%We also call this stage a 'separated stage'. 
Since the distance between impurity tensors is reduced by half for every two iterations, after $2N$ iterations the distance becomes $M$, which is odd. At this stage, although the two impurity networks are separated, the change of their configurations is no longer independent of each other. We call this stage the 'intermediate stage'.

As seen Fig.~\ref{fig:TRG_M2N}(b), we find that the number of impurity tensors of the network on the right part increases through TRG operations. Hence, it is no longer appropriate to define a 'distance' between the corresponding impurity tensors in the two impurity tensor networks. Instead, we define $L_{k} $, the number of columns of tensors, which contains all the impurity tensors and ensures the application of TRG at the same time. 
For example, at the beginning of the intermediate stage, $L_{0} = M+3 $. After two TRG steps, we first compute $L_{1} = L_{0}/2 +1 $. If $L_{1}$ is even, that means $L_{1}$ is enough to perform TRG on the impurity tensor network and contain all the impurity tensors. For odd $L_{1}$, we should include one more column to ensure the application of TRG, $L_{1} = L_{0}/2 +2 $.
The determination of general $L_{k} $ is shown in Fig.~\ref{fig:TRG_M2N}(c). When the number of columns is reduced to $L_{f}=6 $, the two parts of the impurity tensor networks begin to combine and after two more TRG steps, the combined tensor network changes its configuration exactly the same way as the combined stage in case $r=2^{N} $.

\begin{figure}[htb]
  \centering
  \includegraphics[width=\linewidth]{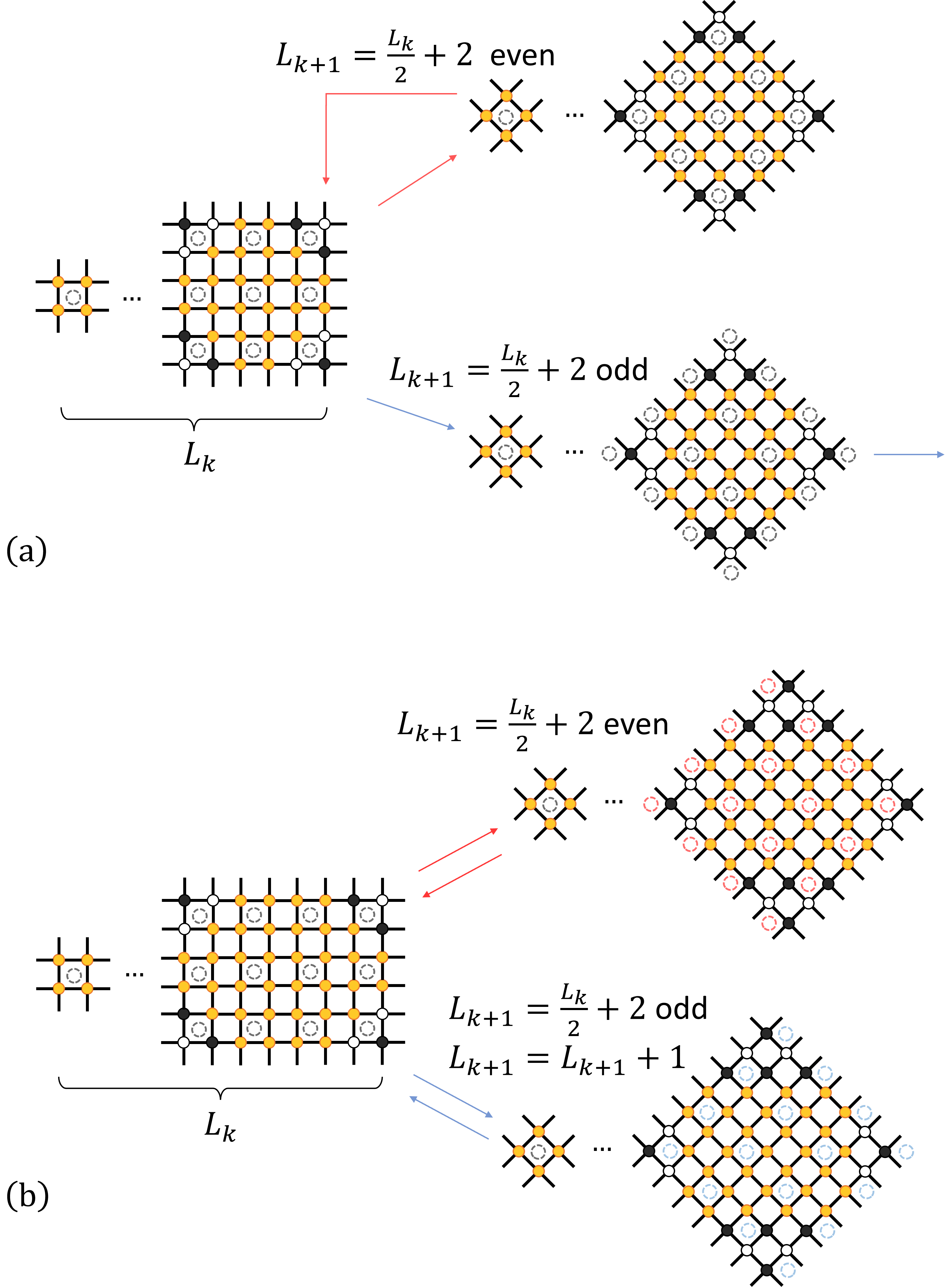}
    \caption{ The change in the configurations of impurity tensor networks when the right part has size 6-by-6. (a) If $L_{k} $ is even, the configuration returns to its previous shape after two Loop-TNR iterations. While for some odd $L_{k}$, the right part of the impurity tensor network grows a larger size of 6-by-8. (b) The right part of the impurity tensor network will no longer enlarge, no matter what the parity of $L_{k}$ is. Hence it reaches a stable configuration under Loop-TNR. }
  \label{fig:loop_rM2N_2}
\end{figure}

\begin{figure}[htb]
  \centering
  \includegraphics[width=\linewidth]{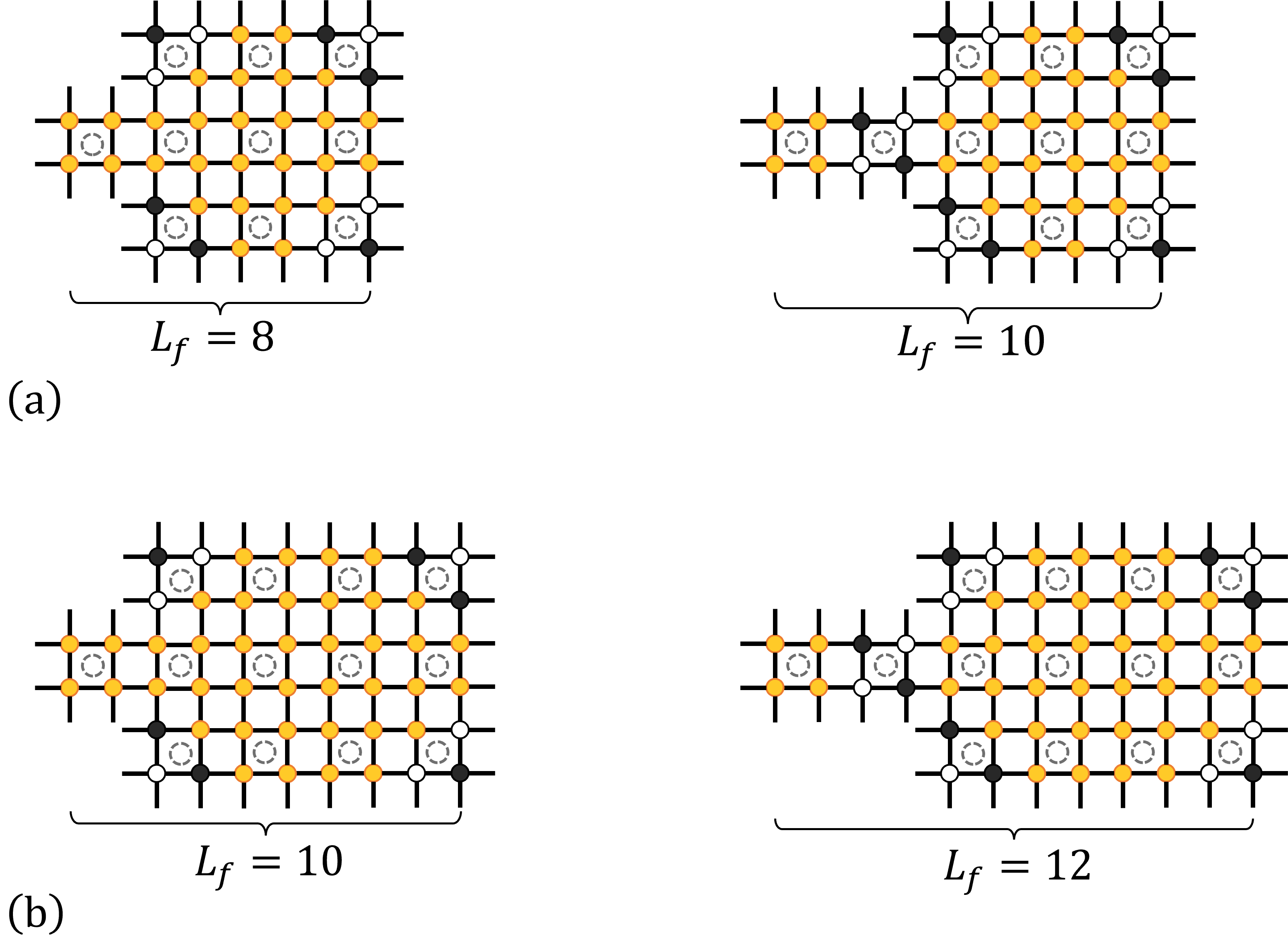}
    \caption{ Different $L_{f}$'s that identifies the end of the intermediate stage. (a) The $L_{f}$'s if the size of the impurity tensor network at the right part is 6-by-6. (b) The $L_{f}$'s if the right part has size 6-by-8. After two more Loop-TNR iterations, the two parts of impurity tensors are fully combined.} 
  \label{fig:loop_rM2N_3}
\end{figure}
%\FloatBarrier

\subsubsection{ Loop-TNR }
The change in the configurations of impurity tensor networks becomes rather complicated when we generalize the algorithm to Loop-TNR. For the separated stage, the configurations change exactly the same as in the previous case. The difference occurs after $(2N-1)$-th iterations, where the tensor network is rotated by 45 degrees. As shown in Fig.~\ref{fig:loop_rM2N_1}, the configuration no longer goes back to its original one if we perform one more iteration of Loop-TNR. Instead, the impurity tensors on the right part proliferate if we fix the number of impurity tensors on the left part, when performing Loop-TNR around the patches marked by gray circles. In the following, we keep this choice and focus on the change in configurations of impurity tensors on the right part. 

We need the $L_{k}$ again for this intermediate stage. From Fig.~\ref{fig:loop_rM2N_1}(a) we find, after $2N$ iterations, $L_{0}=M+3$. Note that the recursion relation between $L_{k} $ and $L_{k-1} $ now becomes $ L_{k} = L_{k-1}/2+2 $, with the application of Loop-TNR. If the newly obtained $L_{k} $ is odd, we should still include one more column. 

We would expect that the configuration of the impurity tensor network on the right will change into the combined tensor network in Fig.~\ref{fig:two_loopr2N}, which eventually grows into a 6-by-6 one and stops changing before combining with the left part. However, this is not the most general case. A different configuration can arise when the right part of the impurity tensor network has already grown into a 6-by-6 one for odd $L_{k} $. For this case, we have a different decomposition on the right part of the impurity tensor network after one more iteration. Consequently, the size increases further to a 6-by-8 one (see Fig.~\ref{fig:loop_rM2N_2}(b)). Fortunately, this configuration is stable under Loop-TNR and we don't need an even larger tensor network to perform Loop-TNR algorithm.

%\FloatBarrier

\begin{figure}[tb]
  \centering
  \includegraphics[width=0.8\linewidth]{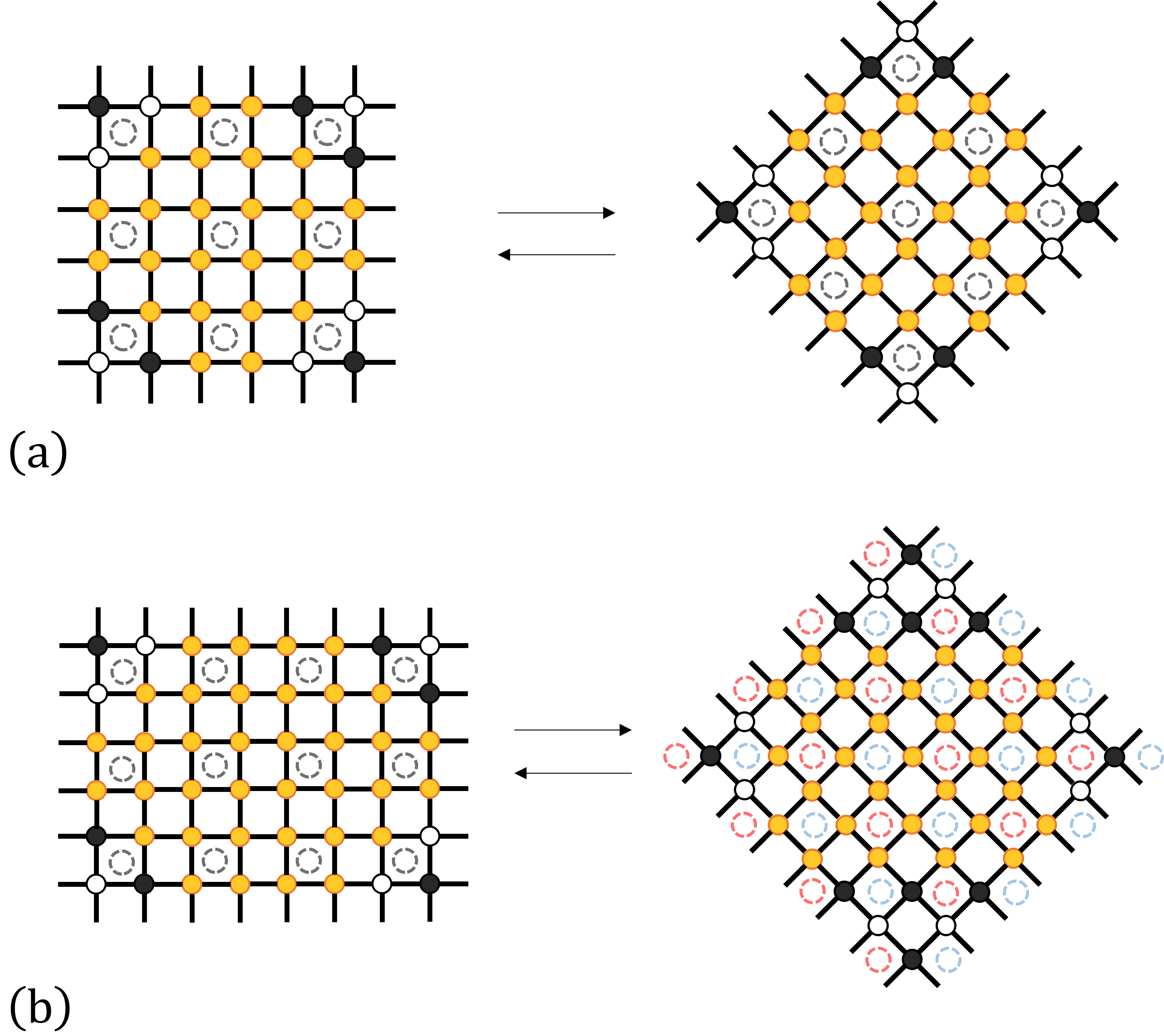}
    \caption{ The change in configurations of combined impurity tensor network. (a) The stable configuration with size 6-by-6, which corresponds to the configurations in Fig.~\ref{fig:loop_rM2N_3}(a) before fully combined. (b) The stable configuration with size 6-by-8, which corresponds to the configurations in Fig.~\ref{fig:loop_rM2N_3}(b) before fully combined. Blue and red circles label two types of decompositions, where the resultant configurations are the same after the application of Loop-TNR. }
  \label{fig:loop_rM2N_4}
\end{figure}

Finally, the left and right parts of the impurity tensor networks start to combine for different $L_{f}$'s, which is determined by different $M$'s, as seen in Fig.~\ref{fig:loop_rM2N_3}. After two more Loop-TNR iterations, the two parts are fully combined. The change in configurations of impurity tensor networks for the combined stage is shown in Fig.~\ref{fig:loop_rM2N_4}.

%\FloatBarrier
The above algorithms complete the computation of correlation with $r=M \times 2^{N} $ by Loop-TNR. In the following, we introduce the way to compute the trace of the combined tensor network.

Similar as before, to compute the trace of the combined tensor network, we should put it into an 8-by-8 tensor network. Here we explain how to trace out a network with size 8-by-8 by several Loop-TNR iterations. Thus the computation of correlation is completed.
\begin{figure}[tb]
  \centering
  \includegraphics[width=\linewidth]{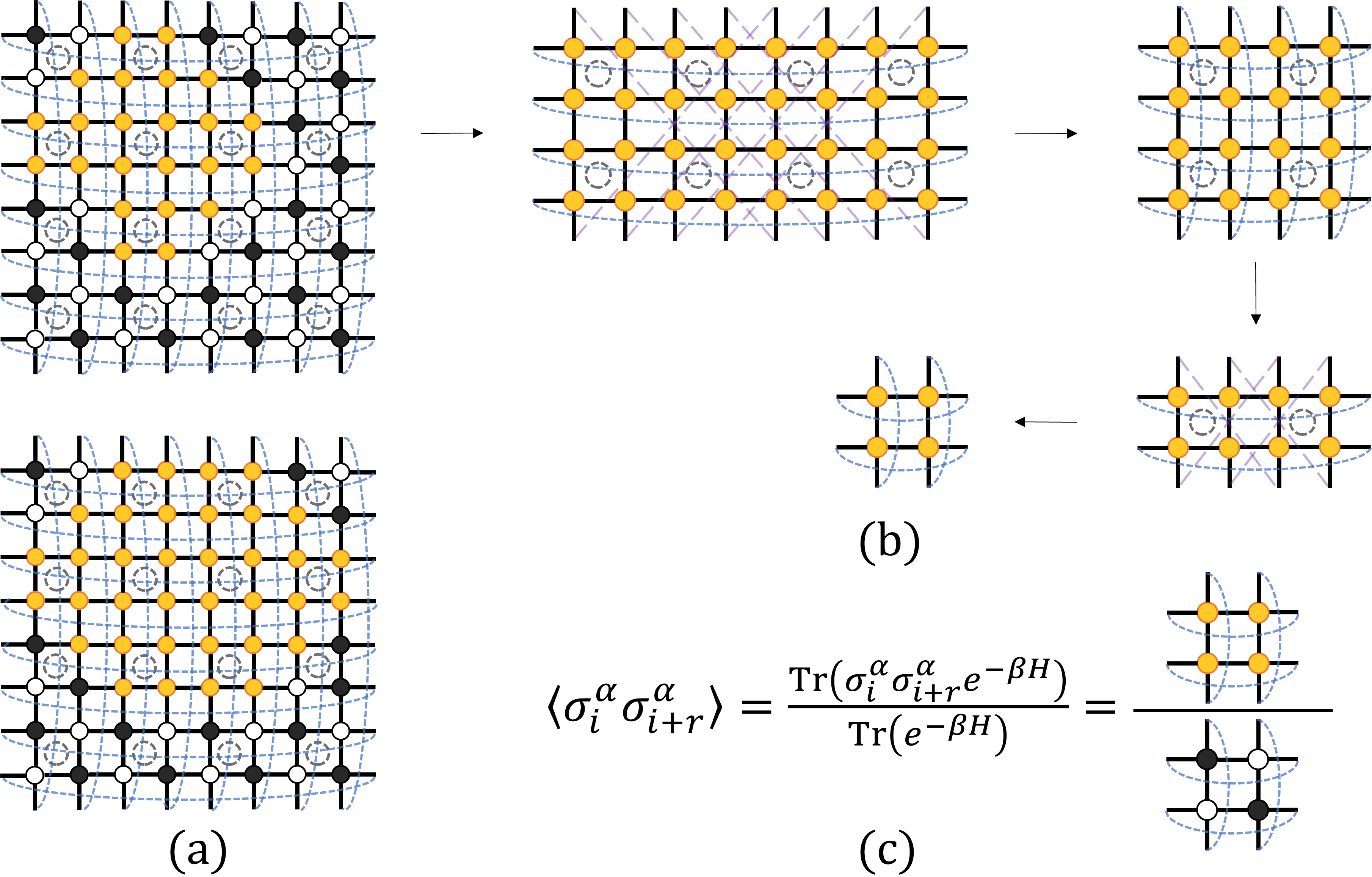}
    \caption{ The method to trace out the 8-by-8 tensor network with periodic boundary condition. (a) The initial configurations, where the 6-by-6 or 6-by-8 impurity tensor network is placed into an 8-by-8 one. (b) The change of configurations of the tensor network under successive applications of Loop-TNR, which are performed around square tensor networks marked by gray circles. Note that the boundary condition can be 'twisted' when we arrive at the configuration of $N$-by-$2N$. (c) The correlation function is easily computed since the large tensor networks are reduced to the 2-by-2 ones. }
  \label{fig:trace8by8}
\end{figure}
%\FloatBarrier

\begin{figure}[h]
  \centering

  \subfloat[]{ \includegraphics[width=0.9\linewidth]{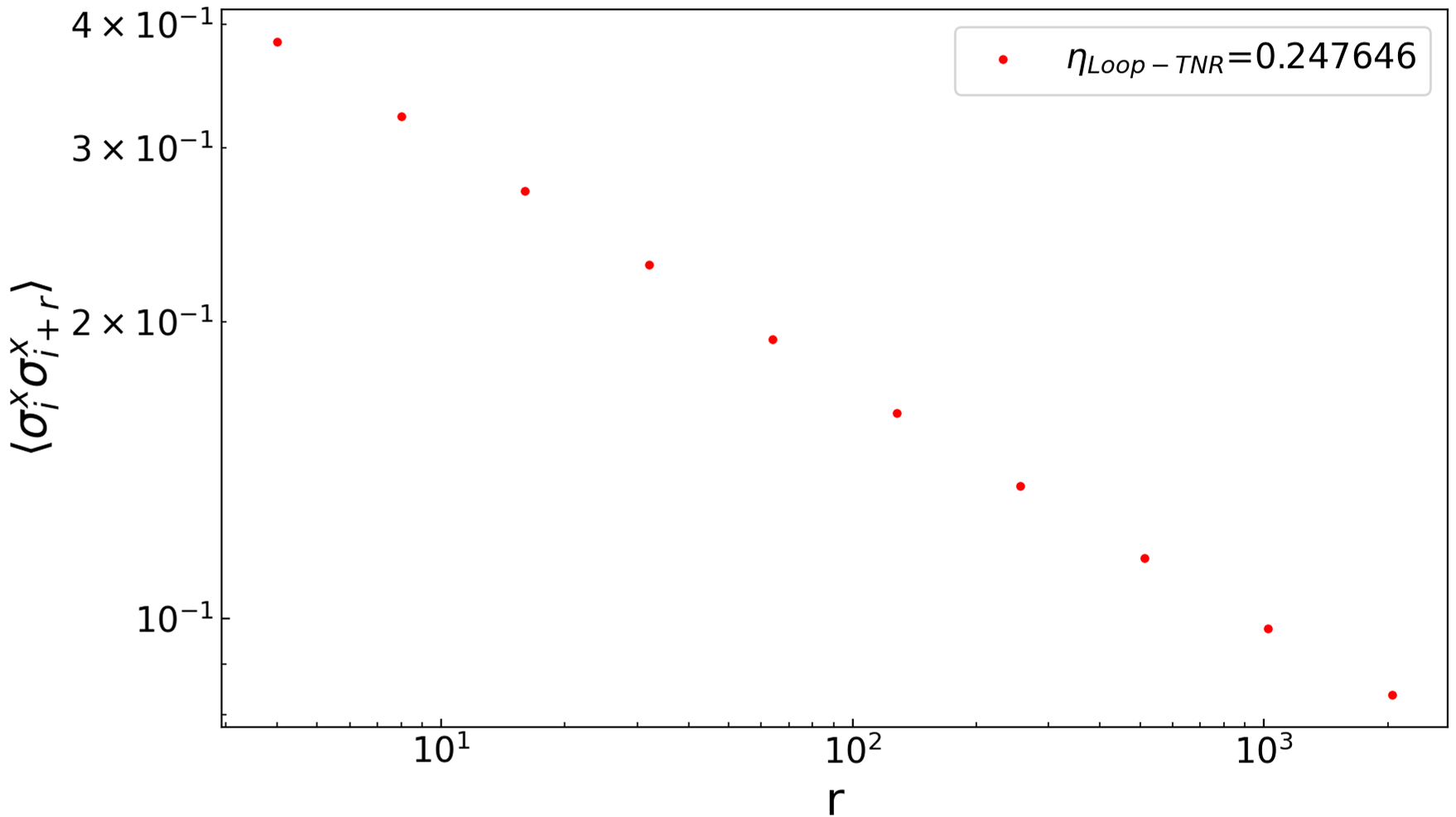} }
  \hfill
  \subfloat[]{ \includegraphics[width=0.9\linewidth]{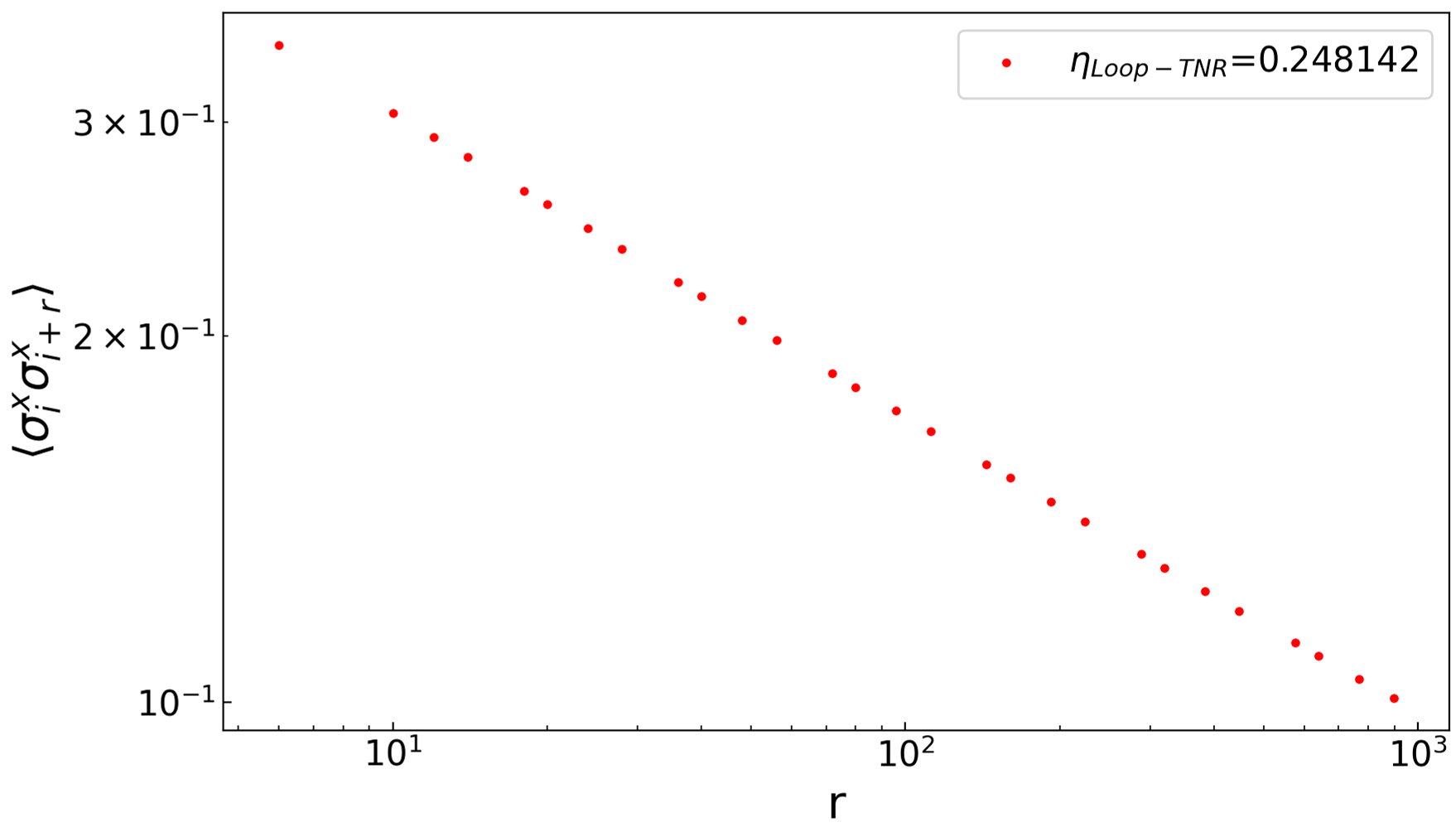} }

  \caption{ The log-log plot for the longitudinal correlation of the quantum Ising chain, with (a) $r=2^{N}$ and (b) $r = M \times 2^{N} $, where $M=3,5,7,9 $ and $r<1000 $. The exact $\eta$ should be 0.25. The results are obtained by Loop-TNR with $D_{cut}=16 $. We performed 26 Loop-TNR iterations for both results. }
  \label{fig:M2Nloop}
\end{figure}

As seen in Fig.~\ref{fig:trace8by8}, we first perform Loop-TNR on the 2-by-2 tensor networks as marked by gray circles. The resultant configuration becomes a 4-by-8 one, with periodic boundary condition in the horizontal direction but with 'twisted' boundary condition in the vertical direction. Such boundary conditions appear again after two more Loop-TNR iterations, where the size is reduced to 2-by-4. After in total four iteration steps, the 8-by-8 tensor network is shrunk to a 2-by-2 one, the trace of which is then easy to compute. Finally, the correlation is computed by tracing out two 2-by-2 tensor networks, see Fig.~\ref{fig:trace8by8}(c).

\subsection{A simple Example}
%\subsubsection{ $r=2^{N}$ case }
% 2^N
% show some results by comparison of TRG or loop-TNR
Fig.~\ref{fig:M2Nloop}(a) displays the results of longitudinal correlation $\langle \sigma^{x}_{i} \sigma^{x}_{i+r} \rangle$, where $r = 2^{N} $, for the critical Ising chain:
\begin{equation}
H = -\sum_{i} \sigma^{x}_{i} \sigma^{x}_{i+1} - \sum_{i} \sigma^{z}_{i}
\label{equ:critical_QI}
\end{equation}
with Loop-TNR method. We performed 26 Loop-TNR iterations in total, which means that the original quantum chain has $2^{15} $ spins. We find Loop-TNR is able to produce accurate critical exponent $\eta \approx 0.247 $ with a relatively small bond dimension $D_{cut}=16$ of the tensors.

%\FloatBarrier

% We also show the transverse correlation

%\subsubsection{ $r=M \times 2^{N}$ case }
In Fig.~\ref{fig:M2Nloop}(b), we present the results of the longitudinal correlation $\langle \sigma^{x}_{i} \sigma^{x}_{i+r} \rangle$ of the critical quantum Ising chain Eq.~(\ref{equ:critical_QI}), for $r = M \times 2^{N} $ with $M=3,5,7,9$. The results are obtained by Loop-TNR with $D_{cut}=16 $ and 26 RG iterations in total. 
% \begin{figure}[htb]
%   \centering
%   \includegraphics[width=0.8\linewidth]{./figures/MN1000_loop.png}
%     \caption{ The log-log plot for the longitudinal correlation of the quantum Ising chain, with $r = M \times 2^{N} $, where $M=3,5,7,9 $ and $r<1000 $. The results are obtained by Loop-TNR with $D_{cut}=16 $. The exact $\eta$ should be 0.25.}
%   \label{fig:MNloop}
% \end{figure}
%\FloatBarrier
% use float barrior

\section{ Imaginary time correlations }
\label{sec:corre_imag}

In previous sections, we presented the algorithms to compute correlation functions for integer separations. For (1+1)D quantum systems, we can also compute correlation functions along the time direction, where the separation can be a fractional number. In Eq.~(\ref{equ:Trotter_expan}) we decompose the evolution operator $ e^{ -\beta H } $ into layers of two-body gates $ e^{ -\epsilon h_{i,i+1} } $ by Suzuki-Trotter expansion. As a result, the time difference between two neighbouring layers is $\epsilon$.
Thus $\epsilon $ defines the minimum time difference that we can compute in time direction. A general time difference can be expressed as $ \tau = k \epsilon + r $, where we denote $\tau$ as the (imaginary) time difference, $r$ as the integer part of the difference and $k \epsilon $ as the decimal part (where $k$ is a non-negative integer and obviously $ 0 \le k \epsilon <1$ should be satisfied). 

\begin{figure}[tb]
  \centering
  \includegraphics[width=\linewidth]{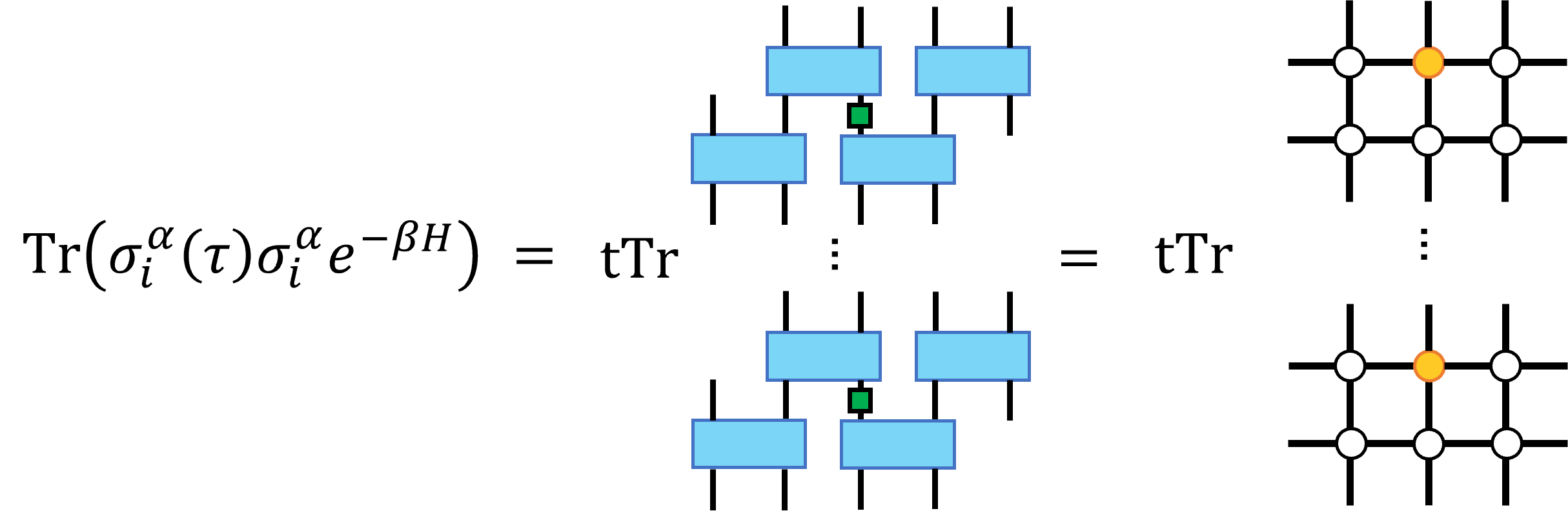}
  \caption{ The numerator of two-point correlation functions in vertical (temporal) direction with fractional separation $\tau=k\epsilon + r $. Right-hand-side (RHS) of the first equality: the tensor network representations obtained directly from Eq.~(\ref{equ:two_point_general}) with spatial separation equals zero. RHS of the second equality: the tensor networks after tensor initialization as introduced in Section~\ref{sec:iniT}. }
  \label{fig:imag_tnp}
\end{figure}

An example of the correlation function in (imaginary) time direction represented by tensor networks is shown in Fig.~\ref{fig:imag_tnp}, where the separation between spins in the time direction is represented as the distance between the single-body operators (green squares) in the vertical direction. After one TRG step, we no longer have odd or even gates, and the time difference between tensors in neighbouring layers now becomes $\epsilon$.

\begin{figure}[tb]
  \centering
  \includegraphics[width=\linewidth]{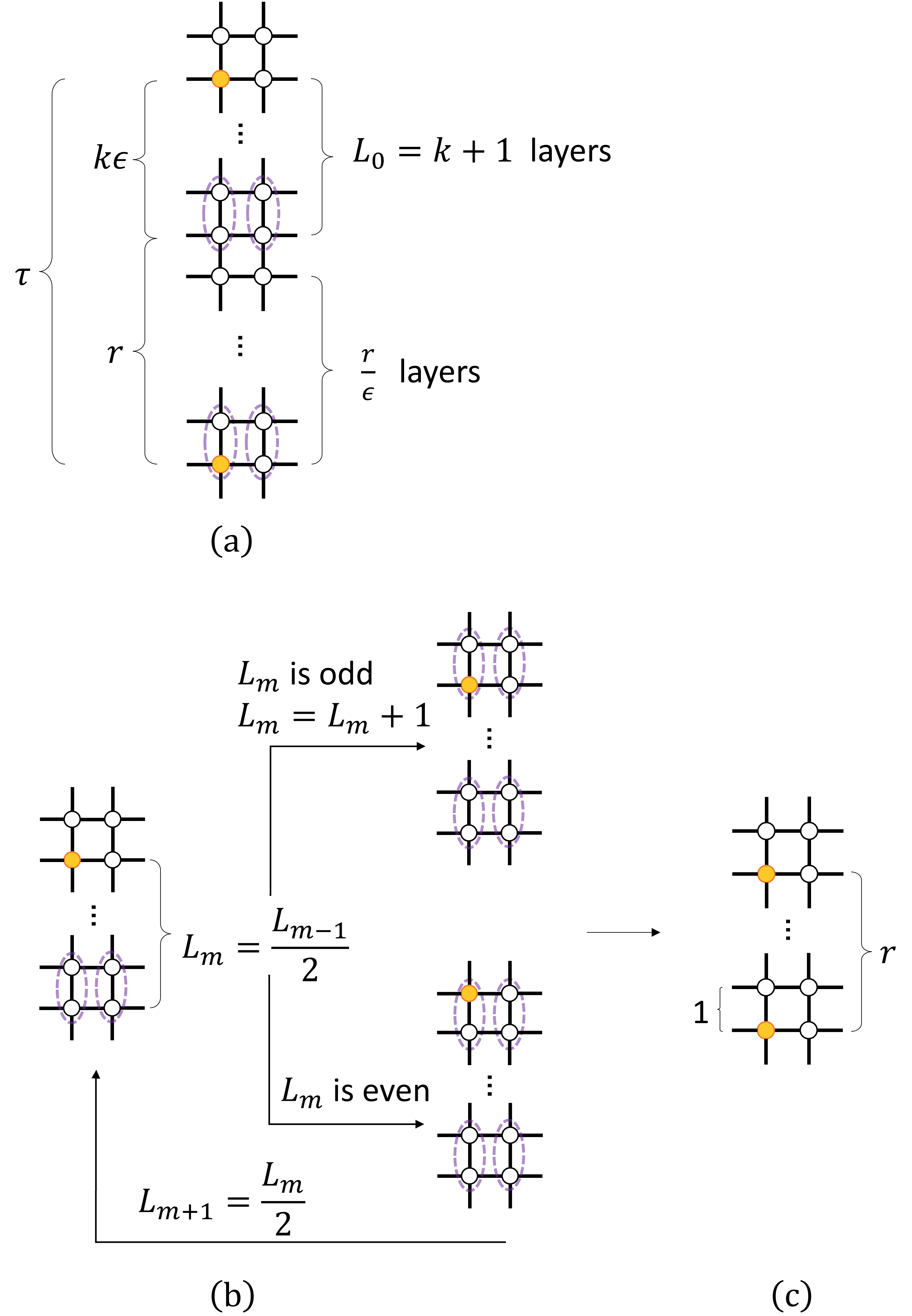}
    \caption{ The algorithm to compress tensors when the separation between impurity tensors takes fractional values. We perform compression on tensors as circled by dashed ellipses. (a) Configuration of impurity tensors when the separation in the time direction is $\tau = k \epsilon + r $. (b) The algorithm for compressing tensors when the number of layers becomes $L_{m} $. (c) The configuration of impurity tensors when the compression step is finished, where the time difference between two neighbouring layers becomes one and the distance of impurity tensors in the vertical direction becomes $r$. }
  \label{fig:imagcorre}
\end{figure}

Since the denominator consists of uniform tensors only, the algorithm for this part is exactly the same as the usual TRG or Loop-TNR for (1+1)D quantum systems. Thus we only need to focus on the numerator, more specifically, the layers between impurity tensors in the temporal direction, as seen in Fig.~\ref{fig:imagcorre}. 

From Section~\ref{sec:iniT} we have already known that, compression steps are necessary to make tensors isotropic. In this step, the algorithm to calculate the imaginary time correlation has already been involved. See Fig.~\ref{fig:imagcorre}(a) for the configuration of impurity tensors with temporal separation $ \tau = k \epsilon +r$
\footnote{The following algorithm is applicable for the quantum Ising model. For generic models, the time difference that the algorithm can compute is modified by a factor $v$: $\tau' = (k \epsilon +r)/v $. See Section~\ref{sec:conformal} for the physical meaning of $v$ and how it is introduced to make the tensor isotropic.  }
. Since the number of layers of tensors is reduced by half after each compression step, it is convenient to control the process of compression by the number of layers between impurity tensors. In Fig.~\ref{fig:imagcorre}(a), the number of layers for the integer part is $\frac{r}{\epsilon}$, which is always even in our algorithm. While the number of layers for the decimal part is $L_{0}= k+1 $. % the bellow needs better illustration 

In the compression step, an impurity tensor can be placed either on top or on the bottom of the uniform tensor. For the lower part impurity tensor as shown in Fig.~\ref{fig:imagcorre}(a), we choose to always place it at the bottom during the compression steps. As a result, the relative position for the upper part impurity tensor is uniquely determined, by the number of layers between two impurity tensors. Moreover, since there are always even layers for the integer part of the separation, the parity is solely determined by $k$, or $L_{0}$.

Fig.~\ref{fig:imagcorre}(b) shows the detail of compressing impurity tensor on the upper part at $m$-th compression step. Namely, when $L_{m} $ is odd, the upper impurity tensor is placed at the bottom and the uniform one is placed at the top for that compression step. It means that we should include one more layer to finish the compression step. So $L_{m} = L_{m}+1 $.
If $L_{m} $ is even, the impurity tensor is to be put on the top and we have no need to add one more layer. After the $m$-th step of compression, the number of layers is reduced by half, $L_{m+1} = \frac{ L_{m} }{2} $, and we start the new compression step until the tensors are isotropic.

When the tensors approach isotropic, the time difference between two neighbouring layers of tensors becomes one. Consequently, the distance between impurity tensors in the vertical direction becomes $r$ (see Fig.~\ref{fig:imagcorre}(c)), indicating that in the following the algorithms are exactly the same as the one for integer separations, which are already introduced in the previous sections. 
%\FloatBarrier % 

\begin{figure}[tb]
  \centering

  \subfloat[]{ \includegraphics[width=0.8\linewidth]{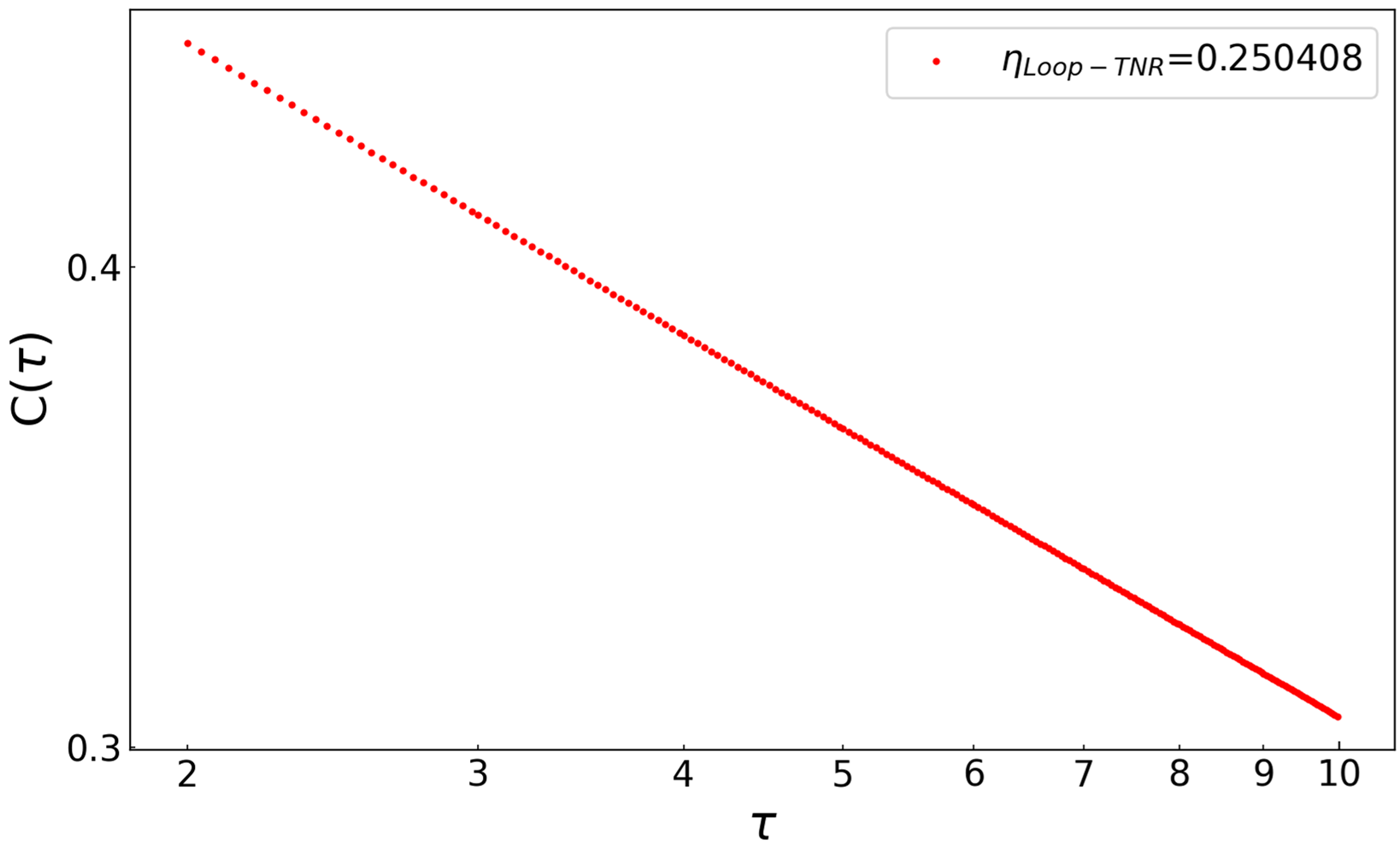} }
  \hfill
  \subfloat[]{ \includegraphics[width=0.8\linewidth]{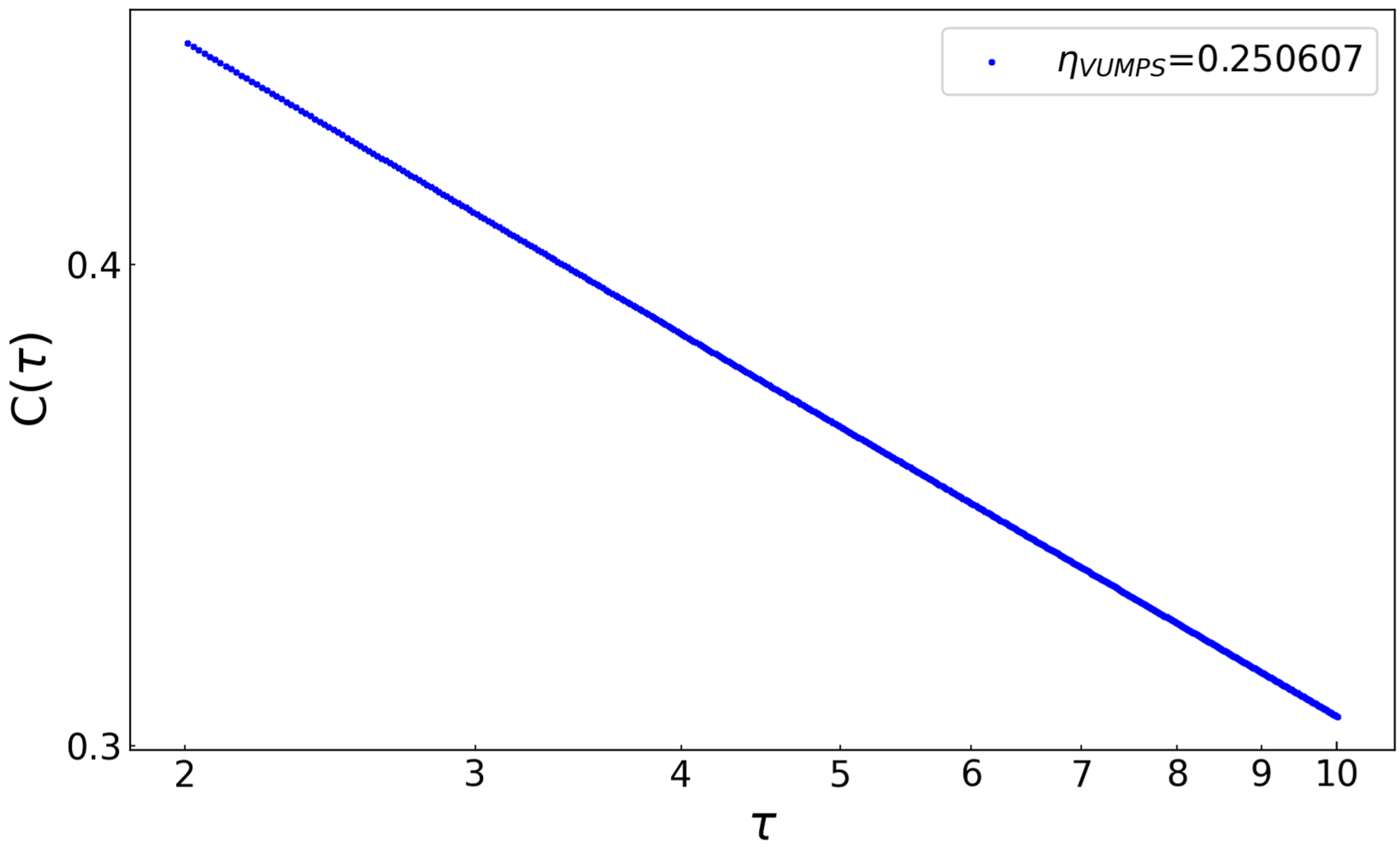} }
  
  \caption{ Log-log plots of the imaginary time correlation function of the quantum Ising model, by (a) Loop-TNR, with $D_{cut}=16 $ and (b) VUMPS, where the virtual bond dimension is $\chi=200 $. For the correlation by Loop-TNR, the temporal difference for the neighboring point is $\frac{5}{128} $. While for the VUMPS method, the temporal difference is chosen as $\frac{1}{64} $ in the present case.}
  \label{fig:it210bench}
\end{figure}

% imaginary time correlation
%\subsection{ Example }
As a simple example, here we again compute the imaginary time correlation function $\langle \sigma^{x}_{i}(\tau) \sigma^{x}_{i} \rangle$ of (1+1)D critical quantum Ising model Eq.~(\ref{equ:critical_QI}) by Loop-TNR, comparing with the results obtained by VUMPS. As shown in Fig.~\ref{fig:it210bench}, we find an accurate critical exponent $\eta$ is fitted with a small $D_{cut}=16$ and a  large size of the tensor network, which is $ 2^{12} $-by-$ 2^{12} $.  In VUMPS calculations, the virtual bond dimension is chosen as $\chi=200 $.
%\FloatBarrier

\section{ Real-time correlations } % maybe set as a section
\begin{figure}[tb]
  \centering
  \includegraphics[width=\linewidth]{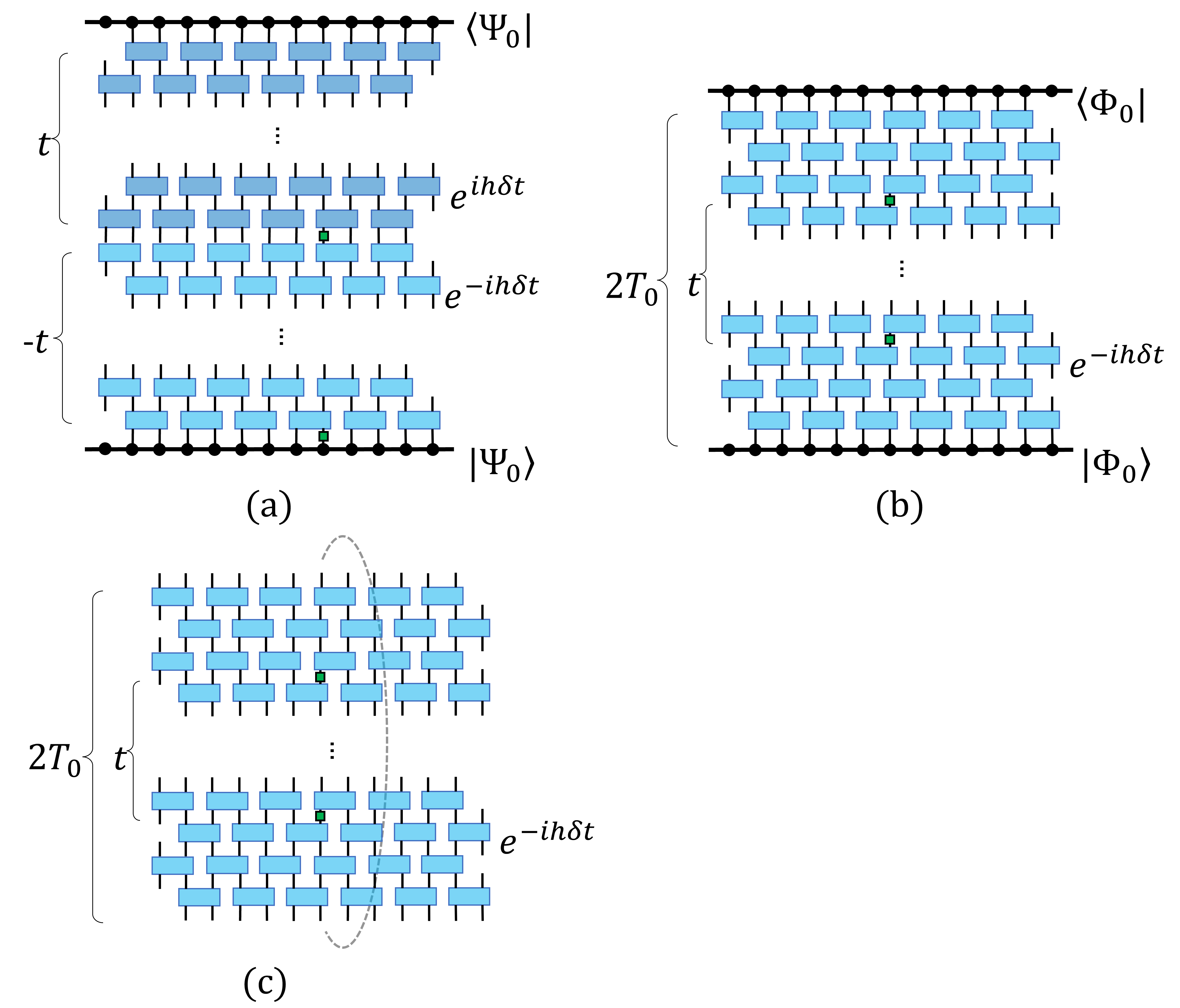}
    \caption{Tensor network representations for real-time correlations. (a) Operator formalism, which is constructed according to Eq.~(\ref{equ:operator}). (b) The numerator of the path-integral formalism of the two-body correlation function, which can be computed by TRG with a boundary. For the denominator, we simply remove the single-body operators. (c) The numerator of the path-integral formalism with slight modification. Here we removed the boundary and the algorithm to compute the real-time correlation function is exactly the same as the one for the imaginary time case, except that we use a complex $\delta t$ in two body gate $e^{-iH\delta t}$ instead of $e^{-\delta \tau H}$ for real $\delta \tau$.  }
  \label{fig:realt_TNrep}
\end{figure}

\label{sec:corre_real}
%%% introduction needed
%\subsection{ Tensor network representation in path-integral formalism }
We can construct a tensor network representation (see Fig.~\ref{fig:realt_TNrep}(a)) from the definition of real-time correlation (in operator formalism): 
\begin{equation}
\langle \sigma^{\alpha}_{i} \left( t \right) \sigma^{\alpha}_{i} \rangle = \langle \Psi_{0} | e^{iHt} \sigma^{\alpha}_{i} e^{-iHt} \sigma^{\alpha}_{i} | \Psi_{0} \rangle
\label{equ:operator}
\end{equation}
where $ | \Psi_{0} \rangle $ is the ground state of the Hamiltonian $H$ and $ \sigma^{\alpha}_{i} $ is a single-body operator at site $i$ (here it is the Pauli matrix where $\alpha$ can be $x,y$ or $z$). In this representation, we can use an MPS to represent the ground state $| \Psi_{0} \rangle$, and the evolution operator $e^{-iHt}$ can be decomposed into local gates through Suzuki-Trotter decomposition. The single-body operator $\sigma^{\alpha}_{i}$ is again represented by green squares in Fig.~\ref{fig:realt_TNrep}.

% Such tensor network representation is the starting point of many methods based on MPS, such as TEBD \cite{Vidal04,Vidal07}, tMPS \cite{Verstraete04,Zwolak04}, TDVP \cite{Haegeman11,Haegeman16}, etc. 
Although such a tensor network representation is the starting point of MPS/DMRG based based methods, it is not convenient for TNR-based methods. When applying TRG or Loop-TNR, we usually assume that there are infinitely many tensors in both space and time directions. While for this case, the length in time direction is finite.
%which is proportional to $t$, the temporal difference.
Furthermore, the TRG or loop-TNR scheme is also ill-defined at the line where the two kinds of gates $e^{-iH \delta t} $ and $e^{iH \delta t} $ merge. Therefore, a tensor network representation suited for TRG or Loop-TNR is very desired.

%\FloatBarrier

\begin{figure}[htb]
  \centering
  \includegraphics[width=0.6\linewidth]{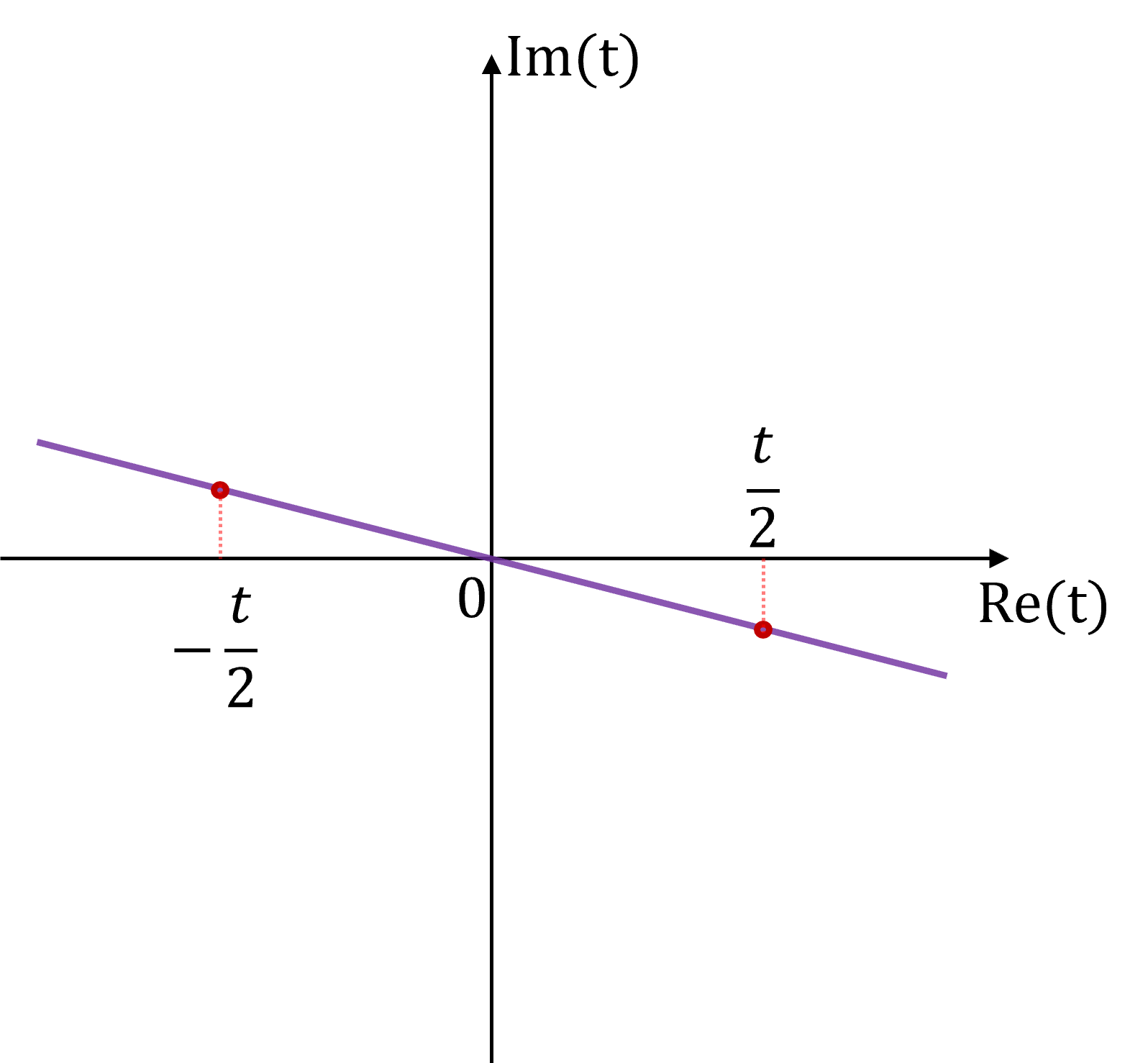} % try different size
    \caption{ The integral contour, marked by a purple line, along which we perform path-integral. A small $\epsilon$ is chosen to ensure convergence. }
    %As a consequence, the contour has a small deviation from the real axis and drives integration away from the exact result. }
  \label{fig:complext}
\end{figure}

It turns out that the path-integral formalism of real-time correlation is a better choice \cite{peskin95}: 
\begin{equation}
\langle \sigma^{\alpha}_{i} \left( t \right) \sigma^{\alpha}_{i} \rangle = \lim_{ T_{0} \to \infty (1-i\epsilon) } \frac{ \langle \Phi_{0} | \hat{ \mathrm{T} } [e^{ -i \int_{ -T_{0}}^{T_{0}} \!H\,\mathrm{d}t' } \sigma_{i}^{\alpha}\left(t \right) \sigma_{i}^{\alpha}] | \Phi_{0}\rangle }{ \langle \Phi_{0} | \hat{ \mathrm{T} } e^{ -i \int_{ -T_{0} }^{T_{0}} \!H\,\mathrm{d}t' }| \Phi_{0} \rangle },
\label{equ:path-int}
\end{equation}
% \begin{equation}
% \langle \sigma^{\alpha}_{i} \left( t \right) \sigma^{\alpha}_{i} \rangle = \lim_{ \beta \to \infty } \frac{ \Tr (\hat{ \mathrm{T} } \sigma_{i}^{\alpha}\left(t \right) \sigma_{i}^{\alpha}(0) e^{ -\beta H })  }{ \Tr (e^{-\beta H}) }
% \label{equ:path-int}
% \end{equation}
% (\textbf{This formulation is incorrect, we need to take the trace in the actual calculation, in general, the tenor network simulation is not a perturbative calculation. Just use the most general fintie temperature formulism and take the limit temperature go to zero, as we discussed long time before.})
where $\hat{ \mathrm{T}}$ is the time-ordering operator and $ | \Phi_{0} \rangle $ can be chosen arbitrarily, as long as it is not orthogonal to $| \Psi_{0} \rangle$, the exact ground state of $H$. 
%Note here that $\sigma_{i}^{\alpha}\left(t \right)$ is not written in the Heisenberg picture. The $t$ in $\sigma_{i}^{\alpha}(t)$ is just a formal label.
To make the path-integral formalism Eq.~(\ref{equ:path-int}) consistent with the standard definition of real-time correlation Eq.~(\ref{equ:operator}), we should perform the integral on complex $t$ plane, i.e., $t = \left(1-i \epsilon  \right)t'$ for a small $\epsilon$ as can be seen in Fig.~\ref{fig:complext} and require $T_{0} $ in Eq.~(\ref{equ:path-int}) to approach $\infty (1-i\epsilon)$. 
% The detailed derivation from Eq.~(\ref{equ:path-int}) to Eq.~(\ref{equ:operator}) is introduced in Eq.~(3.8) in Ref.\cite{negele95} and is similar to the derivation in Eq.~(\ref{equ:pbc_pathint}).

%\FloatBarrier
Similar to the imaginary time case, now we can construct a tensor network representation from the path-integral formalism, as seen in Fig.~\ref{fig:realt_TNrep}(b) for the numerator. For the denominator, we just remove the single-body operators (represented by green squares) in the representation of the numerator. Since now we require $ |T_{0}| $ to go to infinity, we have infinitely many tensors in the time direction. Besides, we only have one type of gate in the tensor network. Therefore a TRG scheme is well defined. 
%Moreover, we no longer require the boundary MPS to represent the exact ground state of the Hamiltonian. This brings convenience for the computation under critical systems, the MPS representation of which is difficult to obtain.
%Although it is straightforward to develop a scheme for TRG with a boundary, it is difficult to apply Loop-TNR to such a tensor network due to the presence of the boundary: the number of impurity tensors would increase a lot through the loop-optimization step on the boundary. Besides, we usually choose a large virtual bond dimension for the MPS tensor to describe the ground state accurately, resulting in an inefficient loop-optimization step.
In addition, we also make a slight modification on Eq.~(\ref{equ:path-int}), by replacing the average over $| \Phi \rangle$ with the trace over a complete set $\{ | \phi_{n} \rangle \}$ that is not orthogonal to the true ground state $| \Psi \rangle $. Such modification can be regarded as imposing periodic boundary conditions for the path integral, which does not affect the results in the thermodynamic limit.
%makes it more complicated for theoretical calculation, but brings convenience for tensor network computation: it removes the boundary, and in time direction the tensors are contracted in periodic boundary condition, which is the same as in space direction.
%It is straightforward to verify such modification reproduces the original definition of real-time correlation as $T_{0}$ approaches infinity.
Fig.~\ref{fig:realt_TNrep}(c) shows the corresponding tensor network representation for the numerator.

With the boundary removed, the algorithm for computing the real-time correlation function by Loop-TNR is identical to the algorithm for imaginary-time correlation, except for the substitution of the imaginary time $\tau$ by the complex time with respect to $t = \left(1-i \epsilon  \right)t'$ in the definition of the two-body gate.

%We note that the present tensor network representation Fig.~\ref{fig:realt_TNrep}(c) is not only suitable for other TRG-based methods, such as TNR \cite{Evenbly15,Evenbly17} and HOTRG \cite{Xie12}, etc, but also applicable to (boundary) MPS based methods. For example, CTM-based approaches \cite{Orus09} and VUMPS \cite{Fishman18}. In Appendix.~\ref{sec:App_VUMPS}, we will show the method to contract the tensor network in Fig.~\ref{fig:realt_TNrep} and compute the real-time correlation using VUMPS.

\begin{figure}[tb]
  \centering

  % \subfloat[]{ \includegraphics[width=0.8\linewidth]{./figures/rt05e006r.png} }
  % \hfill
  \subfloat[]{ \includegraphics[width=0.9\linewidth]{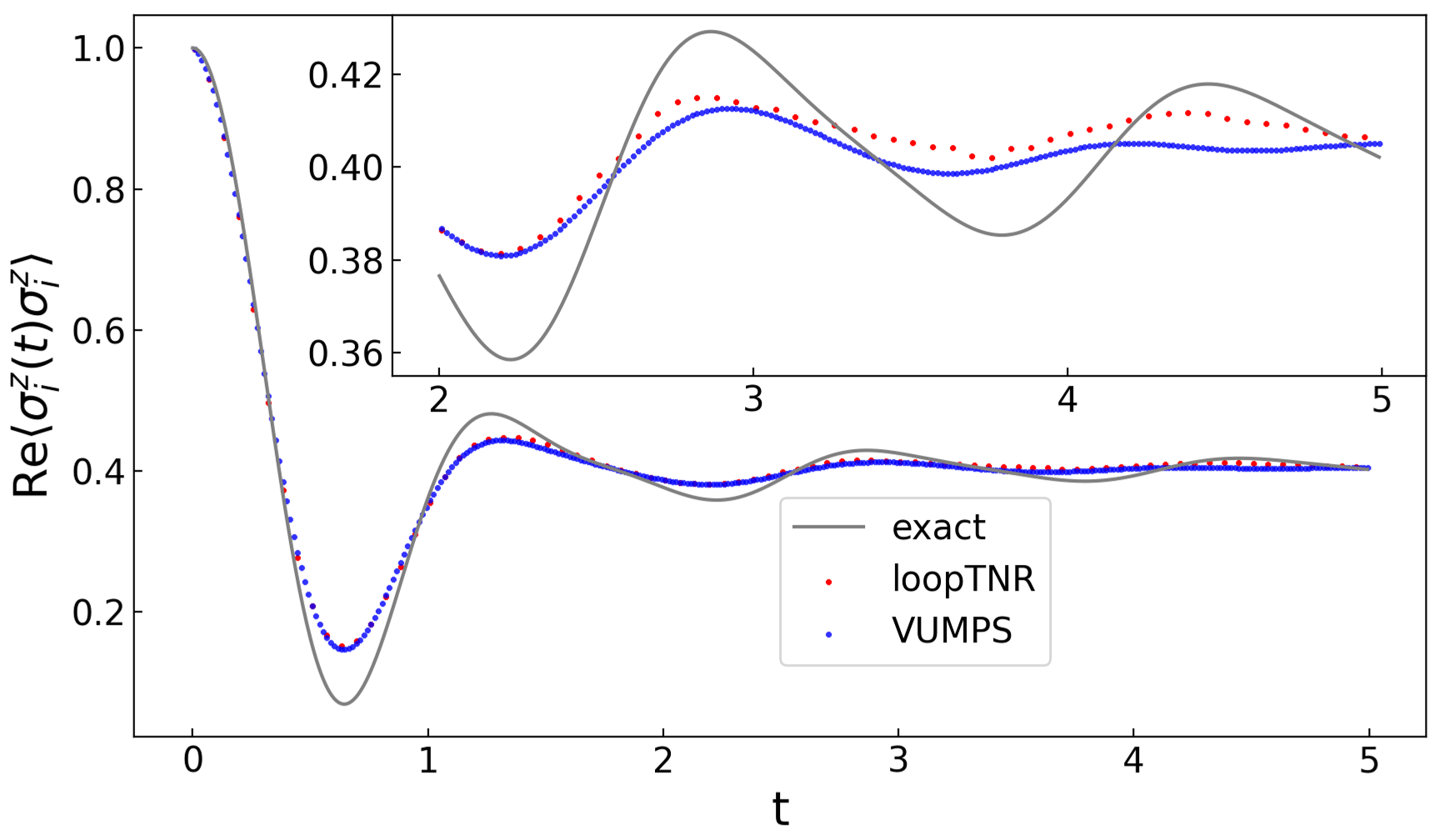} }
  \hfill
  \subfloat[]{ \includegraphics[width=0.9\linewidth]{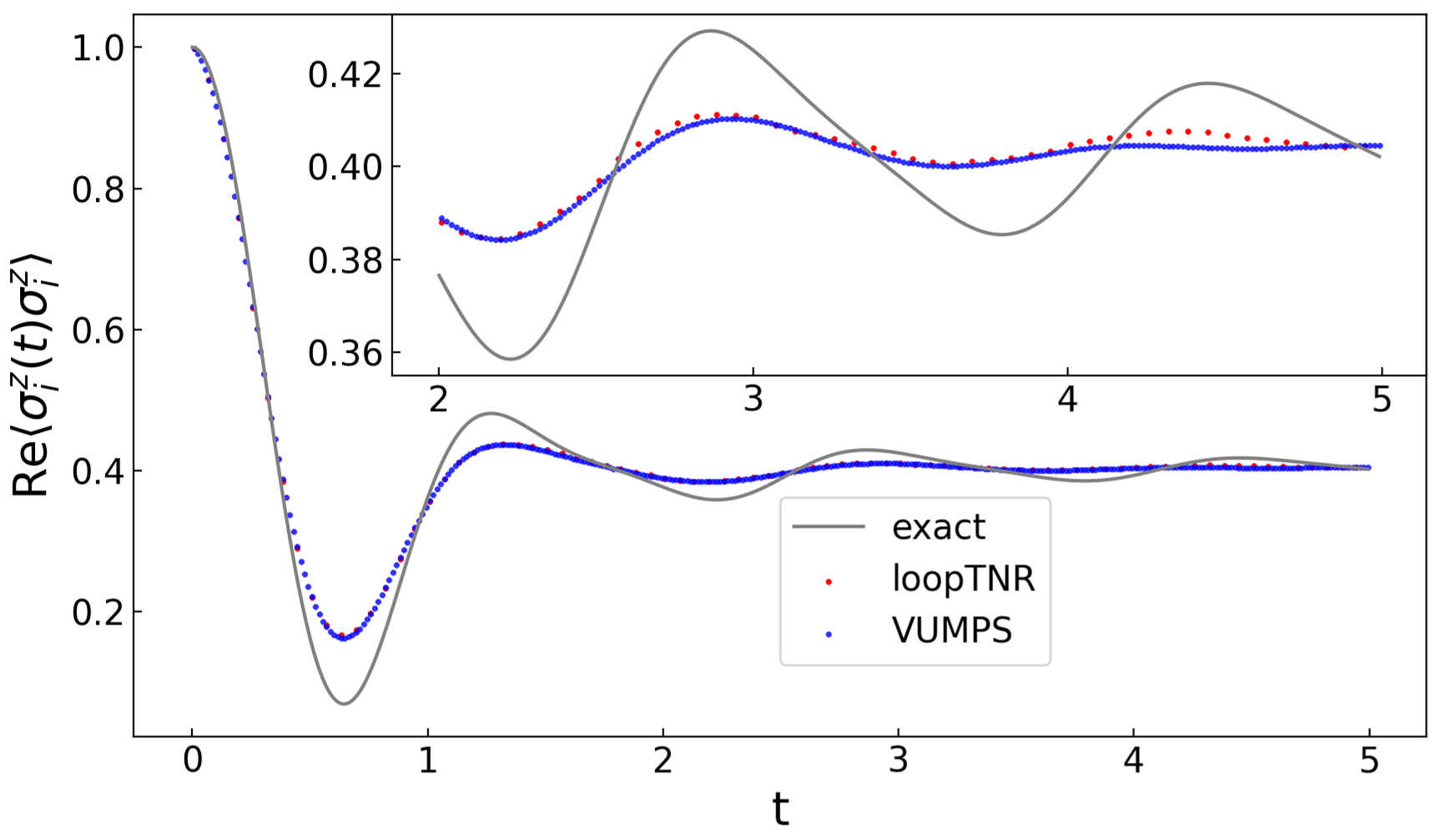} }
  \hfill
  \subfloat[]{ \includegraphics[width=0.9\linewidth]{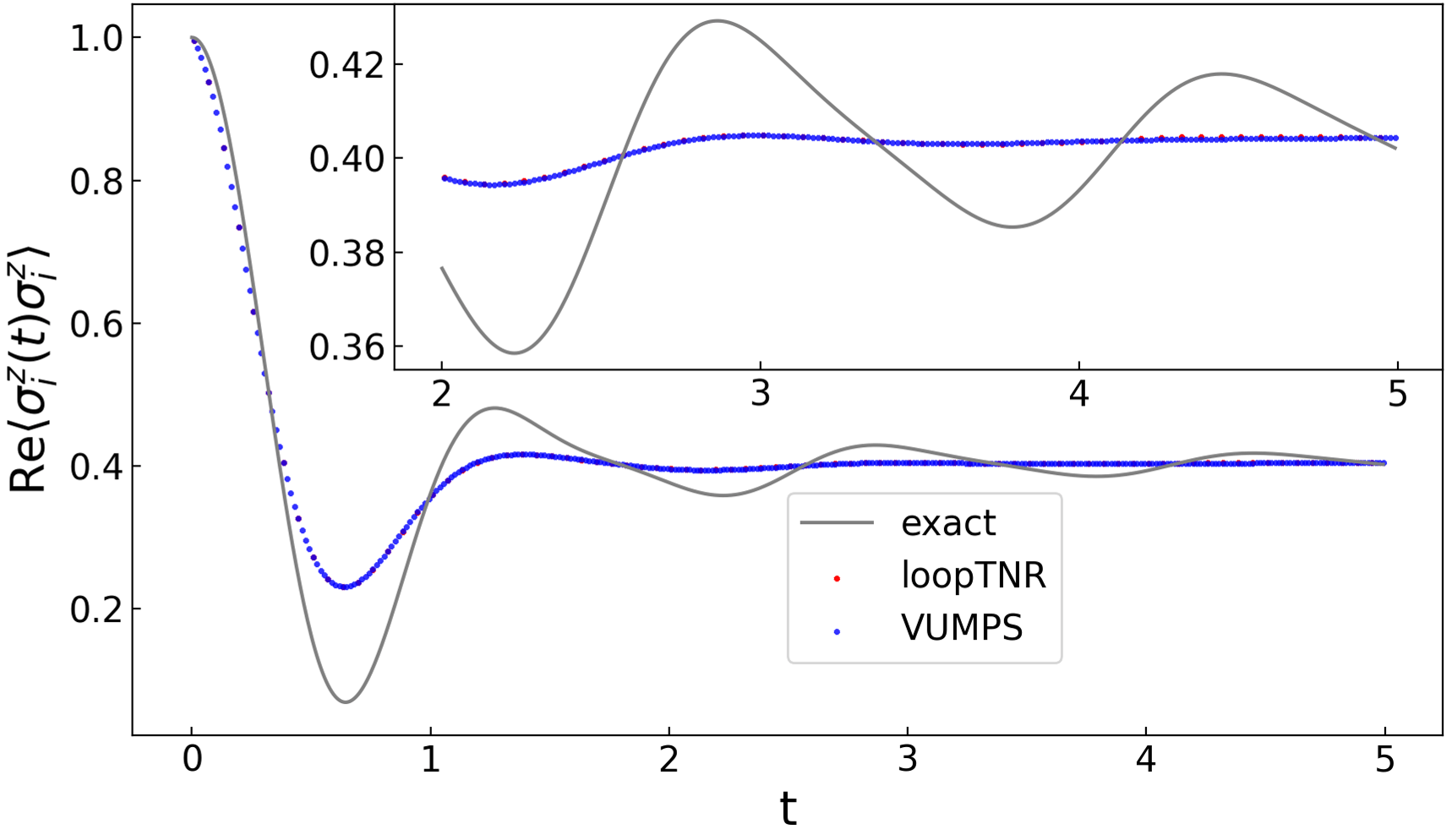} }

  \caption{ Real part for real-time correlation function obtained by Loop-TNR with $D_{cut}=32 $ and VUMPS with $\chi=200 $ for (a) $\epsilon=0.08 $, (b) $\epsilon=0.1 $, and (c) $\epsilon=0.2 $. }
  \label{fig:rt05epsr}
\end{figure}

\begin{figure}[tb]
  \centering

  % \subfloat[]{ \includegraphics[width=0.8\linewidth]{./figures/rt05e006i.png} }
  % \hfill
  \subfloat[]{ \includegraphics[width=0.9\linewidth]{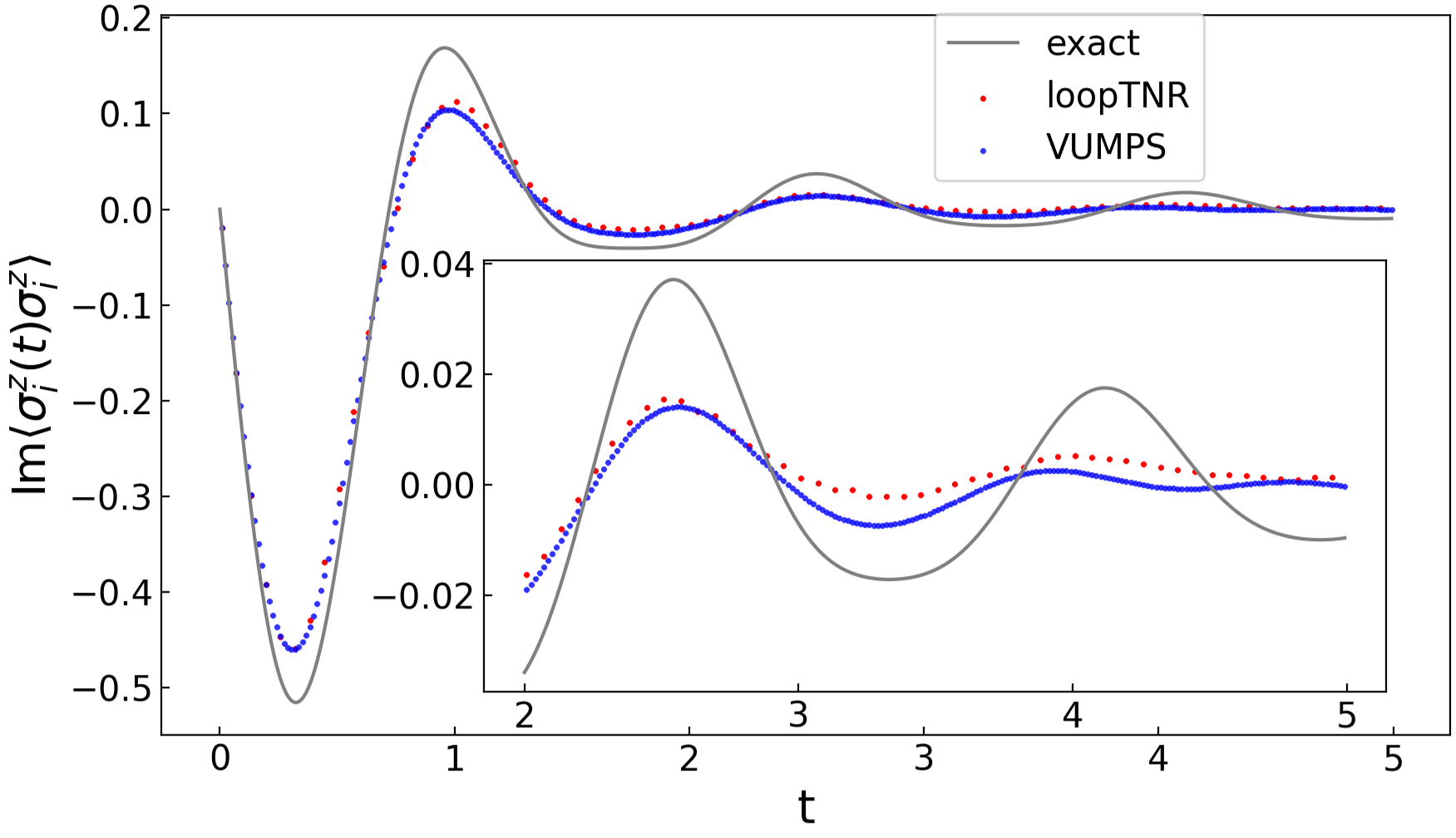} }
  \hfill
  \subfloat[]{ \includegraphics[width=0.9\linewidth]{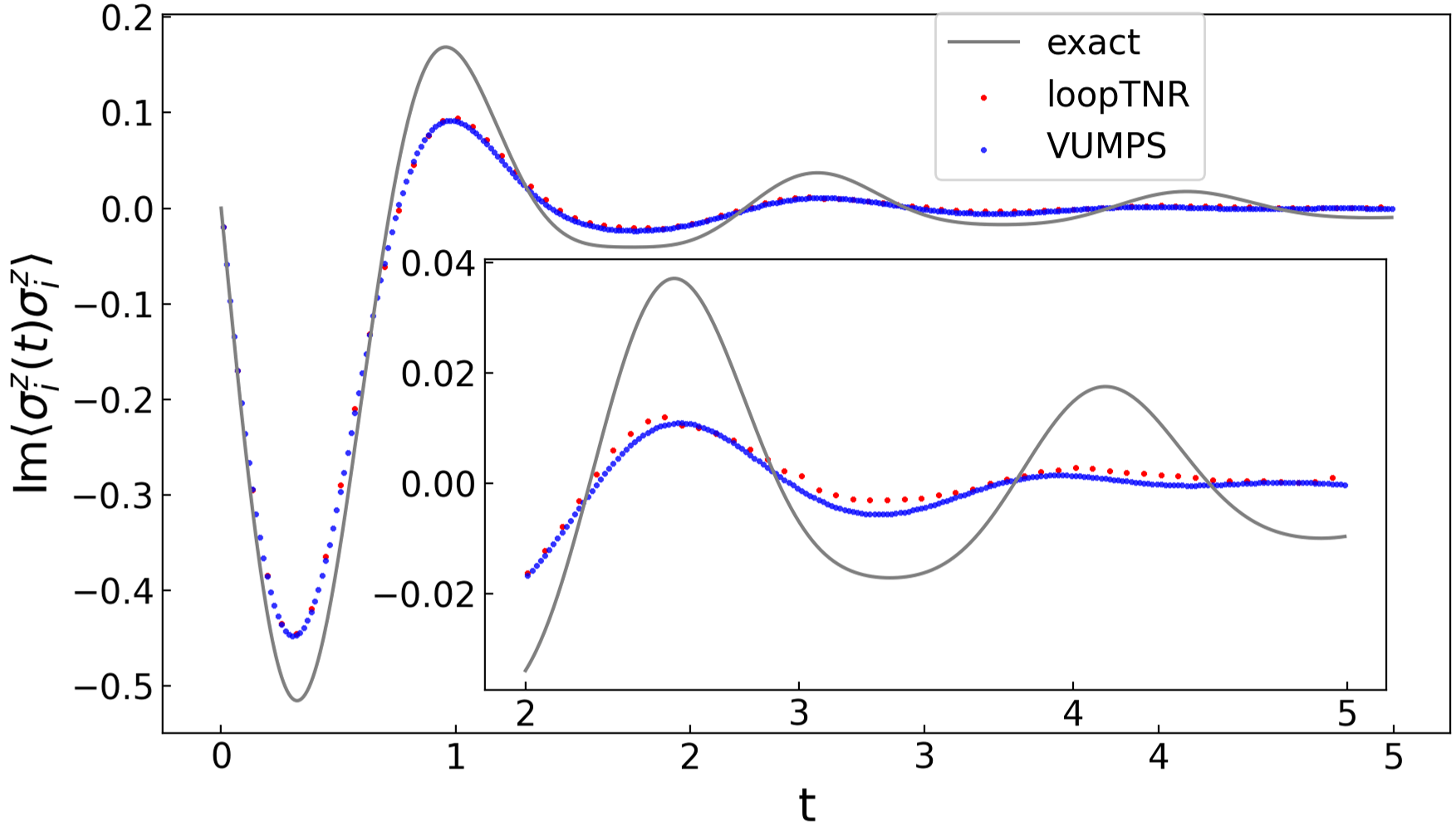} }
  \hfill
  \subfloat[]{ \includegraphics[width=0.9\linewidth]{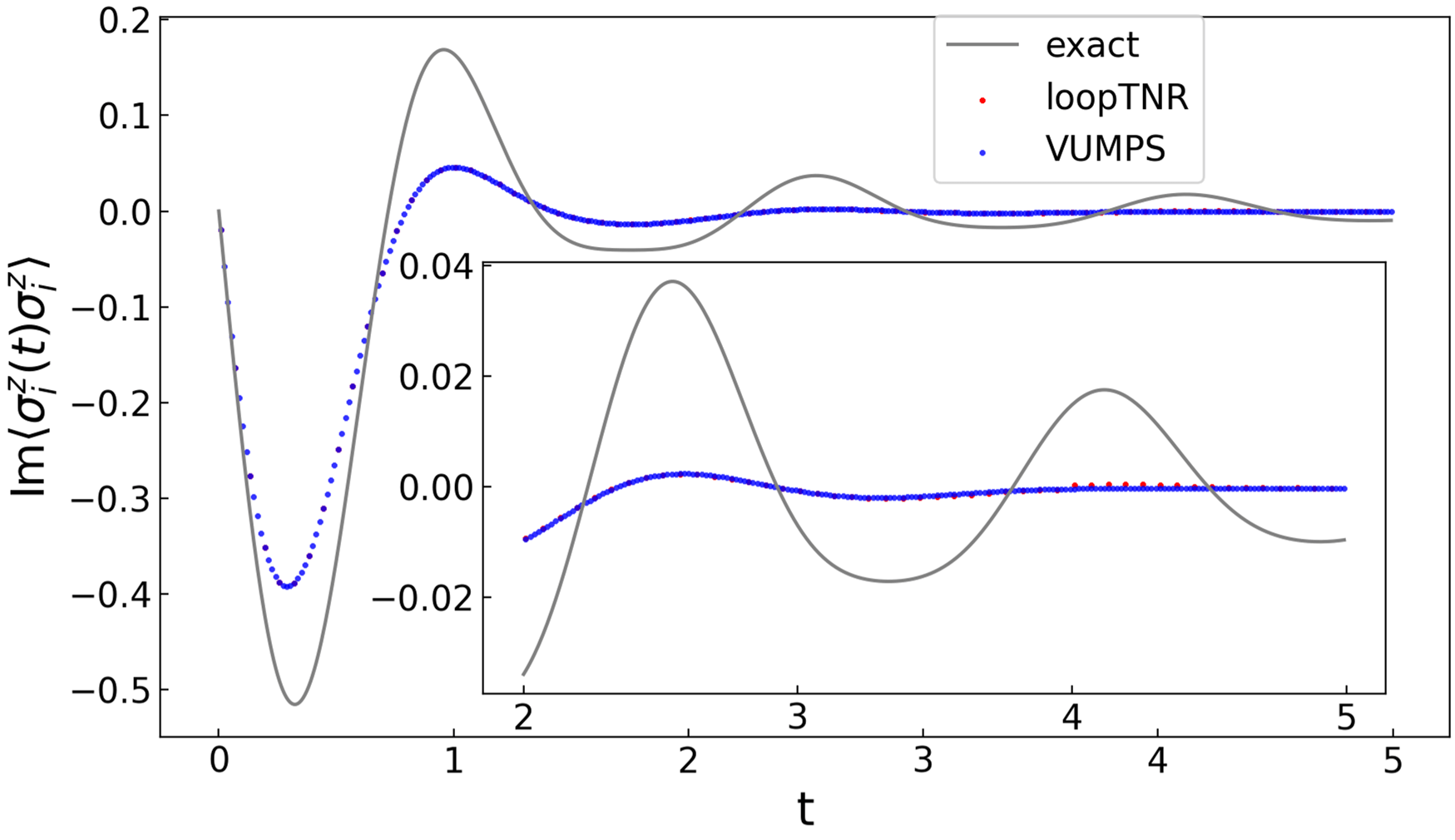} }
  
  \caption{ Imaginary part for real-time correlation function obtained by Loop-TNR with $D_{cut}=32 $ and VUMPS with $\chi=200 $ for (a) $\epsilon=0.08 $, (b) $\epsilon=0.1 $, and (c) $\epsilon=0.2 $.}
  \label{fig:rt05epsi}
\end{figure}

% Real-time correlation
 Finally, we present the results of transverse correlation $\langle \sigma^{z}_{i}(t) \sigma^{z}_{i} \rangle$ in the quantum Ising model at the critical point. The results are obtained by Loop-TNR with $D_{cut}=32$ and compared with the results by VUMPS, where the virtual bond dimension is chosen as $\chi=200$. See Fig.~\ref{fig:rt05epsr} and Fig.~\ref{fig:rt05epsi} for the real part and imaginary part with different $\epsilon $'s, respectively. We performed 24 Loop-TNR iterations to obtain the results, which correspond to a quantum Ising chain with $2^{14} $ spins.

 From these results shown in Fig.~\ref{fig:rt05epsr} and Fig.~\ref{fig:rt05epsi}, we see that for smaller $\epsilon $, the oscillation of the curve is much more bigger. While for a large $\epsilon=0.2$, the results are almost flat and featureless. 
 %Moreover, the results from Loop-TNR have more discrete points as $\epsilon $ gets smaller. 
 In general, the truncation error also increases for small $\epsilon$.
%defined in Eq.~(\ref{equ:errorloop}) as $\epsilon$ gets smaller. 
For example, the truncation error for $\epsilon=0.2$ is about $\mathcal{O}(10^{-2} ) $. When we decrease $\epsilon$ to 0.08, the error becomes $\mathcal{O}(10^{-1})$, which is much larger.  
% looperror v.s. itereation step for different epsilon should be includeed
%Although there are small  in the result of 
We note that even for small $\epsilon$, the Loop-TNR results still qualitatively follow the correct tendency as the exact ones even for large $t \approx 5$, e.g., it gives rise to the correct peak and dip features for the oscillated curve. In contrast, the VUMPS results cannot capture the peak and dip features for small $\epsilon$ at large $t$.

\begin{figure}[h]
  \centering
  
  \subfloat[]{ \includegraphics[width=1\linewidth]{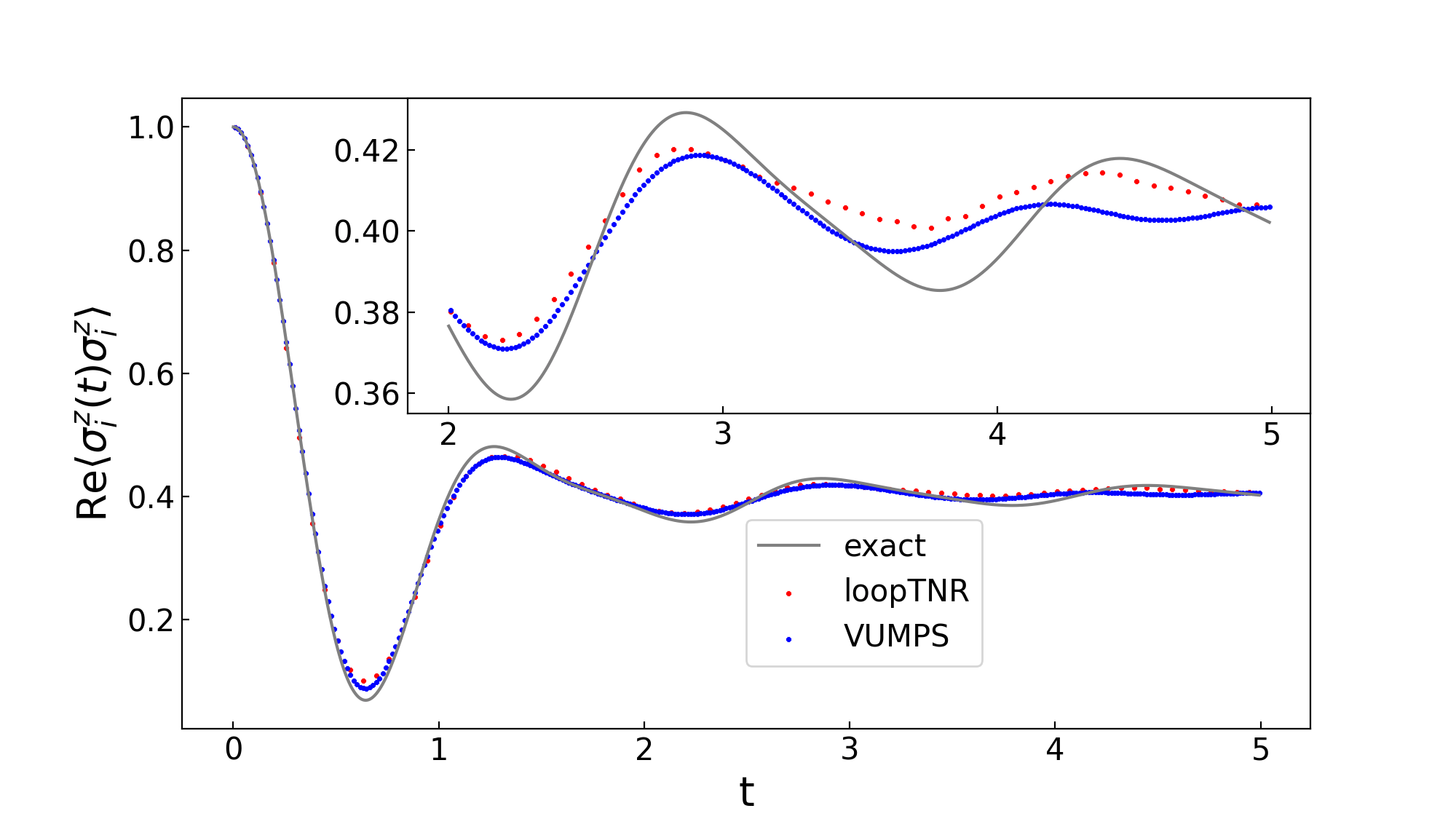} }
  \hfill
  \subfloat[]{ \includegraphics[width=1\linewidth]{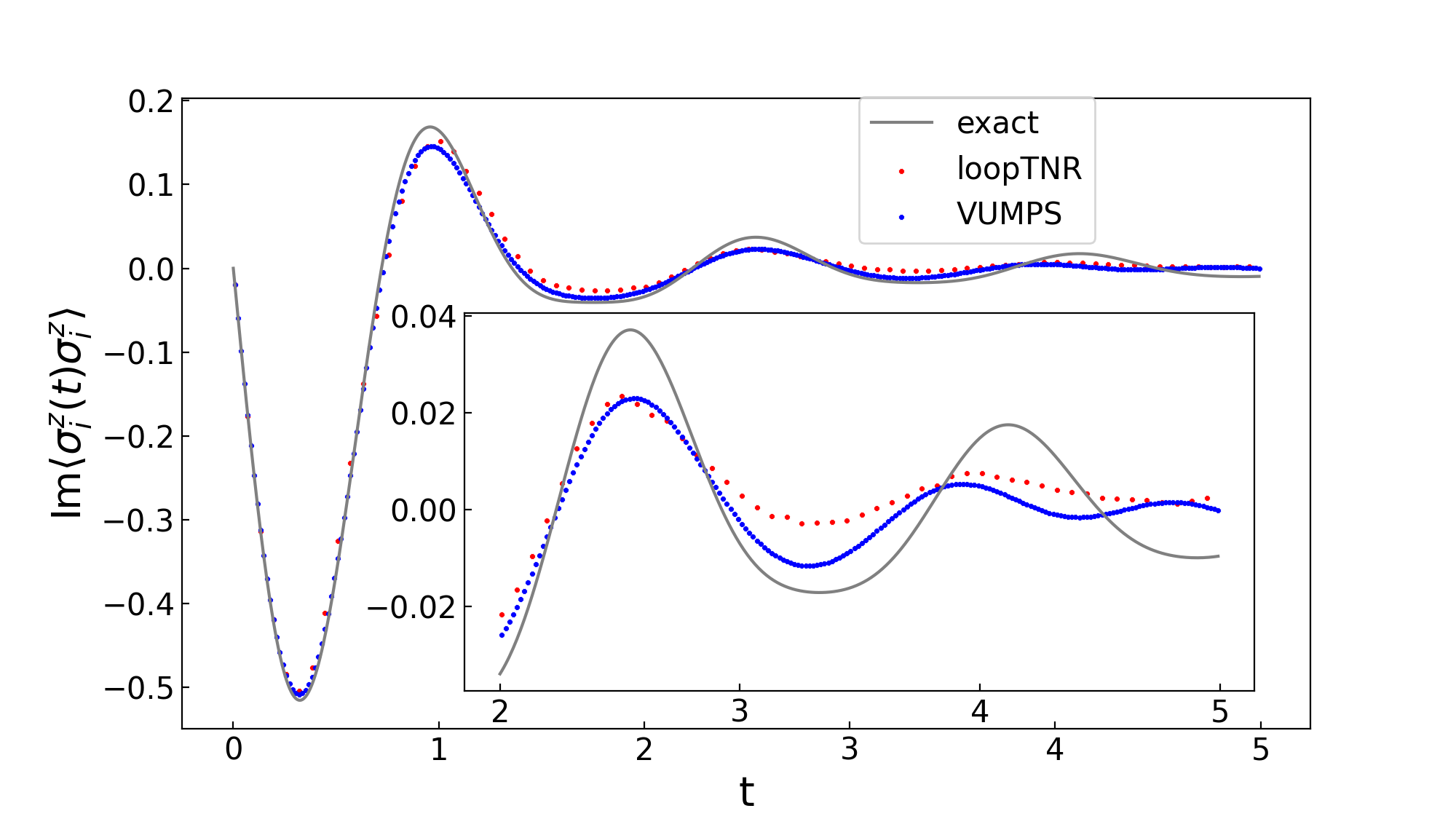} }

  % \subfloat[]{ \includegraphics[width=0.8\linewidth]{./figures/rt05e0r_4pt.png} }
  % \hfill
  % \subfloat[]{ \includegraphics[width=0.8\linewidth]{./figures/rt05e0i_4pt.png} }
    \caption{ $\epsilon \rightarrow 0 $ result of the real-time correlation obtained by the linear fitting method. (a) Real part. (b) Imaginary part. }
  \label{fig:rt05eps0}
\end{figure}

In the theoretical calculation, we should choose to set $\epsilon \rightarrow 0 $ in the end. Computationally, this can be done by using a linear fitting of $\epsilon $. That is, we assume for small $\epsilon $'s, the correlation varies linearly with $\epsilon $:
\begin{equation}
C(\epsilon,t) = A \epsilon + C(0,t)
\label{equ:liearFitting_eps}
\end{equation}
where $A$ is a constant and the intercept $C(0,t)$ denotes the 'true' correlation function. The $\epsilon \rightarrow 0 $ result by linear fitting is shown in Fig.~\ref{fig:rt05eps0}, where the data obtained with $\epsilon=0.08 $, $0.1$ and $0.2$ are utilized.

%\FloatBarrier

%%%%%%%%%%%%%%%%%%%%%%%%%%%%%%%% Conformal data
% Extraction of conformal data connecting part should be added: the above only for quantum Ising model, for general model to make the tensor isotropic, we should utilize \tau, where distance between neighouring layers of tensors after compression is changed, but can be directly identified.
\section{ Conformal data for non-Hermitian systems} % needs modification
\label{sec:conformal}

\subsection{A review of computing conformal data using Loop-TNR algorithm}
In this section, we present the method to extract conformal data (specifically, central charge, scaling dimension, and conformal spin) from the fixed-point tensor by Loop-TNR. 
It should be noted that similar methods have been introduced previously \cite{Zcgu09,Bao19}. Here we attempt to provide more details.

\begin{figure}[h]
  \centering
  \includegraphics[width=\linewidth]{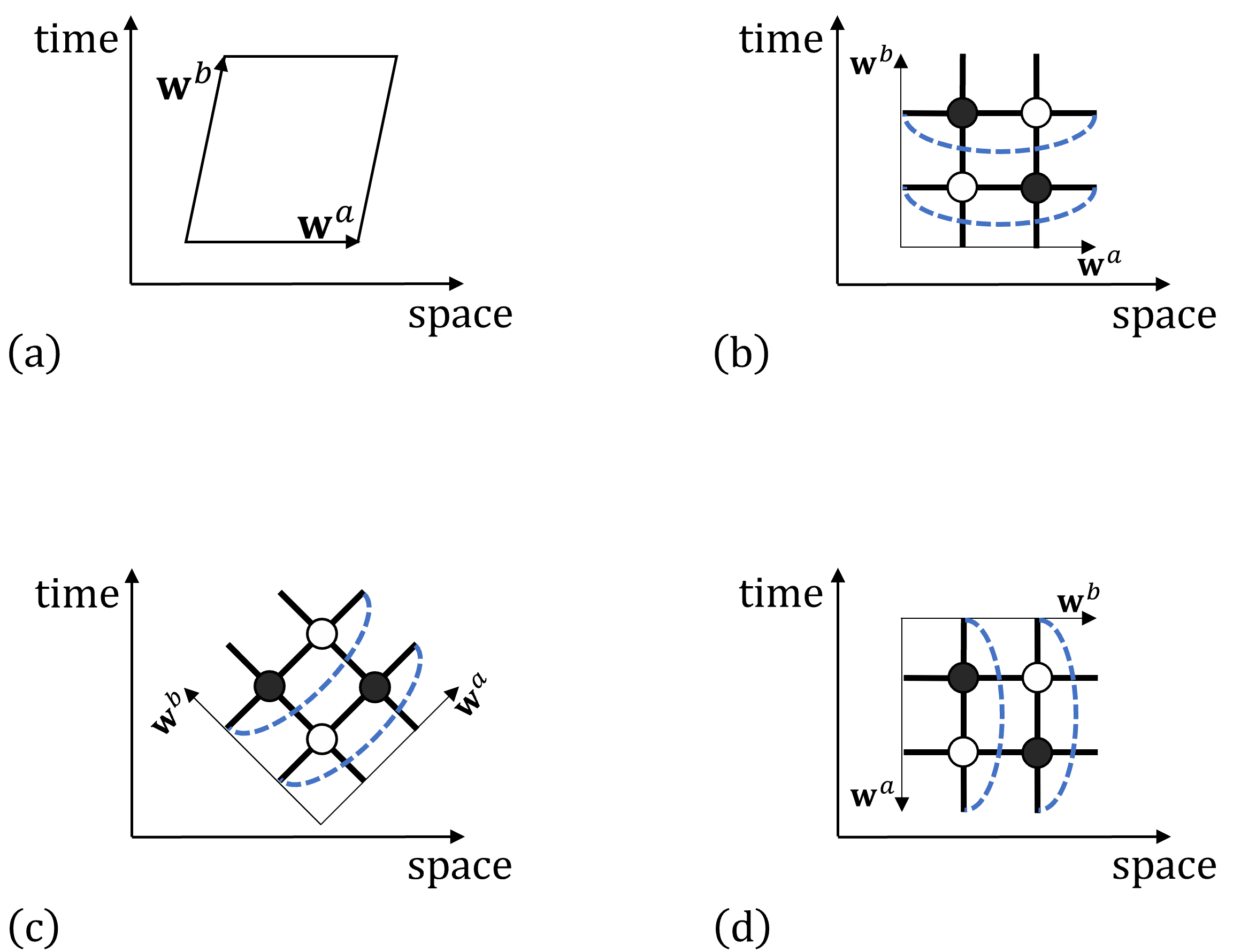}  
    \caption{Modular parameter $\tau$ for (a) a general space-time region; (b) $L=2$ transfer matrix $M^{ud}$ at even RG steps; (c) $L=2$ transfer matrix $M^{ud}$ at odd RG steps and (d) $L=2$ transfer matrix $M^{lr}$ at even RG steps }
  \label{fig:modu_para}
\end{figure}

% theory part
According to conformal field theory (CFT) \cite{Francesco97}, a transfer matrix on a cylinder, which generates the translation along time and space direction, is defined as:
\begin{equation}
\mathcal{T}_{m,n} = e^{ - \frac{m}{n} [ \textrm{Im} (\tau) H + \mathrm{i} \textrm{Re} (\tau) S ] }
\label{equ:trans_mat}
\end{equation}
where $H$ corresponds to the energy operator and $S$ is the spin operator of the system. $m$ and $n$ denote the translations in time and space directions, respectively. The modular parameter $\tau$ describes the geometry of the space-time:
\begin{equation}
\tau = \frac{ \mathbf{w^{b}} }{ \mathbf{w^{a}} } = \frac{ w^{b}_{x} + i w^{b}_{y} }{ w^{a}_{x} + i w^{a}_{y} }
\label{equ:modu_para}
\end{equation}
where the vectors $ \mathbf{w^{a}} $ and $ \mathbf{w^{b}}$ are the basis for the space-time, as shown in Fig.~\ref{fig:modu_para}.
 
\begin{figure}[h]
  \centering
  \includegraphics[width=\linewidth]{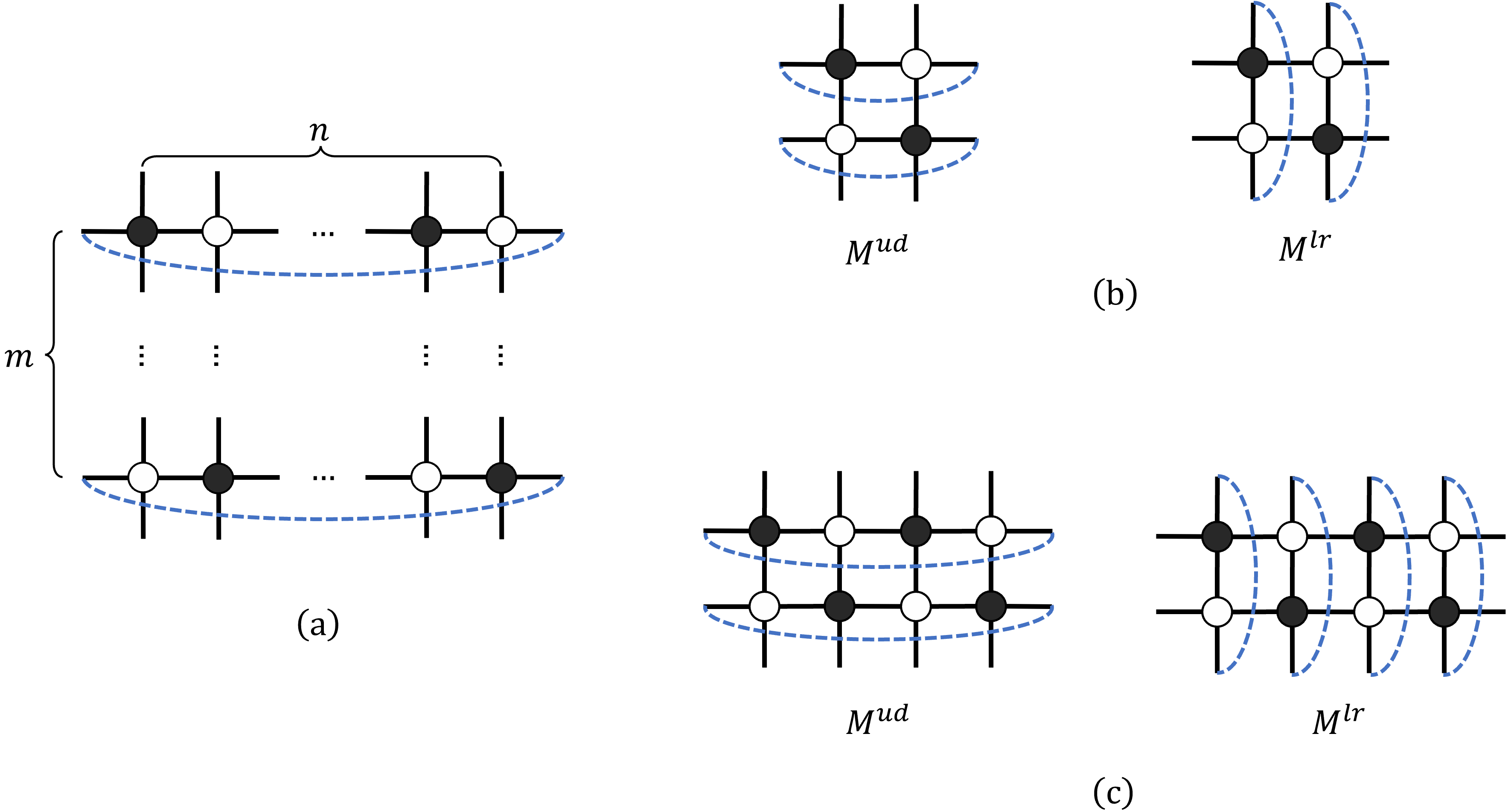}  
    \caption{ Transfer matrices represented by tensor networks. (a) A transfer matrix that is represented as a tensor network with m-by-n tensors. Note that m and n are even integers since for Loop-TNR, there are $T_{a}$, $T_{b}$ sublattices. (b) $L=2$ transfer matrices, with horizontal bonds contracted (left), denoted as $M^{ud} $ and vertical bonds contracted (right), denoted as $M^{lr} $. (c) $L=4$ transfer matrices, with horizontal bonds contracted (left), denoted as $M^{ud} $ and vertical bonds contracted (right), denoted as $M^{lr} $.  }
  \label{fig:trans_mats}
\end{figure}

% numerical part
Note that the trace of the transfer matrix Eq.~(\ref{equ:trans_mat}) corresponds to the partition function of a system. While the partition function has already been described as the trace of a tensor network previously, for example, in Fig.~\ref{fig:iniTensor}(b) and in the denominator of Fig.~\ref{fig:sin_op}(c). Therefore, a generic m-by-n transfer matrix can be represented by a tensor network as depicted in Fig.~\ref{fig:trans_mats}(a). In practical computations, we usually use $L=2$ and $L=4$ transfer matrices to obtain conformal data, as shown in Fig.~\ref{fig:trans_mats}(b) and (c), respectively.
 
The eigenvalues of the transfer matrix correspond to the eigenvalues of $H$ and $S$, hence to the central charge, scaling dimensions, and conformal spins as:
\begin{equation}
\lambda_{i} = e^{ -\frac{m}{n} [ 2\pi \textrm{Im} (\tau) ( \Delta_{i}- \frac{c}{12} ) +  2 \pi \mathrm{i} \textrm{Re} (\tau) s_{i} ] }
\label{equ:eigen_conformal} % m/n is added
\end{equation} % the eigenvalues should be arraged in descending order.
if the transfer matrix is properly normalized. Otherwise, there will be correction factors. In even RG iterations, the modular parameter $\tau^{ud} $ is purely imaginary, as can be seen in Fig.~\ref{fig:modu_para}. Consequently, the relations of eigenvalues and conformal data Eq.~(\ref{equ:udlr_transfer}) at even RG iterations, without proper normalization of the transfer matrix, should be:
\begin{equation}
\lambda_{i} = e^{ -2\pi \frac{m}{n} \textrm{Im} (\tau ) ( \Delta_{i}- \frac{c}{12} )-mn\varepsilon  }
\label{equ:no_normalize}
\end{equation}
where $\varepsilon$ represents the scaled energy density. To calculate the last term, we need both $L=2$ and $L=4$ transfer matrices:
\begin{equation}
\begin{aligned}
\lambda_{i}^{L=2} =& e^{ -2\pi \textrm{Im} (\tau ) ( \Delta_{i}- \frac{c}{12} )-4\varepsilon } \\
\lambda_{i}^{L=4} =& e^{ -2\pi \frac{1}{2} \textrm{Im} (\tau ) ( \Delta_{i}- \frac{c}{12} )-8\varepsilon }
\label{equ:normalize_transfer}
\end{aligned}
\end{equation}
Once $\varepsilon$ is solved, we normalize the transfer matrix as $\mathcal{T}_{m,n}/e^{-mn \epsilon} $. Then the eigenvalues of the normalized transfer matrix produce correct conformal data as in Eq.~(\ref{equ:eigen_conformal}).

\begin{figure}[tb]
  \centering
  \includegraphics[width=\linewidth]{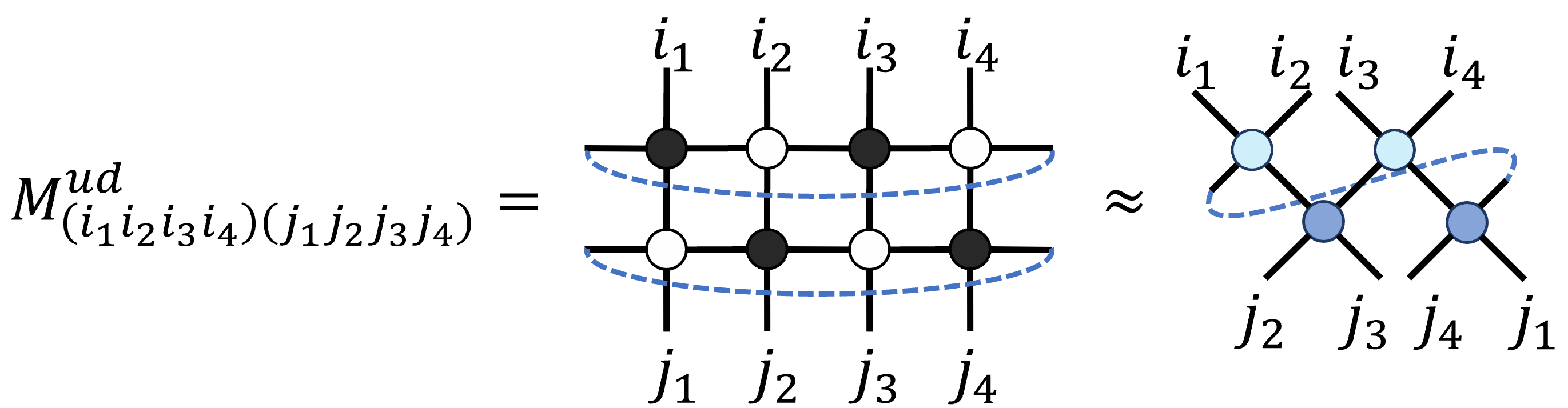}  
    \caption{ Approximate the original $L=4$ transfer matrix by using the tensors in the next RG iteration. The indices denote the order of bonds when grouped into a matrix. }
  \label{fig:L4approx}
\end{figure}

Note that, the computational cost will be high if we directly construct the $L=4$ transfer matrix $M^{ud}$ and compute the eigenvalues $\lambda_{i}^{L=4}$ in Eq.~(\ref{equ:normalize_transfer}). The computational cost can be reduced if we use the tensors in the next RG iteration to approximate the original $L=4$ transfer matrix, as depicted in Fig.~\ref{fig:L4approx}. The computation of eigenvalues for the transfer matrix can be made implicit with the Arnoldi algorithm to reduce the computational cost. 
Compared with the $L=2$ transfer matrix, the $L=4$ transfer matrix produces more distinguishable scaling dimensions and conformal spins.
% \FloatBarrier
% %
% which can be calculated by the eigenvalues of the transfer matrices in Fig.~\ref{fig:trans_mats}, up to a proper normalization (we will resolve the normalization problem later). Note that, in most cases, the eigenvalues are arranged in descending order to reveal the low energy property of the system at the critical point.
% %

We can either contract the horizontal bonds and leave vertical ones open, or the reverse, to construct the transfer matrix, see Fig.~\ref{fig:trans_mats}(b) and (c). In the last choice, we exchange the role of space and time. The two choices bring two sets of equations:
\begin{equation}
\begin{aligned}
\lambda_{i}^{ud} =& e^{ -\frac{m}{n} [ 2\pi \textrm{Im} (\tau^{ud} ) ( \Delta_{i}- \frac{c}{12} ) +  2 \pi \mathrm{i}\textrm{Re} (\tau^{ud}) s_{i} ]}  \\
\lambda_{i}^{lr} =& e^{ -\frac{n}{m} [ 2\pi \textrm{Im} (\tau^{lr} ) ( \Delta_{i}- \frac{c}{12} ) +  2 \pi \mathrm{i}\textrm{Re} (\tau^{lr}) s_{i} ] }
\label{equ:udlr_transfer}
\end{aligned}
\end{equation}
where we have the relation $ \tau^{ud} = - \frac{1}{\tau^{lr} } $, as can be seen in Fig.~\ref{fig:modu_para}(b) and (d). Note that, the eigenvalues are arranged in descending order to reveal the low energy property of the system at the critical point. 
Furthermore, the scaling dimension corresponding to the identity operator is 0, that is, $\Delta_{0}= 0$. Based on the equations and results listed above, the conformal data can be solved below.

\subsubsection{Central charge and scaling dimensions}
For even RG iterations, the modular parameter $\tau^{ud} =iv $ is purely imaginary (see Fig.~\ref{fig:modu_para}(b) and (d) for an example of $L=2$ transfer matrices). Therefore we can determine scaling dimensions and central charge at even RG iterations:
\begin{equation}
c =\frac{n}{m} \frac{6}{\pi} \frac{ \ln \lambda_{0}^{ud} }{ \textrm{Im} ( \tau^{ud} ) } 
\label{equ:central_charge}
\end{equation}
\begin{equation}
\Delta_{i} = \frac{n}{m} \frac{1}{2 \pi \textrm{Im} (\tau^{ud} ) } \ln \frac{ \lambda_{0}^{ud} }{ \lambda_{i}^{ud} } 
\label{equ:scaling_dimension}
\end{equation}
where the modular parameter $\tau$ at even RG iterations is computed as:
\begin{equation}
\textrm{Im} ( \tau^{ud} ) =v =\frac{n}{m} \sqrt{ \frac{ \ln  \lambda_{i}^{ud} }{ \ln  \lambda_{i}^{lr} }  }
\label{equ:imag_modu_para}
\end{equation}
% % %

In Loop-TNR simulations, we should perform tensor initializations (see Section~\ref{sec:iniT}) to produce isotropic tensors. 
As a result, $\tau$ should be $i$. One way to achieve such an aim is to first set $\delta \tau=\frac{1}{2^{M} }$ in the two-body gate of Eq.~(\ref{equ:tensor_gate}) as a trial run, where $M$ is the number of compression steps. Then we compute the modular parameter $\tau$ at even RG iterations by Eq.~(\ref{equ:imag_modu_para}), which can be expressed as $\tau=iv$. 

For the next run of the Loop-TNR algorithm, we absorb $v$ into the new $\delta \tau' = \delta \tau / v$, and the two-body gate now becomes $e^{-\delta \tau' h_{i,i+1} } $. Consequently, the modular parameter is $i$ after compression, which indicates the tensors are isotropic.

% % for normalization factor
% We should note that the conformal data is correctly computed only when the transfer matrix is properly normalized. In even RG iterations, the relations of eigenvalues and conformal data Eq.~(\ref{equ:udlr_transfer}), without proper normalization of the transfer matrix, should be:
% \begin{equation}
% \lambda_{i} = e^{ -2\pi \frac{m}{n} \textrm{Im} (\tau ) ( \Delta_{i}- \frac{c}{12} )-mn\epsilon  }
% \label{equ:no_normalize}
% \end{equation}
% where $\epsilon$ represents the scaled energy density. To calculate the last term, we need both $L=2$ and $L=4$ transfer matrices:
% \begin{equation}
% \begin{aligned}
% \lambda_{i}^{L=2} =& e^{ -2\pi \textrm{Im} (\tau ) ( \Delta_{i}- \frac{c}{12} )-4\epsilon } \\
% \lambda_{i}^{L=4} =& e^{ -2\pi \frac{1}{2} \textrm{Im} (\tau ) ( \Delta_{i}- \frac{c}{12} )-8\epsilon }
% \label{equ:normalize_transfer}
% \end{aligned}
% \end{equation}
% Once we solve $\epsilon$, we normalize the transfer matrix as $\mathcal{T}_{m,n}/e^{-mn \epsilon} $. Then the eigenvalues of the normalized transfer matrix produce correct conformal data as in Eq.\ref{equ:eigen_conformal}.

\subsubsection{Conformal spins}
The conformal spins are determined at odd RG iterations:
\begin{equation}
s_{i} = \frac{n}{m} \frac{1}{4 \pi \textrm{i} \textrm{Re}(\tau^{ud} )} \ln \frac{ \lambda_{i}^{ud} }{ \bar{ \lambda }_{i}^{ud} }
\label{equ:conformal_spin}
\end{equation}
where $\bar{ \lambda }_{i}$ is the complex conjugate of $\lambda_{i}$.

\begin{figure}[tb]
  \centering
  \includegraphics[width=0.6\linewidth]{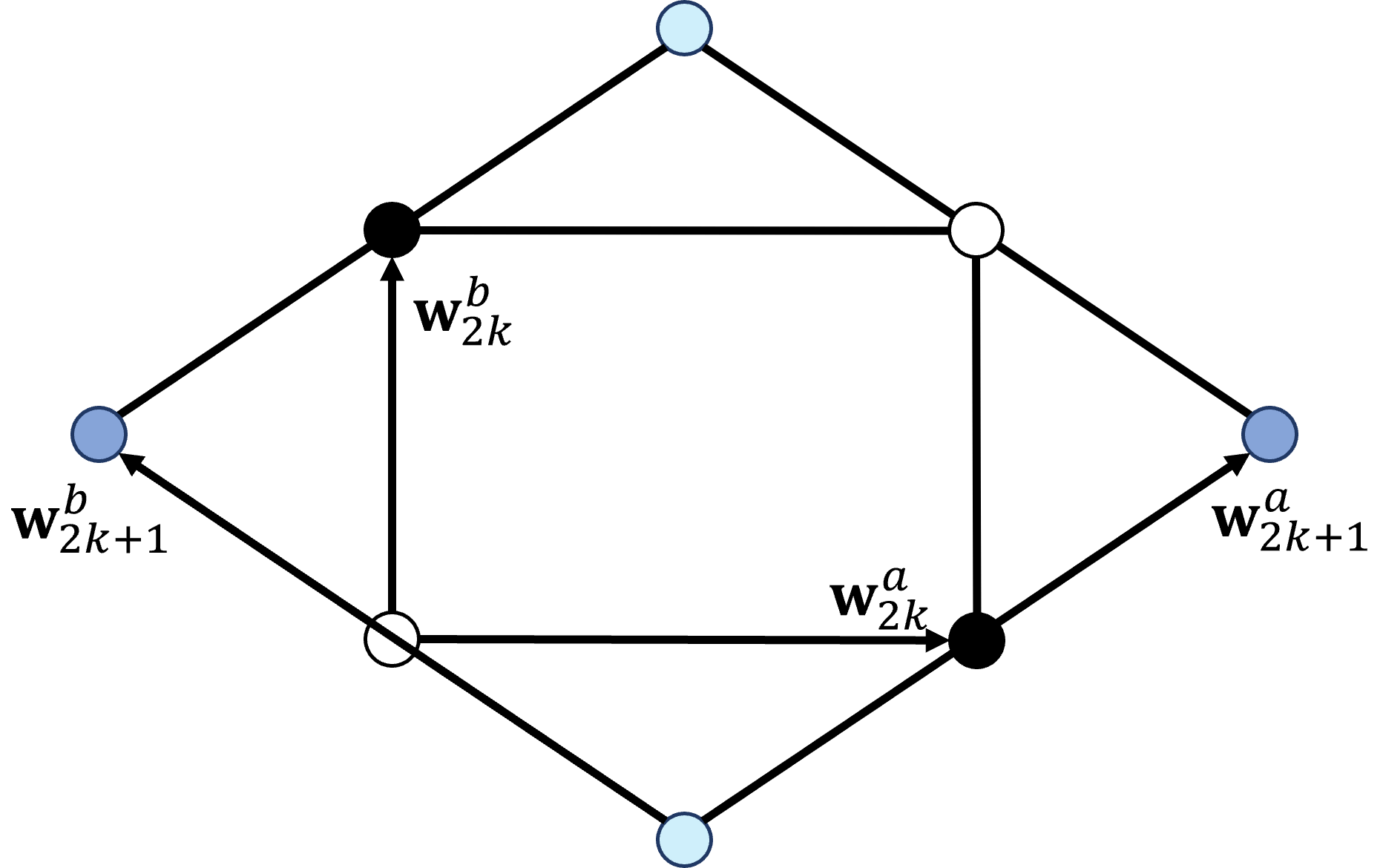}  
    \caption{ The relation of modular parameter $\tau$'s at successive RG iterations. We use light or dark blue circles to represent the tensors for $(2k+1)$-th RG iterations. Whereas for tensors at $(2k)$-th iteration, they are represented by white or black circles. We ignore some legs of the tensors to present the relation Eq.~(\ref{equ:tau_RG}) more clearly. }
  \label{fig:tau_2RG}
\end{figure}

Since at odd RG iterations, modular parameter $\tau$ contains non-zero real part and imaginary part, it is not straightforward to determine $\tau$. One way to circumvent the direct computation is to utilize the relation of $\tau$'s on neighbouring RG iterations:
\begin{equation}
\tau^{2k+1} = \frac{ \tau^{2k}-1 }{ \tau^{2k}+1 }
\label{equ:tau_RG}
\end{equation}
where the superscript in $\tau^{k}$ denotes the $k$-th RG iteration. The relation can be shown graphically, see Fig.~\ref{fig:tau_2RG}. The conformal spin can be directly computed by Eq.~(\ref{equ:conformal_spin}), once the modular parameter $\tau^{2k+1} $ is obtained.

Note that to compute conformal spin, we cannot set modular parameter $\tau$ to be exactly $i$ at even steps. Otherwise, $\tau$ will have no real part for all RG iterations, as can be seen in Eq.~(\ref{equ:tau_RG}). A practical method to compute conformal spin is to set $\tau \approx 0.9i$ at even RG iterations. So that in the next step, the modular parameter will have a non-zero real part, which is necessary to compute conformal spins. Another way to compute conformal spin without modifying $\tau$ is shown in Ref.\cite{Zcgu09}, which is more complex yet more general.

In the following, we will apply the method introduced in this section in the study of a non-Hermitian system, namely, the Yang-Lee edge singularity.

\subsection{A simple Example: Yang-Lee edge singularity}
\label{sec:conformal_YL}
% In Section~\ref{sec:corre_real}, we introduce a complex time $t = \left(1-i \epsilon  \right)t'$. Consequently, the evolution operator $e^{-iHt} $ is no longer Hermitian. 
Here we consider a simple example of non-Hermitian lattice model in (1+1)D, which is the quantum Ising chain with an imaginary longitudinal field:
\begin{equation}
H = - \sum_{i} ( \lambda \sigma^{x}_{i} \sigma^{x}_{i+1} + ih \sigma^{x}_{i} + \sigma^{z}_{i} )
\label{equ:YLquan}
\end{equation}

For a given $0 <\lambda < 1 $, there exist $h = \pm h_{c} $ (for $h_{c}>0$) such that the density of partition zero diverges. Such a singular behavior is called the Yang-Lee edge singularity, and the pair $(\lambda,h_{c})$ is called the Yang-Lee edge. Yang-Lee edge singularity is described by the non-unitary minimal model $\mathcal{M}_{5,2}$ in CFT, with central charge $c=-\frac{22}{5}$ and lowest scaling dimension $\Delta_{\mathrm{min}} = -\frac{2}{5} $.

\begin{figure}[h]
  \centering
  \includegraphics[width=\linewidth]{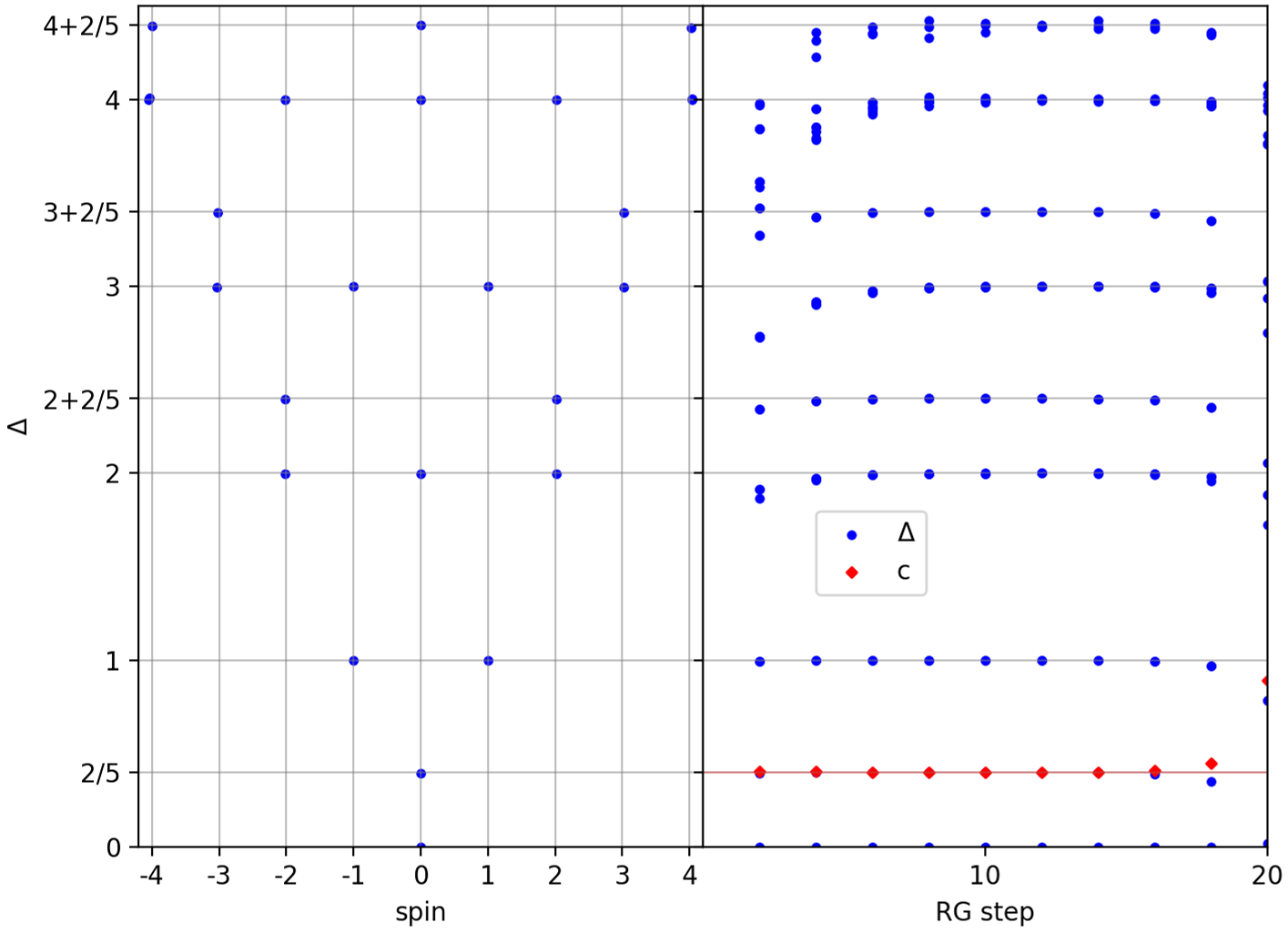}  
    \caption{ Conformal data of Eq.~(\ref{equ:YLquan}) with the parameters chosen as $ ( \lambda=0.4,h=0.23201634 ) $, obtained by Loop-TNR with $D_{cut}=32 $ and $L=4 $ transfer matrix. The gray and red lines mark the exact values for effective scaling dimensions and central charge, respectively. }
  \label{fig:lam04sdsp}
\end{figure}

Here we determine the positions of the Yang-Lee edge and compute conformal data by Loop-TNR. In Table~\ref{table:YLedge} we compare the critical pairs $(\lambda,h_{c})$ determined by Loop-TNR and other methods. This result is consistent with the previous studies. 
\begin{table}[h]
  \caption{ Positions of Yang-Lee edge $(\lambda,h_{c})$ determined by Loop-TNR and other methods. The first row displays several methods that are to be compared. Different $\lambda$'s are listed in the first column. The remaining columns are filled with positive $h_{c}$'s determined under different $\lambda$'s and methods. }
  \begin{ruledtabular}
  \begin{tabular}{ l| l | l | l  } 
     & MPRG \cite{yamada22} & Finite-size scaling \cite{Gehlen91} & Loop-TNR \\ 
    \hline
    $0.1$ & 0.636 & 0.636640 & 0.636638123 \\ 
    \hline
    $0.2$ & 0.457 & 0.458498 & 0.45849517 \\ 
    \hline
    $0.3$ & 0.328 & 0.330031 & 0.3300305 \\ 
    \hline
    $0.4$ & 0.230 & 0.23202 & 0.23201634 \\ 
    \hline
    $0.5$ & 0.154 & 0.15620 & 0.15620181 \\ 
    \hline
    $0.6$ & 0.095 & 0.09807 & 0.09806663 \\ 
    \hline
    $0.7$ & 0.052 & 0.05483 & 0.05482308  \\ 
    \hline
    $0.8$ & 0.021 & 0.02468 & 0.02467332 \\ 
    \hline
    $0.9$ & - & 0.0065 & 0.00649513 \\ 
  \end{tabular}
  \end{ruledtabular}
  \label{table:YLedge}
\end{table}
In addition, Fig.~\ref{fig:lam04sdsp} shows the conformal data obtained by Loop-TNR, with $\lambda=0.4$ chosen as an example. By using the Loop-TNR method, we are able to obtain accurate and stable conformal data with a moderate bond dimension $D_{cut}=32$.   

Note that, in our computations, we shift the ground state energy such that the lowest scaling dimension becomes 0 again, which corresponds to defining effective central charge and effective scaling dimensions \cite{Itzykson86}
\begin{equation}
\begin{aligned}
c_{\mathrm{eff}} =& c - 12 \Delta_{\mathrm{min}} \\
\Delta_{n,\mathrm{eff}} =& \Delta_{n} - \Delta_{\mathrm{min}}
\end{aligned}
\label{equ:YLeff}
\end{equation}
As a result, we have $c_{\mathrm{eff}} = \frac{2}{5} $ and $ \Delta_{\mathrm{eff}}=0, \frac{2}{5},... $ 

\begin{figure}[h]
  % \centering
  \includegraphics[width=\linewidth]{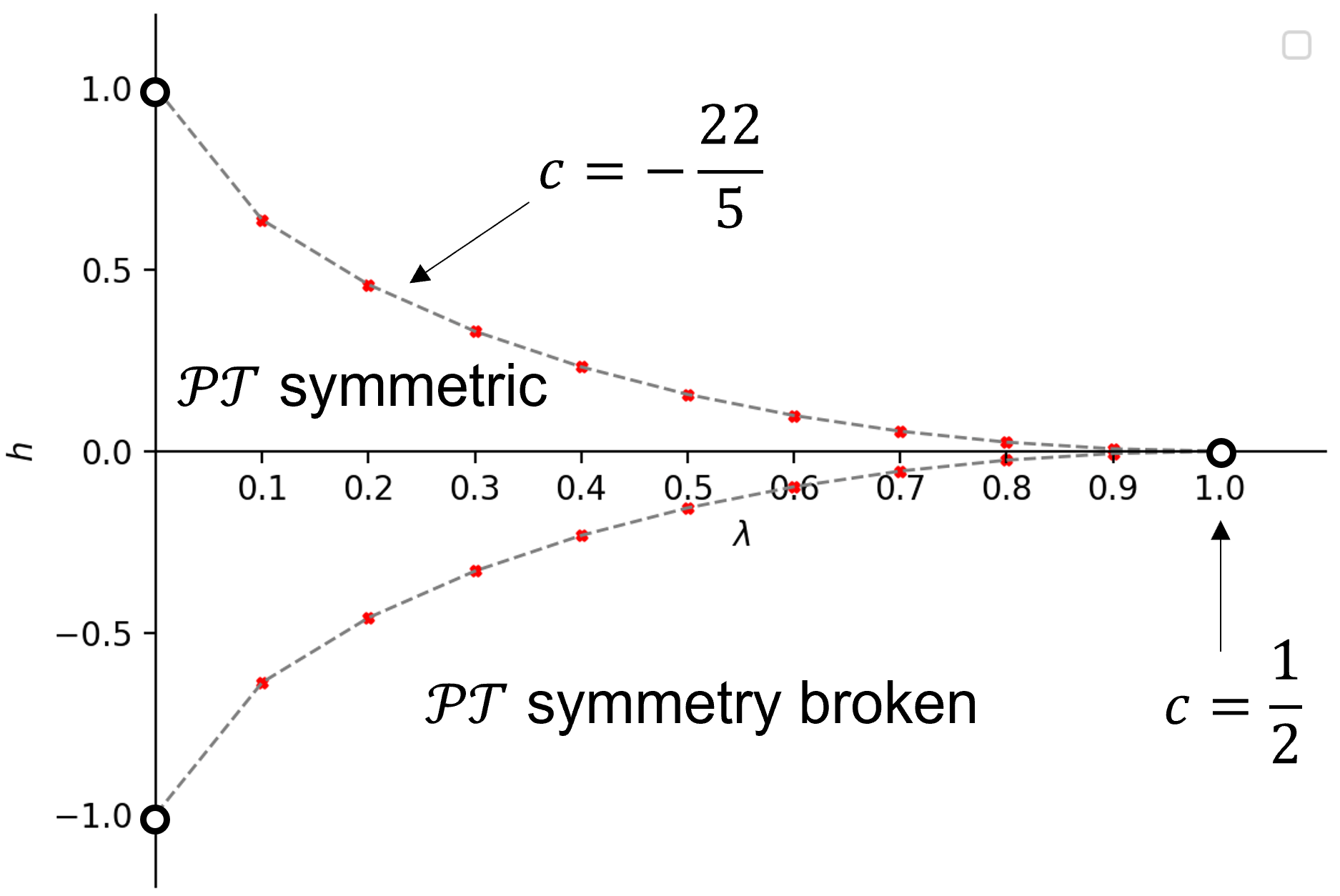}  
    \caption{ The phase diagram of the model Eq.~(\ref{equ:YLquan}). The gray dashed lines represent the phase boundary. The red points represent the positions of the Yang-Lee edge determined by the Loop-TNR method, as listed in Table~\ref{table:YLedge}. The central charge for the edge points is $-\frac{22}{5} $. At $\lambda \rightarrow 1$, the universality class reduces to Ising type, with central charge $c=0.5$. For the area enclosed by edge singularities and $h$-axis, the system lies in $\mathcal{P}\mathcal{T}$ symmetric state, with real energy spectrum. While outside the area, the $\mathcal{P}\mathcal{T}$ symmetry is broken, and the energy spectrum can take complex values.}
  \label{fig:YLq_phase}
\end{figure}

Based on the $( \lambda,h_{c} )$ obtained by Loop-TNR, we can plot the phase diagram of the model Eq.~(\ref{equ:YLquan}), as shown in Fig.~\ref{fig:YLq_phase}. Note that $( \lambda,\pm h_{c} )$ are both Yang-Lee edge points.
The phase diagram is consistent with the result introduced in Ref.\cite{Gehlen94}.

% % Conclusion & Acknowledgements % the version polished by chat-gpt 3.5
\section{Conclusion and discussion}
\label{sec:Conclusion}
In this paper, we systematically employ the Loop-TNR algorithm to compute correlation functions. Our approach enables the computation of not only spatial correlation functions with arbitrary integer separations but also temporal correlations with fractional-time differences.
Furthermore, we computed the real-time correlation functions in the path-integral formalism.
%The tensor network representation constructed from the modified path-integral formalism can be computed not only by TRG-based methods, but also by MPS-based methods, such as VUMPS\cite{Fishman18,Li20}, iTEBD\cite{Orus08} and CTM-methods\cite{Orus09}. 
The present method can be regarded as an attempt to circumvent the problem of the entanglement barrier. Although there are still some small deviations compared to the exact results, we believe that these errors mainly originate from the initialization step, where iMPS-based algorithms are used. Thus, it is very important to implement the entanglement filtering process at the very beginning, and we will leave this problem in our future work.

Moreover, we present a detailed method for computing conformal data from the fixed-point tensors for non-hermitian systems. As a simple example, we apply this method to investigate Yang-Lee edge singularity. Loop-TNR produces extremely accurate conformal data, highlighting its potential for applications in systems that cannot be accessed by DMRG/MPS based methods. 

\begin{acknowledgments}
 This work is supported by funding from Hong Kong’s Research Grants Council (GRF no.14301219, GRF no. 14303722, CRF C7012-21GF) and Direct Grant no. 4053578 from The Chinese University of Hong Kong.
\end{acknowledgments}

% appendix
\appendix
\section{Details of compression steps}
\label{sec:comp}

\begin{figure}[h]
  \centering
  \includegraphics[width=\linewidth]{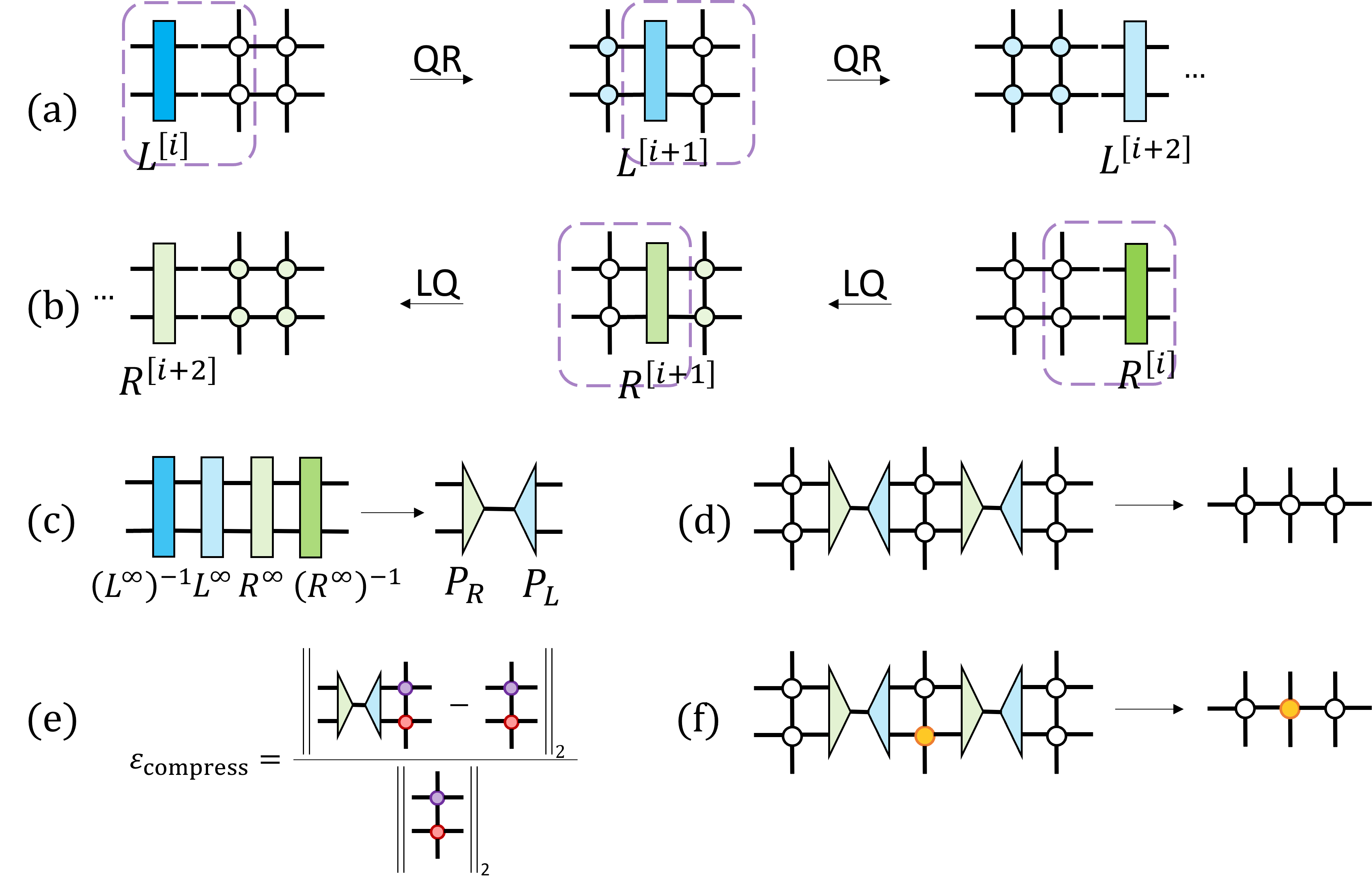}
    \caption{ Details in the compression step. (a) $L^{[i]} $ is contracted with the two-layer tensor, as marked by dashed rounded corners. QR decomposition is performed on the resultant tensor, where the upper triangular matrix is the updated $L^{[i+1]}$. The same operations are performed iteratively until the $L$ matrix is converged, which we denote as $L^{[\infty]} $. (b) $R^{[i]} $ is contracted with the two-layer tensor, as marked by dashed rounded corners. LQ decomposition is performed on the resultant tensor. The lower triangular matrix is the updated $R^{[i+1]} $. Such operations are performed until $R$ is converged, which is denoted as $R^{[\infty]} $. (c) Projectors are obtained according to Eq.~(\ref{equ:projs}), which is approximately an identity. (d) Two layers of tensors are compressed into one when the projectors are contracted. (e) The relative error in the compression step. (f) For two-layer tensors including impurity tensors, we just use the projectors obtained above to compress.  }
  \label{fig:compress}
\end{figure}

In this section, we introduce the algorithm to compress two layers of tensors into one. The basic idea of the compression step is to insert projectors between the two-layer tensors (see Fig.~\ref{fig:compress}(d) and (f)). One compression step is finished when the two-layer tensor is contracted with the projectors, where truncation in the bond dimension is involved. 
In the following, we illustrate the method to obtain the projectors that aims to minimize the truncation error shown in Fig.~\ref{fig:compress}(e).

Such projectors are obtained from the $L$ and $R$ matrices that are converged under successive QR or LQ decompositions. We start with an initial matrix $L^{[i=0] } = \mathbb{I} $. The matrix is contracted with the two-layer tensor and a QR decomposition is performed correspondingly. The upper triangular matrix is the updated $L^{[i=1]}$. The new $L^{[1]}$ matrix is to be contracted with the two-layer tensor to the right, followed by a QR decomposition. This completes another iteration. Such operations are iterated until we reach a converged $L^{[\infty]}$, which corresponds to performing QR decomposition along an infinite chain. The procedure is shown graphically in Fig.~\ref{fig:compress}(a).

For the $R^{[\infty]}$ matrix, similar operations are performed except we now need to start from the right side and perform LQ decompositions, see Fig.~\ref{fig:compress}(b). 
% \FloatBarrier

The projectors are constructed from an identity matrix, which follows a similar approach in Ref.\cite{wang11}:
\begin{equation}
\begin{aligned} % try aligned
\mathbb{I} = \left( L^{[\infty]} \right)^{-1} \cdot L^{[\infty]} \cdot R^{[\infty]} \cdot \left( R^{[\infty]} \right)^{-1}
\label{equ:ident}
\end{aligned}
\end{equation}
Then we perform SVD on $ L^{[\infty]} \cdot R^{[\infty]} = U \Sigma V^{\dagger} $, where $U$ and $V^{\dagger}$ are unitary matrices and $\Sigma$ is diagonal. The inversion of matrices in Eq.~(\ref{equ:ident}) can be avoided:
\begin{equation}
\begin{aligned}
\left( L^{[\infty]} \right)^{-1} &=  R^{[\infty]} \cdot V \cdot \left( \Sigma \right)^{-1} \cdot U^{\dagger} \\
\left( R^{[\infty]} \right)^{-1} &=  V \cdot \left( \Sigma \right)^{-1} \cdot U^{\dagger} \cdot L^{[\infty]}
\label{equ:inv_of_LR}
\end{aligned} 
\end{equation} 

Hence the identity can be rewritten as:
\begin{equation}
\mathbb{I} = R^{[\infty]} \cdot V \cdot \frac{1}{\Sigma }  \cdot U^{\dagger} \cdot L^{[\infty]}
\label{equ:ident2}
\end{equation}

From above we can define the projectors $P_{R} $ and $P_{L} $ as follows:
\begin{equation}
\begin{aligned}
P_{R} &= R^{[\infty]} \cdot V \cdot \frac{1}{\sqrt{ \Sigma } } \\
P_{L} &= \frac{1}{\sqrt{ \Sigma } } \cdot U^{\dagger} \cdot L^{[\infty]}
\label{equ:projs}
\end{aligned} 
\end{equation}
Note that here we should make truncations by keeping the largest $D_{cut}$ singular values in $\Sigma$. Otherwise, the bond dimension in horizontal directions will increase exponentially as more compression steps are performed.

\section{Computation of correlations with VUMPS}
\label{sec:App_VUMPS}
Variational Uniform Matrix Product State (VUMPS) \cite{Stauber18} is an efficient tool to find the ground state of a (1+1)D Hamiltonian with infinite length. It is also useful in the contraction of a 2D tensor network in the thermodynamic limit. For example, we can apply VUMPS to compute the partition function of a (1+1)D system.

In this part, we review the algorithm of VUMPS for tracing out a 2D tensor network \cite{Fishman18,Li20} and apply it to compute correlation functions.

\subsection{Obtain the fixed-point MPS tensors}
\begin{figure}[htb]
  \centering
  \includegraphics[width=0.8\linewidth]{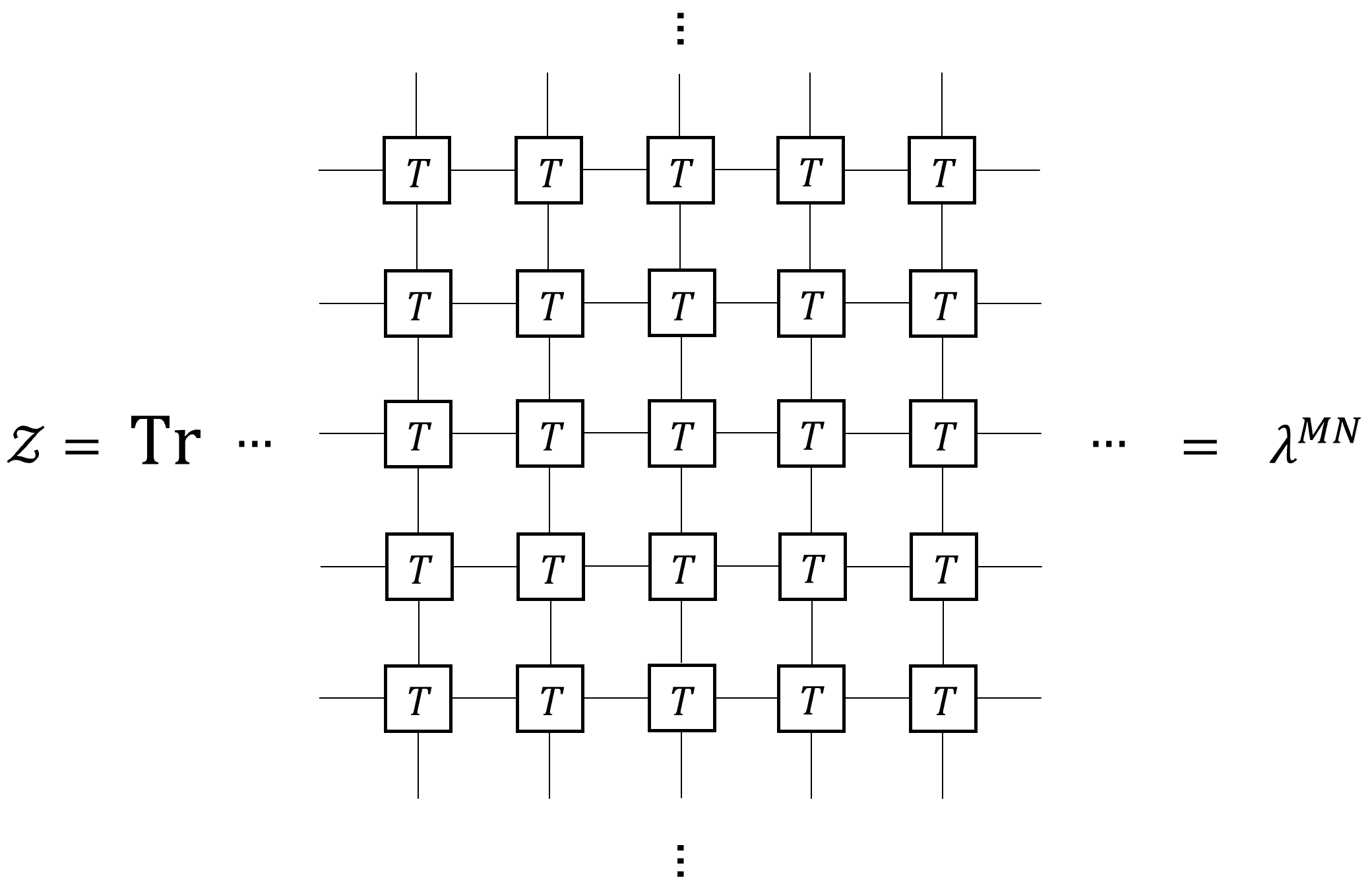}
    \caption{ The tensor network representation of a partition function with M-by-N tensors. $\lambda$ is defined to be the partition function per site, which is related to the free energy per site. Uniform tensors are represented by squares marked by $T$. }
  \label{fig:VUMPS_TNrep}
\end{figure}

Suppose we want to compute the partition function of a (1+1)D system. The tensor network representation can be seen in the denominator of Fig.~\ref{fig:sin_op}(c). Here we replot the tensor network in Fig.~\ref{fig:VUMPS_TNrep}. Note that in Fig.~\ref{fig:VUMPS_TNrep}, the size of the tensor network is M-by-N where M and N are assumed to be infinite. $\lambda$ corresponds to the partition function per site.
% \FloatBarrier

\begin{figure}[htb]
  \centering
  \includegraphics[width=\linewidth]{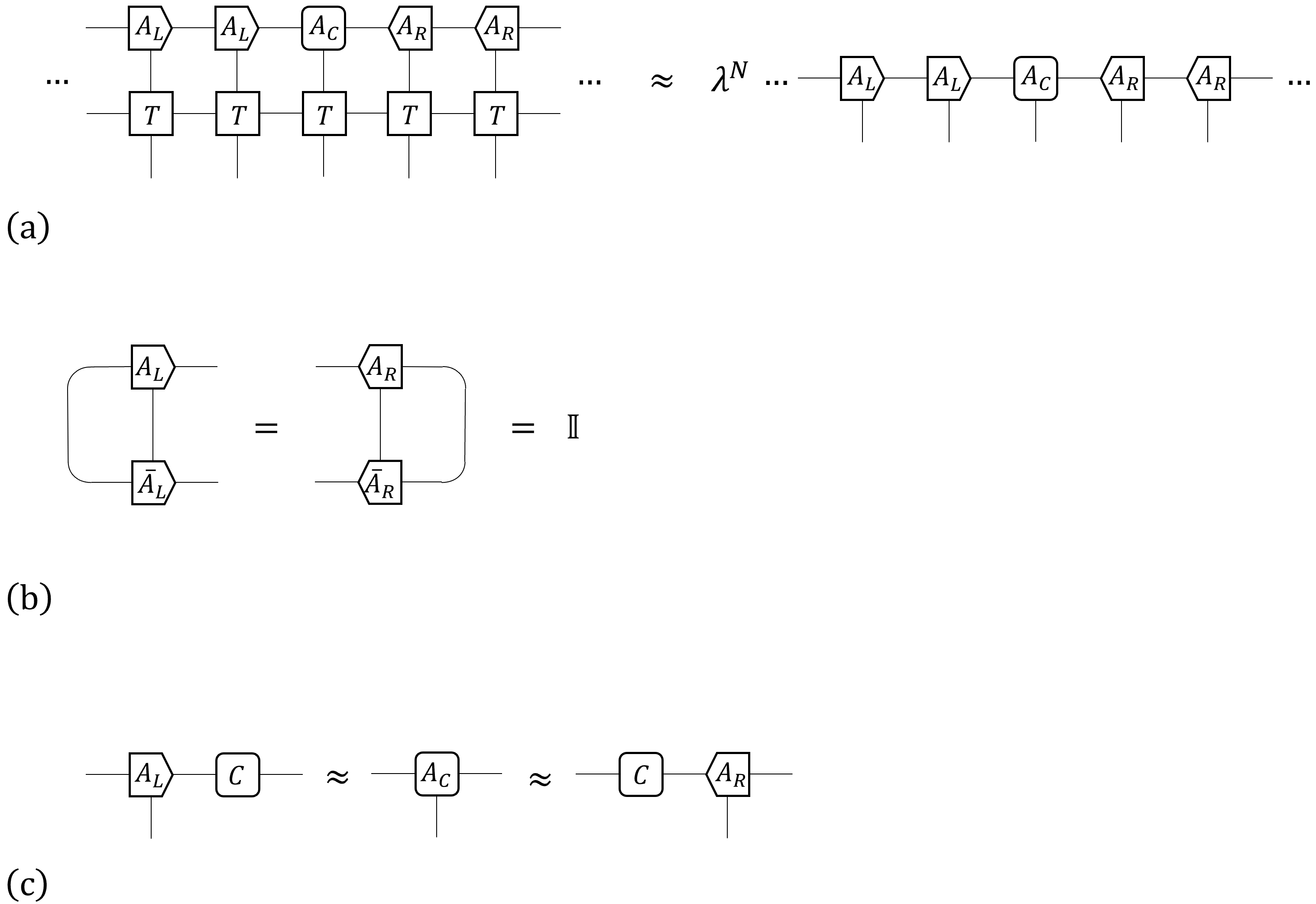}
    \caption{ MPS in VUMPS. (a) The leading eigenvector of the row-to-row transfer matrix is represented by an MPS in the mixed canonical form, which consists of $A_{L}$, $A_{R} $, and $A_{C} $ tensors. (b) Unitarity of $A_{L} $ and $A_{R} $ as isometric tensors. (c) Relations of tensor $A_{C} $ to $A_{L} $, $A_{R} $ and $C$ when approaching to the fixed point.  }
  \label{fig:VUMPS_tensors}
\end{figure}

For boundary MPS methods, the contraction of such tensor network corresponds to finding the leading eigenvectors of the row-to-row transfer matrix (Fig.~\ref{fig:VUMPS_tensors}(a)) \cite{Fishman18}. Fixed-point tensors in the MPS correspond to contracting infinite rows of tensors. In Fig.~\ref{fig:VUMPS_tensors}, the leading eigenvectors are represented in terms of an MPS, constructed in the mixed canonical form. $A_{L} $ and $A_{R} $ are isometric tensors. $C$ is a diagonal matrix that stores the singular values of the MPS. See Fig.~\ref{fig:VUMPS_tensors}(b) and (c) for the properties of $A_{L} $, $A_{R} $ tensors and their relations to $C $ via $A_{C} $.
\FloatBarrier % maybe commented

The fixed-point MPS tensors can be found by VUMPS algorithm, which is stated as follows:
\begin{enumerate}
  \item Solve the left and right environment tensors $E_{L} $ and $E_{R} $ (Fig.~\ref{fig:VUMPS_Elr}):
  \begin{figure}[htb]
    \centering
    \includegraphics[width=\linewidth]{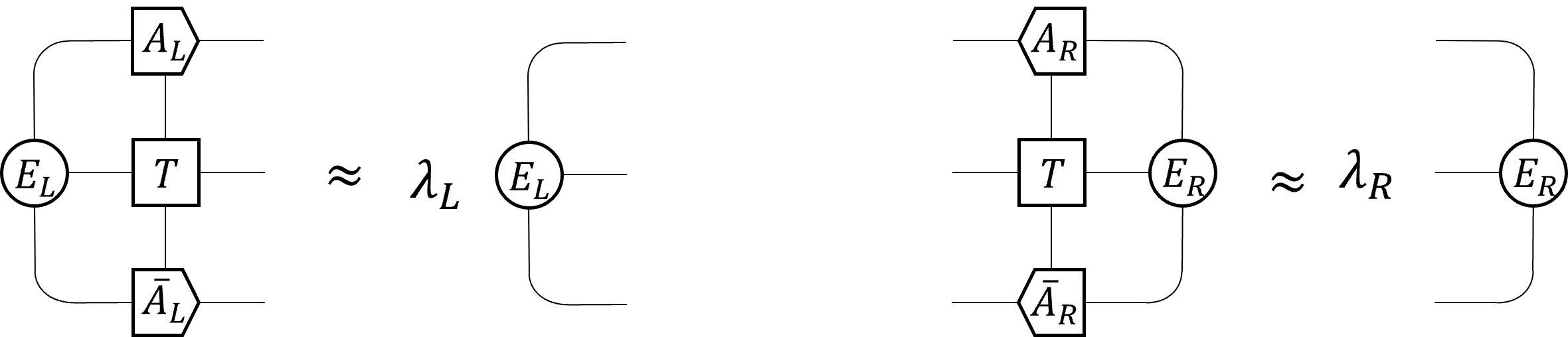}
      \caption{ Solve the environment tensors $E_{L} $ or $E_{R} $ as the leading eigenvector of the matrix constructed from $A_{L} $, $T$ and $\bar{A}_{L} $, or the matrix from $A_{R} $, $T$ and $\bar{A}_{R} $. The $E_{L} $ and $E_{R} $ can be solved efficiently by the Arnoldi method. The 'leading eigenvector' corresponds to the eigenvector of which the eigenvalue has the largest magnitude. }
    \label{fig:VUMPS_Elr}
  \end{figure}
  \FloatBarrier

  \item Find the central tensors $A_{C} $ and $C$ (Fig.~\ref{fig:VUMPS_AcC}):
  \begin{figure}[htb]
    \centering
    \includegraphics[width=\linewidth]{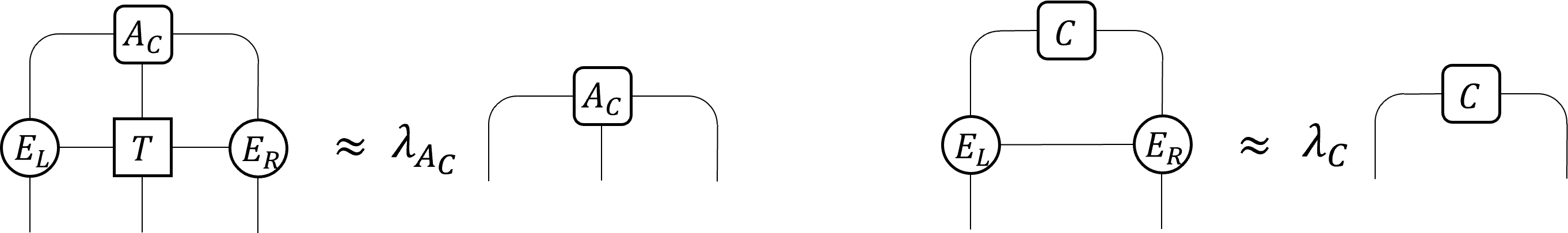}
      \caption{ Solve the central tensors $A_{C} $ and $C$ as the leading eigenvectors, by Arnoldi method. }
    \label{fig:VUMPS_AcC}
  \end{figure}
  \FloatBarrier
  where we have $\lambda_{A_{C} } / \lambda_{C} \approx \lambda_{L} \approx \lambda_{R} $ near the fixed point. 

  \item Update the isometric tensors $A_{L} $ and $A_{R} $ from $A_{C} $ and $C$. This can be done by various methods, such as QR decomposition, SVD, and polar decomposition. Here we introduce the way to update $A_{L} $ and $A_{R} $ by QR decomposition (Fig.~\ref{fig:VUMPS_ALR}):
  \begin{figure}[htb]
    \centering
    \includegraphics[width=\linewidth]{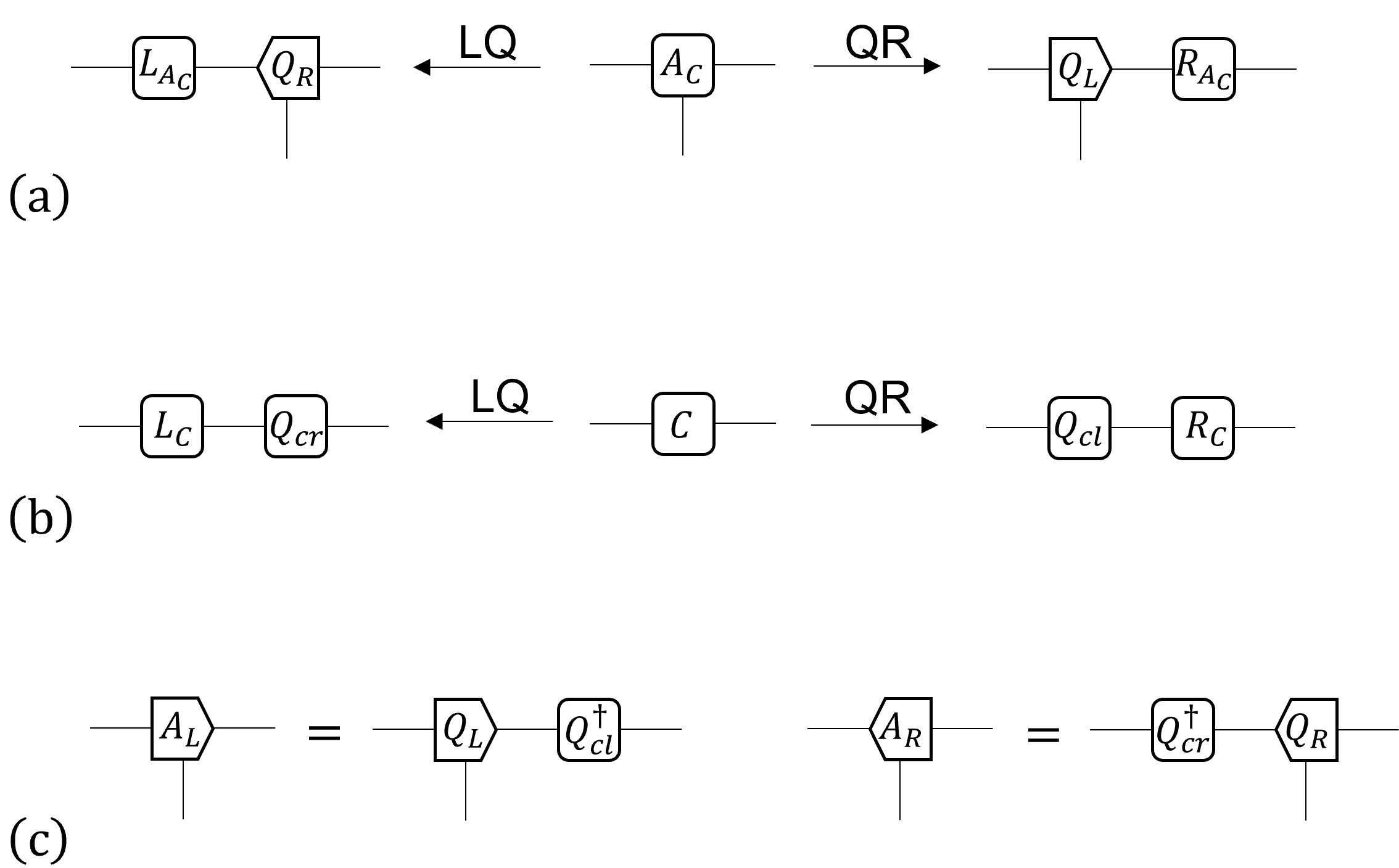}
      \caption{ The update scheme for $A_{L} $ and $A_{R} $ tensors. (a) QR and LQ decompositions for $A_{C} $ tensor. (b) QR and LQ decompositions for $C$. (c) Update $A_{L} $ and $A_{R} $ by the unitary matrices obtained in previous steps.  }
    \label{fig:VUMPS_ALR}
  \end{figure}
  \FloatBarrier
  Note that we should ensure the uniqueness of QR or LQ decomposition, such that the diagonal elements in $L_{A_{C} } $, $R_{A_{C} } $, $L_{C} $ and $R_{C} $ are all non-negative (for complex elements, one choice is to make the real part to be non-negative).

  \item Repeat steps 1 to 3 until the MPS reaches its fixed point, where the relations in Fig.~\ref{fig:VUMPS_tensors}(c) should be satisfied.

\end{enumerate}

Note that we don't require the rank-4 tensors to be isotropic, as did in TRG or Loop-TNR computations. Therefore we can just use the tensor obtained by one TRG step on the two-body gates (see Fig.~\ref{fig:TRG_comp}(c)) without further compression, which is free of the compression error.

Once the fixed-point MPS tensors are found, we can use them to compute physical quantities. In the following, we will introduce the application in computing two-point correlations.

\subsection{Computation of correlations}
Having obtained the fixed-point tensors $ \{ A_{L}, A_{R}, A_{C}, C, E_{L}, E_{R} \}$, we now apply them to compute correlation functions in spatial (horizontal) and temporal (vertical) directions, respectively.

\begin{figure}[htb]
  \centering
  \includegraphics[width=\linewidth]{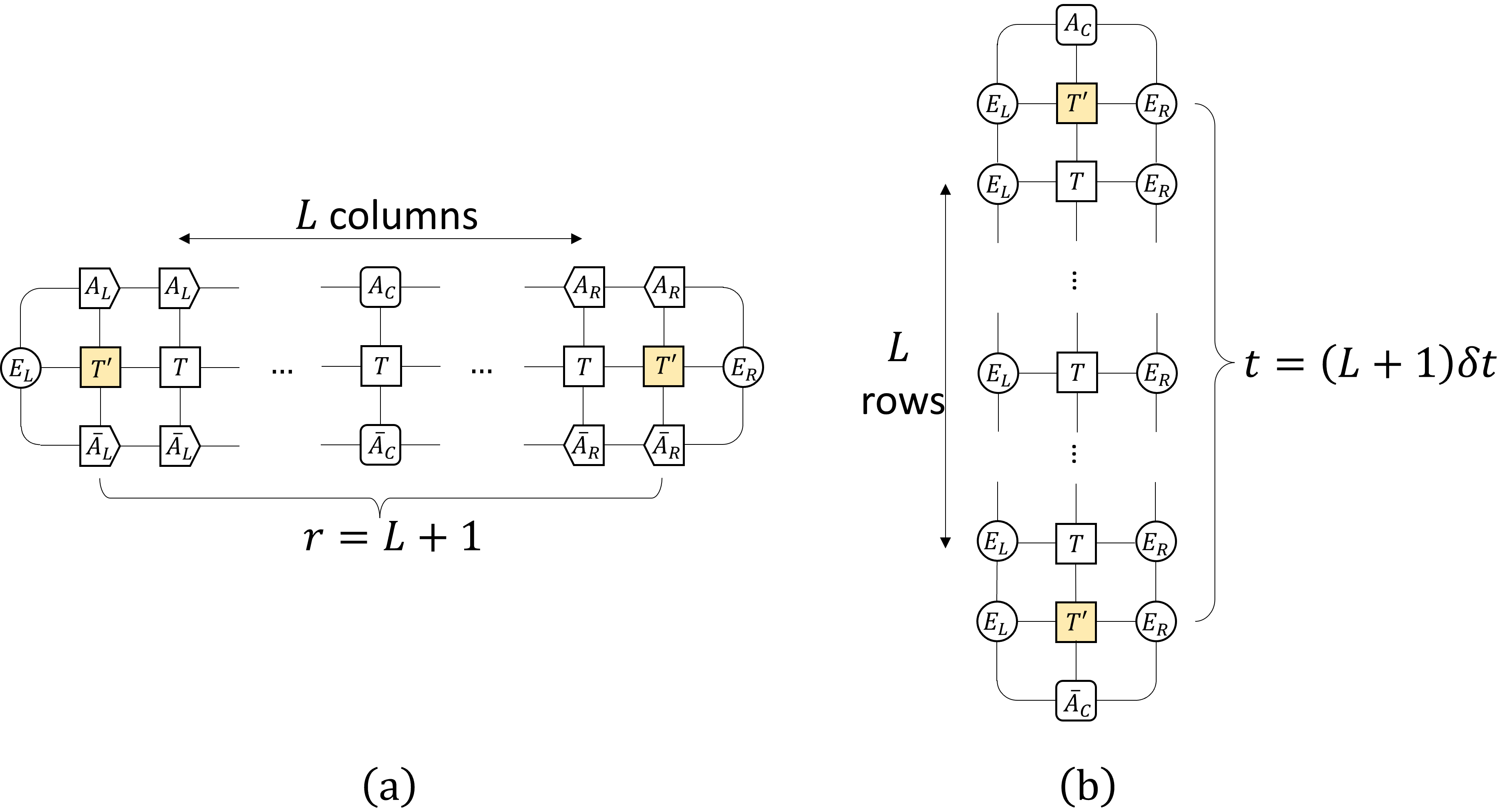}
    \caption{ The computation of the numerator of the correlations in (a) spatial and (b) temporal directions, respectively. $L$ denotes the number of columns/layers of tensors between the impurity tensors. }
  \label{fig:VUMPS_corre}
\end{figure}

As shown in Fig.~\ref{fig:VUMPS_corre}, if the number of columns of tensors between the two impurity tensors is $L$, the spatial separation for the impurity tensors is simply $r=L+1$, in the horizontal direction. Similarly in vertical directions, if $L$ denotes the number of rows of tensors between the impurity tensors, the temporal difference is $t= (L+1)\delta t $, since the time difference for neighbouring rows is $\delta t$. Direct contraction of the tensor networks in Fig.~\ref{fig:VUMPS_corre} gives the numerator of the correlations. For the denominator, we just replace the impurity tensors with the uniform ones.
\FloatBarrier

% % The \nocite command causes all entries in a bibliography to be printed out whether or not they are actually referenced in the text. 
% % This is appropriate for the sample file to show the different styles of references, but authors most likely will not want to use it.
% \nocite{*}

\bibliography{main}

\end{document}